\documentclass[preprint]{aastex}

\received{}
\accepted{}

\slugcomment{11 February 2002 Revision, accepted by \aj}

\shortauthors{Richards et al.}
\shorttitle{SDSS Quasar Target Selection}

\title{Spectroscopic Target Selection in the Sloan Digital Sky Survey:
The Quasar Sample}

\author{
Gordon T. Richards\altaffilmark{1},
Xiaohui Fan\altaffilmark{2},
Heidi Jo Newberg\altaffilmark{3},
Michael A. Strauss\altaffilmark{4},
Daniel E. Vanden Berk\altaffilmark{5},
Donald P. Schneider\altaffilmark{1},
Brian Yanny\altaffilmark{5},
Adam Boucher\altaffilmark{3},
Scott Burles\altaffilmark{5,6},
Joshua A. Frieman\altaffilmark{5,6},
James E. Gunn\altaffilmark{4},
Patrick B. Hall\altaffilmark{4,7},
\v{Z}eljko Ivezi\'{c}\altaffilmark{4},
Stephen Kent\altaffilmark{5,6},
Jon Loveday\altaffilmark{8},
Robert H. Lupton\altaffilmark{4},
Constance M. Rockosi\altaffilmark{6},
David J. Schlegel\altaffilmark{4},
Chris Stoughton\altaffilmark{5},
Mark SubbaRao\altaffilmark{6}, and 
Donald G. York\altaffilmark{6,9}
}

\altaffiltext{1}{Department of Astronomy and Astrophysics, The Pennsylvania State University, University Park, PA 16802}
\altaffiltext{2}{Institute for Advanced Study, Olden Lane, Princeton, NJ 08540}
\altaffiltext{3}{Physics Department, Rensselaer Polytechnic Institute, SC1C25, Troy, NY 12180}
\altaffiltext{4}{Princeton University Observatory, Princeton, NJ 08544}
\altaffiltext{5}{Fermi National Accelerator Laboratory, P.O. Box 500, Batavia, IL 60510}
\altaffiltext{6}{The University of Chicago, Department of Astronomy and Astrophysics, 5640 S. Ellis Ave., Chicago, IL 60637}
\altaffiltext{7}{Pontificia Universidad Cat\'{o}lica de Chile, Departamento de Astronom\'{\i}a y Astrof\'{\i}sica, Facultad de F\'{\i}sica, Casilla 306, Santiago 22, Chile}
\altaffiltext{8}{Astronomy Centre, University of Sussex, Falmer, Brighton BN1 9QJ, United Kingdom}
\altaffiltext{9}{The University of Chicago, Enrico Fermi Institute, 5640 S. Ellis Ave., Chicago, IL 60637}

\begin{document}

\begin{abstract}

We describe the algorithm for selecting quasar candidates for optical
spectroscopy in the Sloan Digital Sky Survey.  Quasar candidates are
selected via their non-stellar colors in $ugriz$ broad-band
photometry, and by matching unresolved sources to the FIRST radio
catalogs.  The automated algorithm is sensitive to quasars at all
redshifts lower than $z\sim5.8$.  Extended sources are also targeted
as low-redshift quasar candidates in order to investigate the
evolution of Active Galactic Nuclei (AGN) at the faint end of the
luminosity function.  Nearly 95\% of previously known quasars are
recovered (based on 1540 quasars in 446 square degrees).  The overall
completeness, estimated from simulated quasars, is expected to be over
90\%, whereas the overall efficiency (quasars:quasar candidates) is
better than 65\%.  The selection algorithm targets ultraviolet excess
quasars to $i^* = 19.1$ and higher-redshift ($z\gtrsim3$) quasars to
$i^*=20.2$, yielding approximately 18 candidates per square degree.
In addition to selecting ``normal'' quasars, the design of the
algorithm makes it sensitive to atypical AGN such as Broad Absorption
Line quasars and heavily reddened quasars.

\end{abstract}

\keywords{quasars: general --- surveys}

\section{Introduction}

The Sloan Digital Sky Survey (SDSS;\citealt{yor+00}) will digitally
map 10,000 square degrees of the Northern Galactic Cap (hereafter, the
Main Survey), and a smaller area ($\sim750\,\deg^2$) in the
Southern Galactic Cap (hereafter, the Southern Survey).  The imaging
survey is done in five broad bands, $ugriz$ \citep{fig+96,sto+01},
that were specially designed for the survey.  In addition to the
imaging data produced by a large CCD mosaic camera \citep{gcr+98}, the
SDSS will conduct a spectroscopic survey of objects selected from the
catalogs derived from the processed images with the goal of obtaining
spectra of approximately one million galaxies and one hundred thousand
quasars.  A detailed description of the overall target selection
algorithm for all classes of spectroscopic targets can be found in
\citet{van+02}; in this paper we describe the spectroscopic target
selection algorithm for the SDSS Quasar Survey.

The primary SDSS quasar science goals are defined by the two ``Key
Projects'' that will be addressed with the final quasar sample.  The
Key Projects are 1) the evolution of the quasar luminosity function
and 2) the spatial clustering of quasars as a function of redshift.
These studies require the assembly of a large sample of quasars
covering a broad range of redshift and chosen with well-defined,
uniform selection criteria.

The SDSS Quasar Survey will increase the number of known quasars by a
factor of 100 over previous surveys such as the Large Bright Quasar
Survey (LBQS; \citealt{hfc95}), which was until recently the largest
complete quasar survey.  In so doing, the SDSS Quasar Survey will also
be approximately four times the size of the concurrent 2dF QSO
Redshift Survey \citep{csb+01}.  The SDSS quasar sample has a high
completeness fraction from $z = 0$ to $z \approx 5.8$; the sample also
has high-quality five-color photometry and high signal-to-noise ratio
spectroscopy at moderate spectral resolution.

The selection of quasars from multi-color imaging data was pioneered
by \citet{sw65} and has continued through the years
\citep[e.g.,][]{kk82,sg83,who91,hfc95,hog+96,csb+01}; a review of the
history of surveys for high-redshift quasars is given by \citet{wh90}.
The approach we adopt when selecting quasars from the imaging data of
the Sloan Digital Sky Survey is similar to previous studies.  However,
certain characteristics of the SDSS (such as the novel filter system,
matching to radio sources, etc., along with the quality and quantity
of the data) make the selection of SDSS quasars unique.  These
features of the Survey require a detailed description of the algorithm
that selects quasar candidates for follow-up spectroscopy, hereafter
the SDSS quasar target selection algorithm.

Our quasar target selection algorithm seeks to explore all the regions
of color space that quasars are known to occupy, avoid most regions of
color space where the quasar density is much lower than the density of
contaminants, and to explore relatively uncharted regions of color
space to the extent possible given the available number of
spectroscopic fibers.  As a result of our exploration of less
populated regions of color space, our algorithm is open to
serendipitous discovery of quasars with unusual colors and thus
unusual properties.  Our formal science requirements are to recover
90\% or more of previously known quasars, while maintaining an
efficiency in excess of 65\%, where the efficiency is given by the
ratio of true quasars to quasar candidates.  The balance between
completeness and efficiency is a delicate one; with so many more stars
than quasars in the data, improvements in efficiency by rejecting
objects in regions of color space in which both stars and quasars lie
will necessarily cut back on completeness.  In addition, unlike many
quasar surveys, it is important to realize that the SDSS quasar target
selection algorithm was finalized in advance of collecting most of the
imaging data.  The algorithm also was required to be completely
automated and operate on a single object at a time, independently of
all other objects.

The wavelength coverage of the SDSS filters allows for the selection
of quasars from $z=0$ to beyond a redshift of 6.  The automated
algorithm described herein requires that there be flux in at least two
bands, which imposes an upper limit to the redshift of $z\lesssim5.8$
(based on the minimum transmission between the $i$ and $z$ filter
curves at $8280\,{\rm \AA}$ straddled by Lyman-$\alpha$ emission).  It
is possible to use SDSS imaging data to discover quasars with even
larger redshifts by investigating objects that are detected in the $z$
filter only.  However, such a search is beyond the capability of the
survey proper as the efficiency of the automated selection for objects
detected only in the $z$-band is much too low; searches for the very
highest redshift quasars require spectroscopy outside of normal SDSS
operations \citep[e.g.,][]{fss+01}.

At the low-redshift end, the design of the $u$ filter and the location
of the gap between the $u$ and $g$ filters were chosen to emphasize
the difference between objects with power-law spectral energy
distributions (SEDs), such as quasars at $z < 2.2$, and objects that
are strongly affected by the Balmer decrement, particularly A stars,
which are historically the prime contaminants in multi-color surveys
for low-redshift quasars.  The filter curves are described in
\citet{fig+96} with modifications as described by \citet{sto+01}.

Briefly, the quasar target selection code works as follows.  1)
Objects with spurious and/or problematic fluxes in the imaging data
are rejected.  2) Point source matches to FIRST radio sources are
preferentially targeted without reference to their colors.  3) The
sources remaining after the first step are compared to the
distribution of normal stars and galaxies in two distinct 3D color
spaces, one nominally for low-redshift quasar candidates (based on the
$ugri$ colors), and one nominally for high redshift quasar candidates
(based on the $griz$ colors).  Stars in particular follow a
one-dimensional locus in the four-dimensional SDSS color space, which
we model explicitly (and keep fixed for the duration of the Survey).
Those objects that are discernible outliers from the regions of color
space populated by stars and non-active galaxies are selected for
spectroscopic follow-up if they meet all of the other criteria,
including the magnitude cuts.  Objects meeting any of the selection
requirements have one or more target flags set, see
Table~\ref{tab:tab1}.

During the color selection process we draw no specific line between
quasars and their less luminous cousins, Seyfert galaxy nuclei;
objects which have the colors of low-redshift AGN are targeted even if
they are resolved.  This policy is in contrast to some other surveys
for quasars; it is not unusual for extended objects to be rejected,
which has the effect of imposing a lower limit to the redshift
distribution of the survey.  Throughout this paper, we will, in fact,
use the word ``quasar'' where we often mean ``AGN''.  The complete
SDSS data set is needed to ultimately determine if the traditional
distinction between quasars and Seyfert galaxy nuclei is warranted.

Once quasar candidates have been identified, we obtain spectra of each
candidate using the SDSS fiber-fed spectrographs.  Each $3^{\circ}$
diameter SDSS spectroscopic plate holds 640 fibers, of which an
average of 80 are assigned to quasar candidates.  Quasar candidates
are allocated approximately 18 fibers per square degree.  Plate
overlaps account for the difference between 18 objects per square
degree and 80 objects per plate (which have an area of $\sim 7$ square
degrees).  Given this constraint on density and the requirements on
completeness and efficiency, we target color-selected quasar
candidates to a Galactic-extinction corrected $i^*$ magnitude of
$19.1$\footnote{SDSS magnitudes in this paper will be quoted as
$u^*g^*r^*i^*z^*$ and not in the notation of the filters ($ugriz$) in
order to indicate that the final calibration to the formal $ugriz$
system is not yet complete.  See the discussion in \citet{sto+01}.}.
We also identify radio-selected quasar candidates by matching
unresolved SDSS point sources to FIRST radio sources \citep{bwh95} to
the same magnitude limit.  Finally, since the density of high-redshift
quasars ($z\gtrsim3$) is relatively low and since the SDSS
spectrographs are capable of obtaining redshifts of quasars that are
much fainter than $i^*=19.1$, we use $i^*=20.2$ as the magnitude limit
for high-redshift quasar candidates.

Section~\ref{sec:commissioning} discusses the role of the
commissioning period to the development of the Quasar Target Selection
Algorithm.  Section~\ref{sec:selection} presents the detailed
selection criteria.  Section~\ref{sec:diagnostics} contains the
diagnostic analysis used to finalize the algorithm.  In
\S~\ref{sec:discussion} and \S~\ref{sec:conclusions}, we present
discussion and conclusions, respectively.
Appendix~\ref{sec:appendixA} describes the creation of the stellar
locus that we use to define outliers, whereas
Appendix~\ref{sec:appendixB} discusses the algorithm by which outliers
from this locus are selected.

\section{Commissioning\label{sec:commissioning}}

It is well-known in the astronomical community that ``first light''
for any new telescope is not synonymous with the commencement of full
science operations.  The SDSS Collaboration recognized from the
beginning that a testing period would be needed prior to the Survey
proper --- particularly for the development of the final quasar target
selection algorithm.  The initial pre-commissioning algorithm rejected
objects with colors consistent with the stellar locus, while selecting
regions of color-space that were known to harbor quasars or that had a
sufficiently low density of sources that they could be explored
efficiently.  As this preliminary approach was inadequate for the main
survey, this algorithm deliberately erred on the side on
inclusiveness, so that we could explore the boundaries of the quasar
locus in color and determine the nature of contaminants.  The primary
goals of the commissioning period were to refine this algorithm to 1)
meet the survey requirements that the quasar sample be at least 90\%
complete and 65\% efficient, and 2) establish the magnitude limits
needed to meet the density requirements.

Although the results of previous quasars surveys reveal approximately
what regions of color space quasars inhabit and where their
contaminants are in color space, prior to commissioning we did not
have all of the knowledge that was needed to achieve our goals.  For
example, we did not know the intrinsic spread in quasar spectral
indices, which translates to a lack of knowledge of the intrinsic
spread in quasar colors.  Compounding this problem is that, prior to
the SDSS, no large samples of quasars with photometric accuracy
comparable to the SDSS existed \citep[see the discussion
in][]{rfs+01}; {\em all} of the large area spectroscopic surveys for
quasars prior to the SDSS have been based on magnitudes measured from
photographic plates.

One of our primary concerns during the commissioning period was that
as stellar populations change between halo and disk, the detailed
position and width of the stellar locus will be a function of
direction on the sky.  As the stellar locus used by the code is fixed
by definition, any shift of the stellar locus as a function of
position on the sky would have the effect of greatly worsening the
target selection efficiency in certain areas of sky.  We tested this
possibility by examining the stellar locus in the Early Data Release
\citep{sto+01} runs 752 and 756, each of which extend over six hours
of right ascension on the Celestial Equator; they extend from low
Galactic latitudes ($b = +25^\circ$) to high ($b = +62^\circ$) over a
large range of Galactic longitudes ($l=230^\circ-360^\circ$).  We used
a simplified version of the stellar locus code described in
\citet{ny97} to fit line segments to various parts of the stellar
locus, and examined its parameters as a function of position on the
sky.  We found that the position of the locus shows a root-mean-square
scatter of 0.015 mag in color, consistent with our errors on absolute
photometric calibration.  No systematic trend was seen with position
on the sky.  Similarly, the effective width of the locus was
essentially constant.  There was a noticeable effect that the blue end
of the stellar locus is a function of magnitude as low metallicity
halo blue horizontal branch stars enter the sample in large numbers
fainter than $r^* \approx 19$.  However, we found empirically that
there is no substantial change in the efficiency of quasar target
selection either as a function of magnitude or as a function of
position on the sky.

Based on the knowledge gained during the commissioning period and the
results of the first 66 plates of data (2555 quasars), we refined the
algorithm to the point where it was deemed to be sufficiently robust
for regular survey operations.  The target selection algorithms,
including the quasar module, were ``frozen'' on 2000 November 3, such
that all imaging runs processed after this date and all spectroscopic
plates tiled after this date were appropriate, in principle, for the
creation of statistical samples of quasars.

However, additional observations revealed that the completeness of the
algorithm at $z\sim3.5$ and $z\sim4.5$ was unacceptably low --- a fact
that was only realized after a sufficient number of $z>3$ objects had
been discovered to realize the deficit.  In addition, changes to the
photometric pipeline which processes the imaging data ({\tt PHOTO};
\citealt{lup+01}), small shifts in the photometric solutions and
better simulated quasar photometry dictated that further refinements
to the code were necessary.  The changes to the code were based upon a
``testbed'' of imaging data from SDSS runs 756, 1035, 1043, 1752, 1755
and part of run 752, in addition to all of the spectra obtained in
these areas prior to 2001 July 20.  (See \citet{sto+01} for
information on the SDSS runs.)  These changes were approved on 2001
August 24; all spectra taken with this final version of the code thus
constitute the basis for the statistical sample of SDSS quasars.

\section{The Quasar Target Selection Process\label{sec:selection}}

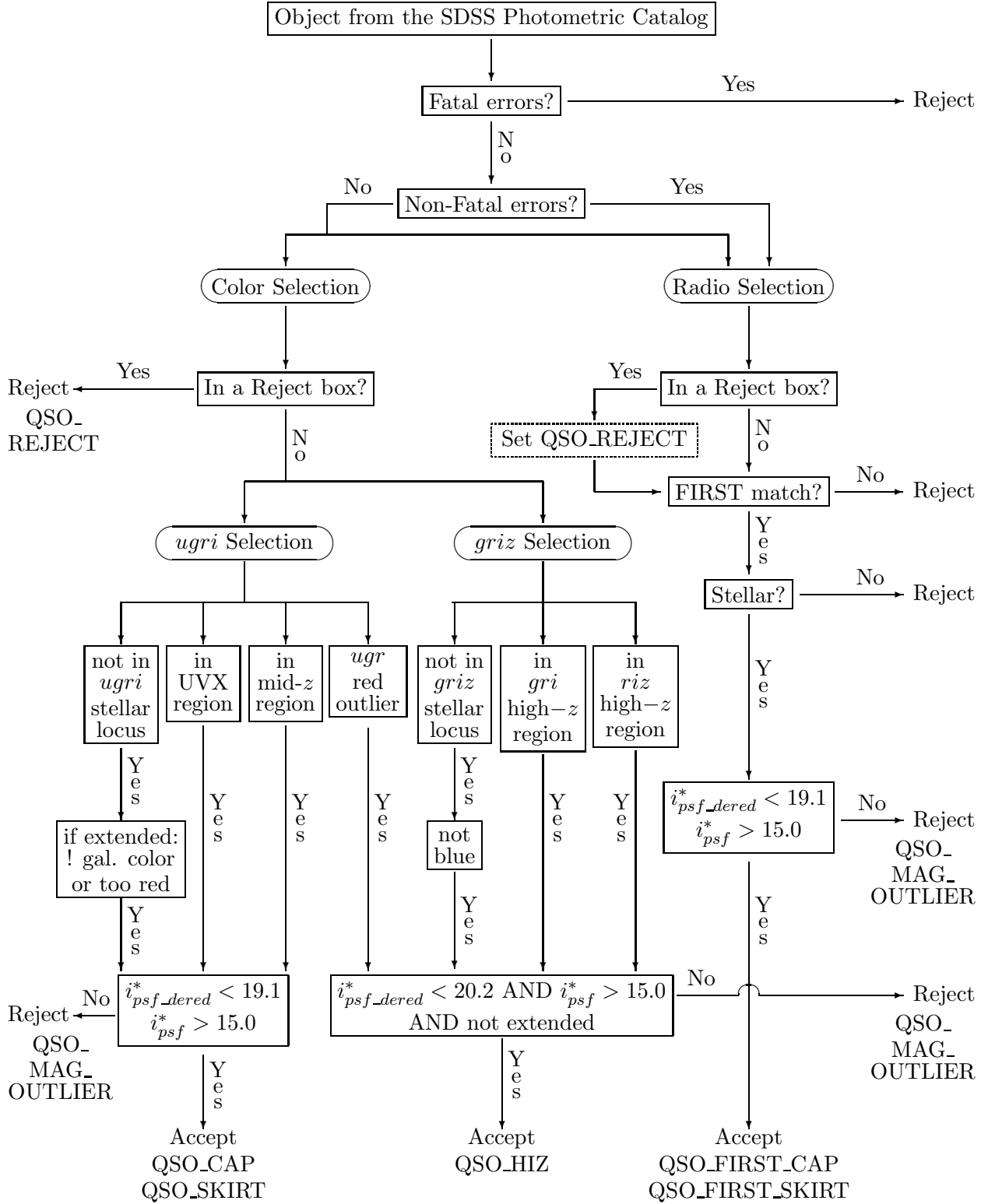
\begin{figure}
\begin{picture}(500,540)(0,20)

\put(235,570){\makebox(0,0){\fbox{Object from the SDSS Photometric Catalog}}}
\put(235,560){\vector(0,-1){20}}

\put(235,530){\makebox(0,0){\fbox{Fatal errors?}}}
\put(272,530){\vector(1,0){163}}
\put(355,535){\makebox(0,0)[b]{Yes}}
\put(455,530){\makebox(0,0){Reject}}
\put(235,520){\vector(0,-1){30}}
\put(238,500){\shortstack{N\\o}}

\put(235,480){\makebox(0,0){\fbox{Non-Fatal errors?}}}
\put(283,480){\line(1,0){87}}
\put(330,485){\makebox(0,0)[b]{Yes}}
\put(370,480){\vector(0,-1){30}}
\put(187,480){\line(-1,0){32}}
\put(170,485){\makebox(0,0)[b]{No}}
\put(155,480){\line(0,-1){15}}
\put(135,465){\line(1,0){215}}
\put(350,465){\vector(0,-1){15}}
\put(135,465){\vector(0,-1){15}}

\put(135,440){\makebox(0,0){Color Selection}}
\put(135,440){\oval(82,16)}
\put(135,430){\vector(0,-1){30}}

\put(360,440){\makebox(0,0){Radio Selection}}
\put(360,440){\oval(82,16)}
\put(360,430){\vector(0,-1){30}}

\put(360,390){\makebox(0,0){\fbox{In a Reject box?}}}
\put(360,380){\vector(0,-1){30}}
\put(363,360){\shortstack{N\\o}}
\put(315,390){\line(-1,0){30}}
\put(300,395){\makebox(0,0)[b]{Yes}}
\put(285,390){\vector(0,-1){15}}
\put(285,365){\makebox(0,0){\dashbox(100,16){Set QSO\_REJECT}}}
\put(285,355){\line(0,-1){15}}
\put(285,340){\vector(1,0){32}}

\put(360,340){\makebox(0,0){\fbox{FIRST match?}}}
\put(402,340){\vector(1,0){33}}
\put(419,345){\makebox(0,0)[b]{No}}
\put(455,340){\makebox(0,0){Reject}}
\put(360,330){\vector(0,-1){30}}
\put(363,305){\shortstack{Y\\e\\s}}

\put(360,290){\makebox(0,0){\fbox{Stellar?}}}
\put(385,290){\vector(1,0){50}}
\put(419,295){\makebox(0,0)[b]{No}}
\put(455,290){\makebox(0,0){Reject}}
\put(360,280){\vector(0,-1){80}}
\put(363,235){\shortstack{Y\\e\\s}}

\put(360,199){\makebox(0,0)[t]{\fbox{\shortstack{$i^*_{psf\_dered}<19.1$\\$i^*_{psf}>15.0$}}}}
\put(405,180){\vector(1,0){30}}
\put(420,185){\makebox(0,0)[b]{No}}
\put(455,180){\makebox(0,0){Reject}}
\put(470,155){\makebox(0,0)[r]{{\shortstack{QSO\_\\MAG\_\\OUTLIER}}}}
\put(360,160){\vector(0,-1){130}}
\put(363,120){\shortstack{Y\\e\\s}}
\put(360,30){\makebox(0,0)[t]{Accept}}
\put(360,17){\makebox(0,0)[t]{QSO\_FIRST\_CAP}}
\put(360,4){\makebox(0,0)[t]{QSO\_FIRST\_SKIRT}}

\put(135,390){\makebox(0,0){\fbox{In a Reject box?}}}
\put(90,390){\vector(-1,0){58}}
\put(61,395){\makebox(0,0)[b]{Yes}}
\put(15,390){\makebox(0,0){Reject}}
\put(0,370){\makebox(0,0)[l]{{\shortstack{QSO\_\\REJECT}}}}
\put(135,380){\line(0,-1){35}}
\put(138,355){\shortstack{N\\o}}
\put(115,345){\line(1,0){145}}
\put(115,345){\vector(0,-1){20}}
\put(260,345){\vector(0,-1){20}}

\put(115,315){\makebox(0,0){$ugri$ Selection}}
\put(115,315){\oval(86,16)}
\put(115,306){\line(0,-1){20}}
\put(55,286){\line(1,0){120}}
\put(55,286){\vector(0,-1){20}}
\put(95,286){\vector(0,-1){20}}
\put(135,286){\vector(0,-1){20}}
\put(175,286){\vector(0,-1){20}}

\put(55,265){\makebox(0,0)[t]{\fbox{\shortstack{not in\\$ugri$\\stellar\\locus}}}}
\put(55,215){\vector(0,-1){35}}
\put(58,188){\shortstack{Y\\e\\s}}
\put(55,180){\makebox(0,0)[t]{\fbox{\shortstack{if extended:\\! gal. color\\or too red}}}}
\put(55,142){\vector(0,-1){35}}
\put(58,115){\shortstack{Y\\e\\s}}

\put(95,265){\makebox(0,0)[t]{\fbox{\shortstack{in\\UVX\\region}}}}
\put(95,226){\vector(0,-1){119}}
\put(98,170){\shortstack{Y\\e\\s}}

\put(135,265){\makebox(0,0)[t]{\fbox{\shortstack{in\\mid-$z$\\region}}}}
\put(135,226){\vector(0,-1){119}}
\put(138,170){\shortstack{Y\\e\\s}}

\put(175,265){\makebox(0,0)[t]{\fbox{\shortstack{$ugr$\\red\\outlier}}}}
\put(175,225){\vector(0,-1){118}}
\put(178,170){\shortstack{Y\\e\\s}}

\put(95,105){\makebox(0,0)[t]{\fbox{\shortstack{$i^*_{psf\_dered}<19.1$\\$i^*_{psf}>15.0$}}}}
\put(95,66){\vector(0,-1){36}}
\put(98,40){\shortstack{Y\\e\\s}}
\put(52,85){\vector(-1,0){20}}
\put(43,90){\makebox(0,0)[b]{No}}
\put(15,85){\makebox(0,0){Reject}}
\put(0,60){\makebox(0,0)[l]{{\shortstack{QSO\_\\MAG\_\\OUTLIER}}}}

\put(95,30){\makebox(0,0)[t]{Accept}}
\put(95,17){\makebox(0,0)[t]{QSO\_CAP}}
\put(95,4){\makebox(0,0)[t]{QSO\_SKIRT}}

\put(260,315){\makebox(0,0){$griz$ Selection}}
\put(260,315){\oval(86,16)}
\put(260,306){\line(0,-1){20}}
\put(216,286){\line(1,0){89}}
\put(217,286){\vector(0,-1){20}}
\put(260,286){\vector(0,-1){20}}
\put(305,286){\vector(0,-1){20}}

\put(217,265){\makebox(0,0)[t]{\fbox{\shortstack{not in\\$griz$\\stellar\\locus}}}}
\put(217,215){\vector(0,-1){35}}
\put(220,188){\shortstack{Y\\e\\s}}
\put(217,180){\makebox(0,0)[t]{\fbox{\shortstack{not\\blue}}}}
\put(217,155){\vector(0,-1){47}}
\put(220,120){\shortstack{Y\\e\\s}}

\put(260,265){\makebox(0,0)[t]{\fbox{\shortstack{in\\$gri$\\high$-z$\\region}}}}
\put(260,212){\vector(0,-1){105}}
\put(263,170){\shortstack{Y\\e\\s}}

\put(305,265){\makebox(0,0)[t]{\fbox{\shortstack{in\\$riz$\\high$-z$\\region}}}}
\put(305,214){\vector(0,-1){107}}
\put(294,170){\shortstack{Y\\e\\s}}

\put(240,105){\makebox(0,0)[t]{\fbox{\shortstack{$i^*_{psf\_dered}<20.2$ AND $i^*_{psf}>15.0$\\AND not extended}}}}
\put(240,74){\vector(0,-1){44}}
\put(243,45){\shortstack{Y\\e\\s}}
\put(325,95){\line(1,0){30}}
\put(360,95){\oval(10,10)[t]}
\put(365,95){\vector(1,0){70}}
\put(337,100){\makebox(0,0)[b]{No}}
\put(455,95){\makebox(0,0){Reject}}
\put(470,70){\makebox(0,0)[r]{{\shortstack{QSO\_\\MAG\_\\OUTLIER}}}}
\put(240,30){\makebox(0,0)[t]{Accept}}
\put(240,17){\makebox(0,0)[t]{QSO\_HIZ}}

\end{picture}
\smallskip
\caption{Schematic flow diagram of the quasar target selection
algorithm.  ``Yes'' branches without a corresponding ``no'' branch
indicate that the selection no longer proceeds along this path, but
that the object is not actually rejected from consideration (i.e., it
can still be selected along another branch).  Note that objects are
evaluated an an object-by-object basis and not as a whole.  See
\S~\ref{sec:selection} for details.}
\label{fig:fig1}
\end{figure}

This discussion of the SDSS quasar target selection algorithm is
ordered largely in the way that events occur.  A flowchart that
describes the selection process is given in
Figure~\ref{fig:fig1}; the order of the flowchart is slightly
different from that of the code for the sake of clarity.  Note that
each object is considered individually and that more than one target
selection flag can be set for each object.  The target selection flags
that are set by this algorithm are given in
Table~\ref{tab:tab1}; see \citet{sto+01}, particularly
Table~27, for additional information on the all of the target
selection flags.

We start our description of the target selection algorithm with a
discussion of the input data, both the photometry
(\S~\ref{sec:input-photometry}) and flags indicating possible problems
(\S~\ref{sec:flags}).  Point sources that have counterparts in the
FIRST survey are targeted, as described in \S~\ref{sec:FIRST}, but the
heart of the target selection algorithm is the selection of outliers
from the stellar locus in color-color space, as described in
\S~\ref{sec:colorselect}.  This algorithm takes into account the
estimated magnitude errors to determine whether it lies within the
stellar locus (Appendix~B).  Finally, this algorithm is supplemented
by the inclusion and exclusion of several special regions in color
space, as described in \S~\ref{sec:hardcuts}.

\subsection{Input Photometry\label{sec:input-photometry}}

The measured SDSS fluxes are converted into asinh magnitudes
\citep{lgs99}; these magnitudes are more robust for color selection at
low flux levels than are traditional logarithmic magnitudes
\citep{pog1856}.  This affects the handling of limiting magnitudes, as
discussed in \S~\ref{sec:preprocess}.  For objects that are detected
at more than $10\sigma$, asinh magnitudes differ from logarithmic
magnitudes by less than 1\%.

Since we allow extended objects to be selected by the algorithm, but
only in specific regions of color space, the algorithm needs to
distinguish between extended and point sources.  In the context of our
algorithm, extended sources are defined as follows.  Each
two-dimensional image of each object in the SDSS is fit with a series
of models, including the locally determined Point Spread Function
(PSF), along with exponential and de Vaucouleurs profiles of arbitrary
scale size, axis ratio, and orientation (convolved with the PSF).  An
extended object is defined as an object with substantial flux on
scales beyond the PSF.  With this in mind, star-galaxy separation is
based on the difference between the PSF and either the exponential or
the de Vaucouleurs magnitude (whichever has a larger likelihood) in
each band; an object is classified as extended in that band if this
difference is greater than 0.145 mag.  The final, overall, morphology
classification is determined by summing the counts in all five bands
and applying the same criterion as for any single band, see \S~4.4.6
of \citet{sto+01} for more details.  \citet{yfn+01} and \citet{scr01}
show that the star-galaxy separation is reliable at least to
$r^*\sim21$ --- typically much fainter than the limit explored by
quasar target selection.

All magnitudes used within the quasar target selection algorithm (and
throughout this paper) are PSF magnitudes, as opposed to fiber
magnitudes or galaxy model magnitudes; see \S~4.4.5 of \citet{sto+01}
for more details.  For point sources, PSF magnitudes yield the most
accurate color information; for extended sources the colors will be
less diluted by starlight than will model, fiber or Petrosian
magnitudes.

Finally, the magnitudes used by the quasar target selection algorithm
have been corrected for Galactic reddening according to \citet{sfd98}.
These have not been updated for the recently discovered shifts of the
effective wavelengths of the SDSS filter curves (compare
\citealt{fig+96} with \citealt{sto+01}), which causes systematic
errors of 7\% of the reddening correction or less --- negligible for
the high-latitude regions of sky covered by SDSS.

\subsection{Photometric Pipeline Flag Checking\label{sec:flags}}

During the process of extracting objects from the images and measuring
their photometric properties, {\tt PHOTO} sets a number of flags
(which can be good or bad) for each detected object, some of these
flags indicate objects whose photometry (and therefore, colors) may be
problematic (e.g., blending of close pairs of objects, objects too
close to the edge of the frame, objects affected by a cosmic ray hit,
etc.).  Some flags indicate problems that are sufficiently serious
that the object in question should not be targeted for spectroscopy
under any circumstances; these will be referred to as ``fatal''
errors.  Other flags are important, but do not indicate a problem that
is serious enough to reject an object outright; these are referred to
as ``non-fatal'' errors.  Objects with non-fatal errors are considered
by the FIRST targeting algorithm (\S~\ref{sec:FIRST}), but not by the
color selection algorithm.  A list of all of the {\tt PHOTO} flags
used {\em directly} by the quasar module of the target selection
algorithm are given in Table~\ref{tab:tab2}.  For more details
regarding all of the {\tt PHOTO} flags, see Table~9 of \citet{sto+01}.

\subsubsection{Fatal Errors}

Fatal errors include objects flagged as {\tt BRIGHT}, {\tt SATURATED},
{\tt EDGE}, or {\tt BLENDED} in any band.  {\tt BRIGHT} objects
duplicate entries of sufficiently high signal-to-noise ratio objects.
The photometry of {\tt SATURATED} objects is clearly not to be
trusted.  {\tt EDGE} objects lie sufficiently close to the edge of a
frame that their photometry is unreliable.  {\tt BLENDED} objects have
several peaks; they are deblended into children (flagged as {\tt
CHILD}, each of which {\em are} considered by the algorithm).  Objects
are further required to have the status flag {\tt OK\_SCANLINE} set,
which avoids duplicate entries for regions of overlap between two
adjacent scans.  An additional constraint is that the morphology, {\tt
objc\_type}, must be either ``3'' or ``6'' (galaxy and stellar,
respectively; see the discussion in \S~\ref{sec:input-photometry}),
since the other types are all error codes.  Objects are required to be
detected at 5$\,\sigma$ in at least one of the five bands; objects
whose magnitude errors are larger than $0.2$ in all five bands are
rejected.

At one point we had considered explicitly rejecting objects that {\tt
PHOTO} deemed to be moving (i.e., asteroids).  Main-belt asteroids
move several arcseconds over the roughly five minutes between imaging
in $r$ and $g$.  {\tt PHOTO} recognizes such moving objects
explicitly, and does proper photometry of them \citep{ive01}.  As
asteroids have colors very close to that of the Sun, they lie in the
stellar locus; thus we have not found the need to reject them
explicitly.

\subsubsection{Non-Fatal Errors}

Some of the flags in the database are more subtle.  We have identified
a number of such flags (or combinations thereof) that effect quasar
target selection.  Some objects have flags that indicate that they
have colors that are unreliable; these objects are not allowed to be
targeted via the color-selection criteria, but radio sources
associated with such objects can be targeted.  The most common problem
is associated with poor deblends of complex objects; the following are
empirical combinations of flags that allow us to reject essentially
all such problematic cases.  In particular, we flag as non-fatal
errors deblended children with {\tt PEAKCENTER}, {\tt NOTCHECKED}, or
{\tt DEBLEND\_NOPEAK} set in any band --- if they are brighter than 23
mag in the same band, and have an error in that band of less than 0.12
mag.  Similarly, children with photometric errors greater than 1.0 mag
are flagged as non-fatal errors, as are children with photometric
errors greater than 0.25 in a detected band but that are {\em not}
{\tt DEBLEND\_NOPEAK}.

Objects with {\tt INTERP\_CENTER} set have a cosmic ray or bad column
within 3 pixels of their center, which has been interpolated over;
empirically, we find that many false quasar candidates are found with
$i^*<16.5$ and this flag set, so we flag all such objects as non-fatal
errors.  The errors of fainter detected objects with {\tt
INTERP\_CENTER} set are occasionally underestimated, so we increase
the photometric errors by $0.1$ mag in that band in quadrature.
Finally, there are a few bad columns that are not properly
interpolated over by the photometric pipeline, and so we reject
objects in CCD columns 1383-1387 and 1452-1460 in dewar 2.  We also
rejected CCD columns 1019-1031 in dewar 5 for imaging runs prior to
run 1635; starting with run 1635 this defect was corrected in the CCD
electronics.

\subsubsection{In the Special Case of CCD Amplifier Failure}

For a few imaging runs, one of the two read-out amplifiers on the $u$
CCD in dewar 3 was not operating; as a result, half of the dewar 3 $u$
images were completely blank.  Both non-detections in $u$ and objects
lying on the amplifier split cause problems for the quasar target
selection code.  In both cases the objects will be flagged as {\tt
BINNED1}, {\tt BINNED2}, or {\tt BINNED4} by PHOTO; objects on the
edge are further flagged as {\tt LOCAL\_EDGE} and will have erroneous
$u$-band photometry, whereas objects that are entirely in the blank
region have {\tt NOTCHECKED\_CENTER} set.

For those objects lying in the undetected region or on the edge we
increase the magnitude error in the problematic band by a full 10 mag.
This increase in error causes nearly all of the objects affected by
this problem to be considered as being consistent with the stellar
locus and thus not selected as a quasar candidate (see
\S~\ref{sec:colorselect}).  This solution is a general one that can
account for problems not only in the $u$, but also in the other
filters should future problems arise.

However, sufficiently blue objects in $g-r$ can still be false
outliers from the stellar locus regardless of their $u$-band fluxes.
Thus the above solution is not sufficient for objects with very blue
$g-r$ colors in the $ugri$ color cube (\S~\ref{sec:ugri}).  As a
result, we explicitly reject objects that are fully or partially in a
region of a CCD with a bad amplifier during the $ugri$ color selection
process.  In addition, we expand the area of color space where white
dwarfs are rejected (see \S~\ref{sec:exclusion}) to include any $u-g$
color.

Fortunately, the failure of the $u$ CCD amplifier in dewar 3 was
limited to a few imaging runs and the problem has been repaired.
Nevertheless, the software patch remains in the algorithm should the
problem recur in the future.

\subsection{FIRST matching\label{sec:FIRST}}

The SDSS catalog is matched against the FIRST catalog of radio sources
\citep{bwh95}; stellar objects ({\tt objc\_type} $= 6$) with $15.0 <
i^*<19.1$ that also have radio counterparts are selected by setting
the target flag {\tt QSO\_FIRST\_CAP}.  An SDSS object is considered
to be a match to a FIRST object if the FIRST and SDSS positions agree
to within $2\arcsec$ (the relative astrometry of the two surveys is
excellent, as discussed in \citealt{ive+02}).  No {\em explicit}
morphology criteria are applied to the FIRST data; however, no attempt
is made to match SDSS sources to double-lobed FIRST sources, for which
the optical position would be expected to be located on the line
between the two radio sources.  This exclusion introduces a bias in
the SDSS radio quasar sample against steep-spectrum, lobe-dominated
quasars, yet it simplifies the matching process considerably and
keeps contamination to a minimum.  Extended optical sources are
excluded from the radio selection (though not the color selection),
since these are mostly moderate redshift galaxies.

\subsection{Color Selection\label{sec:colorselect}}

\subsubsection{The Distribution of Stars and Quasars in SDSS Color Space}

Since stars far outnumber quasars to the SDSS magnitude limits, the
first step to selecting quasars in an efficient manner is to remove
from consideration the region inhabited by stars.  The colors of
ordinary stars occupy a continuous, almost one-dimensional region in
$(u-g),(g-r),(r-i),(i-z)$ color-color-color-color space
\citep{ny97,fan99,fif+00}, where temperature is the primary parameter
that determines position along the length of the ``locus'' of stars.
Whereas stars have a spectrum that is roughly blackbody in shape,
quasars have spectra that are characterized by featureless blue
continua and strong emission lines, causing quasars to have colors
quite different from those of stars.  As a result of their distinct
colors, quasar candidates can be identified as outliers from the
stellar locus.  We define a region of multicolor space which contains
this locus of stars.  This stellar locus does not include {\em all}
types of stars, but rather is limited to ordinary F to M stars that
dominate the stellar density in the Galaxy (at high latitude).  The
quasar target selection algorithm models this stellar locus, following
\citet{ny97}, as a two-dimensional ribbon with an elliptical
cross-section (see Appendix A for details).  In practice, our
definition of the stellar locus is done in two stages, once for the
$(u-g),(g-r),(r-i)$ color cube (hereafter $ugri$), and once for the
$(g-r),(r-i),(i-z)$ color cube (hereafter $griz$).  Splitting 4D color
space into 3D color spaces was a choice made to simplify the code:
there is no physical basis for this separation.  The application of
this code to select stellar locus outliers is described in detail in
Appendix B, and is briefly outlined below.

The algorithm chooses objects that lie more than $4\sigma$ from the
stellar locus, where the quantity $\sigma$ is determined from the
errors of the object in question and the width of the stellar locus at
the nearest point.  This procedure gives a well-defined, reproducible
color-cut for quasar target selection (as we define the stellar locus
a priori, and do not dynamically adjust it as the survey progresses),
and as we will see, allows us to meet our completeness and efficiency
goals.

The overall shape of the featureless continua of quasars is well
approximated by a power-law \citep{vrb+01}, although the continuum
need not be a power-law physically.  Since a redshifted power-law
remains a power-law with the same spectral index, quasar colors are
only a weak function of redshift for $z\le 2.2$, as emission-lines
move in and out of the filters \citep{rfs+01}.  However, quasar
spectra deviate dramatically from power-laws at rest wavelengths below
1216\AA, where the Lyman-$\alpha$ forest systematically absorbs light
from the quasar \citep{lyn71}; the net effect is that quasars become
increasingly redder with redshift as the Lyman-$\alpha$ forest moves
through the filter set.  Modeling of this effect by \citet{fan99} and
empirical evidence from \citet{rfs+01} shows that, in the SDSS
filters, the quasar locus is well-separated from the stellar locus for
both relatively low ($z\le2.2$) and relatively high ($z>3.0$)
redshifts, but that at intermediate redshifts, quasars have broad-band
SDSS colors that are often indistinguishable from early F and late A
stars.  We handle the region of color space in which intermediate
redshift quasars lie separately, as described in
\S~\ref{sec:hardcuts}.

\subsubsection{Color Selection Pre-processing\label{sec:preprocess}}

Before color selection begins, we first remove from consideration
those objects with non-fatal errors, since their colors may be
suspect.  Since we allow such objects to be selected by the FIRST
algorithm, we can statistically correct for their loss during color
selection.  Next, to account for systematic errors in the photometric
calibration, we add 0.0075 mag in quadrature to the estimated PSF
magnitude errors from {\tt PHOTO}.  As discussed above, stellar
outliers are defined as those more than $4\sigma$ from the stellar
locus, so this process in effect allows for 0.03 mag calibration
errors.  Furthermore, since we correct the magnitudes for Galactic
reddening according to the reddening map of \citet{sfd98}, we also add
to the errors (in quadrature) 15\% of the reddening values, which is
their estimate of the systematic errors in this reddening map.

Finally, objects at or below the $5\sigma$ detection limit in any band
are treated by the outlier algorithm somewhat differently from objects
that are detected in all five bands (see Appendix~B).  The non-linear
relationship between count errors and magnitude errors requires
special treatment in determining what ``$4\sigma$ from the stellar
locus'' means.  For such objects, we determine the final magnitudes by
converting the asinh magnitude to counts, adding 4 times the estimated
error in counts, and converting back to asinh magnitude.  These new
asinh magnitudes (and the resulting colors) are used to test whether
an object is consistent with being in the stellar locus.

\subsubsection{Color-Selection Target Flags}

With these preprocessing steps completed, the actual color outlier
selection begins.  The colors of each object are examined in turn, and
asked whether they are consistent with lying within either the $ugri$
or $griz$ stellar loci, incorporating the photometric errors.  The
details of this process are described in Appendix~B.  In short, quasar
candidates are chosen to be those objects which lie more than
$4\sigma$ from either stellar locus.  Outliers from the $ugri$
color-cube have the quasar target selection flag {\tt QSO\_CAP} set,
whereas outliers from the $griz$ color-cube are flagged as {\tt
QSO\_HIZ} (see Table~\ref{tab:tab1}).  Objects which are
located in regions of color space where the contamination rate is
expected to be high are flagged as {\tt QSO\_REJECT} and are not
targeted even if they are outliers from either of the stellar loci
(see \S~\ref{sec:hardcuts}).

Quasar target selection is limited to objects fainter than $i^*=15$;
the wings of the spectra of the brighter objects tend to overwhelm the
spectra of objects in adjacent fibers as seen on the CCD detector of
the SDSS spectrographs.  The $ugri$-selected objects are targeted to
$i^*=19.1$, and the $griz$-selected objects are targeted to
$i^*=20.2$, but our photometry is sufficiently accurate both brighter
and fainter than these limits that outliers from the stellar locus are
of interest for follow-up studies.  We therefore mark such objects
outside these magnitude limits as {\tt QSO\_MAG\_OUTLIER}; these
objects are not targeted for spectroscopy under routine operations.

\subsubsection{Low-redshift ($ugri$) Selection\label{sec:ugri}}

\begin{figure}[p]
\epsscale{1.0}
\plotone{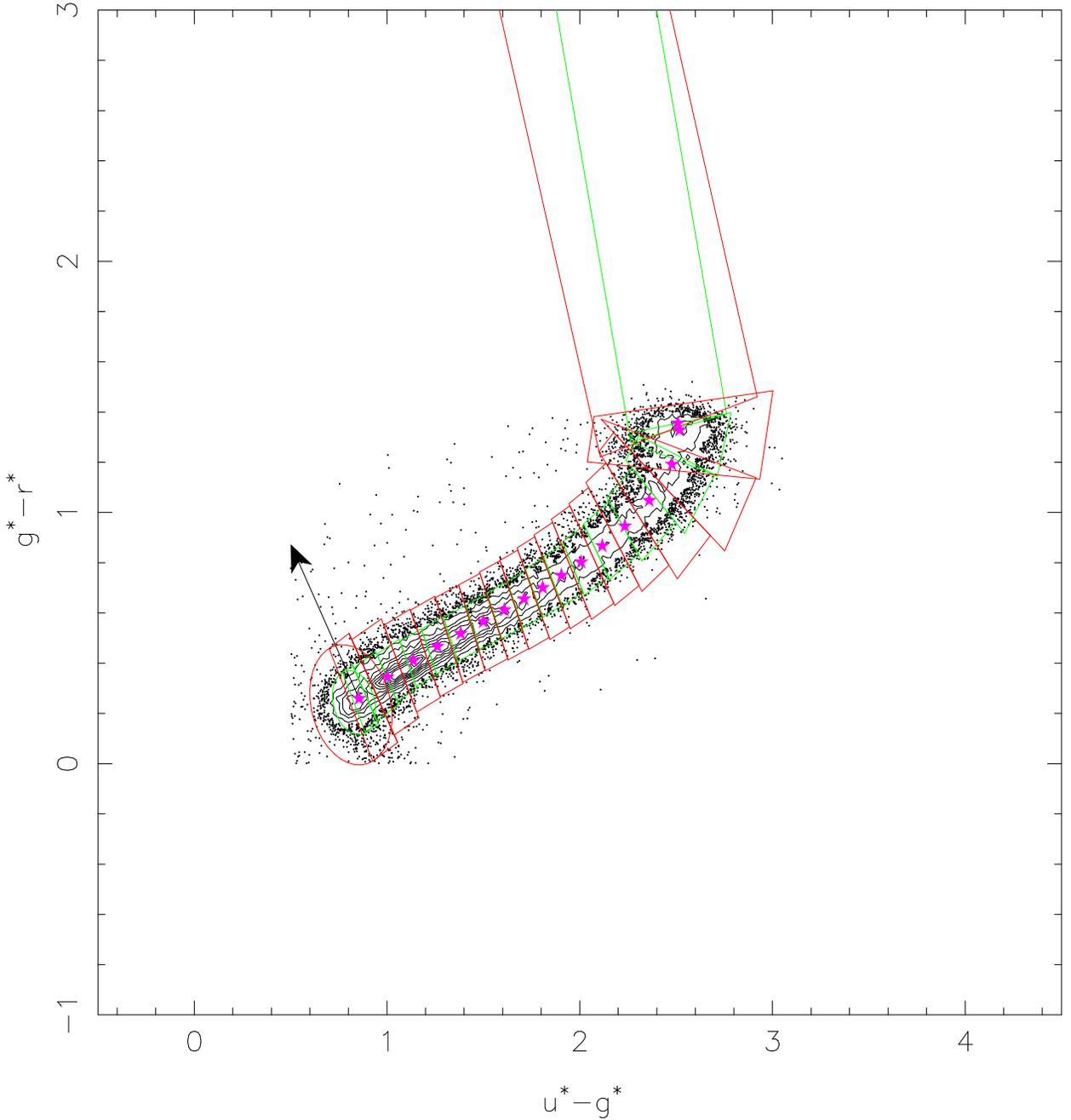}
\caption{$u^*-g^*$ vs. $g^*-r^*$ color-color diagram showing the
projection of the $ugri$ stellar locus in this plane.  Black contours
and black points show the distribution of 50000 point sources with
small errors that are used to define the stellar locus.  Point sources
with $(u^*-g^*)<0.5$ or $(g^*-r^*)<0$, which are likely to be quasars
and A stars, respectively, have been excluded.  The magenta
star-shaped symbols are the trace of the stellar locus as projected
into this color plane (see Table~\ref{tab:tab3}).  The red lines show
the outline of a $4\sigma$ error surface (which need not be smooth)
surrounding the central stellar locus points.  The round ellipse at
the blue end serves to ``cap'' the locus region.  Objects with zero
error will be selected up to this surface, whereas non-zero errors
will cause this region to grow outward.  We emphasize that these locus
curves are actually projections of a 3D region onto 2D.  The green
lines show the $2\sigma$ locus region that is used for mid-$z$ quasar
selection (see \S~\ref{sec:inclusion}).  The black vector shows the
blue extent of the region where extended objects are rejected (see
Figure~\ref{fig:fig4}). \label{fig:fig2}}
\end{figure}

\begin{figure}[p]
\epsscale{1.0}
\plotone{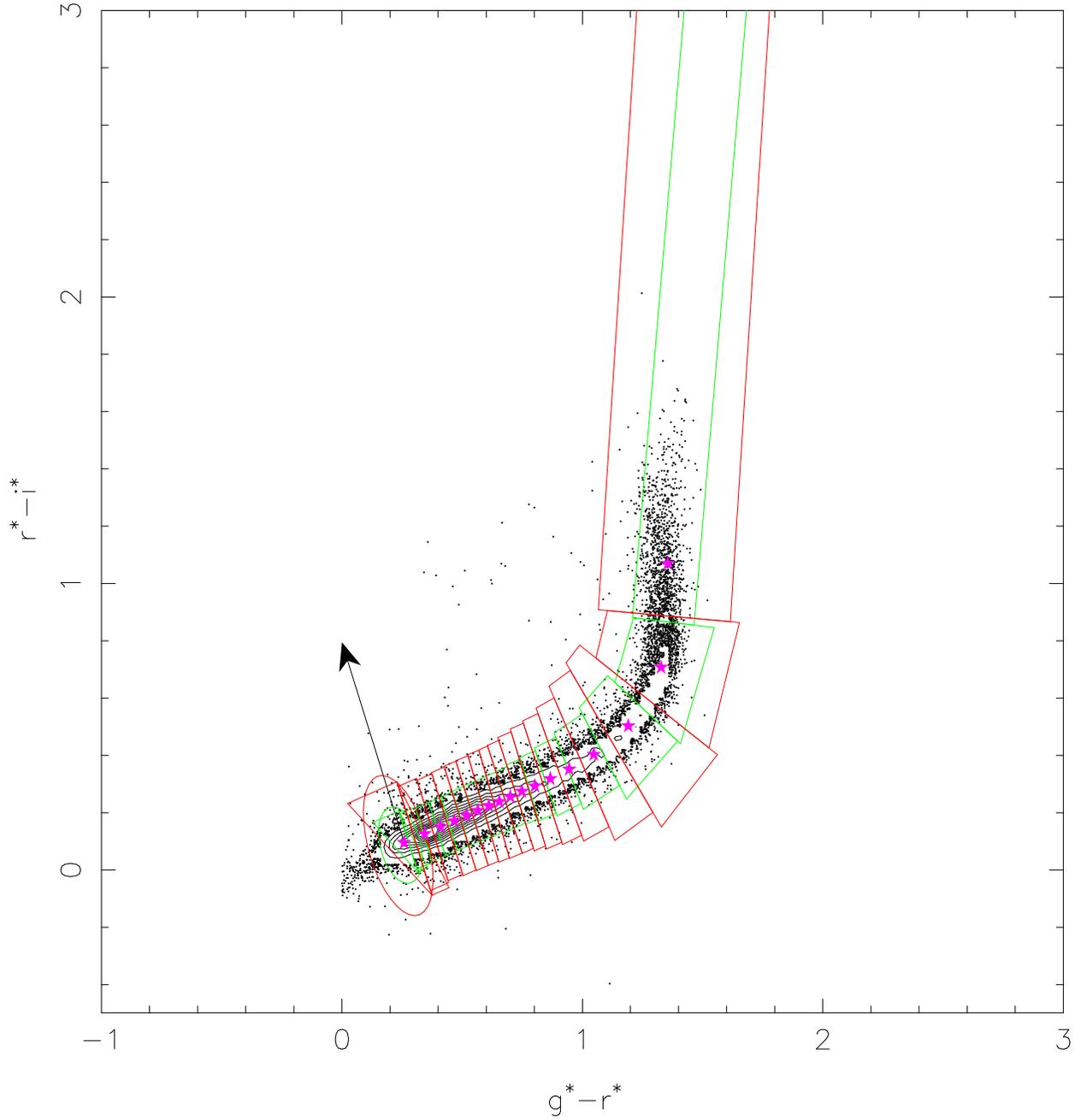}
\caption{$g^*-r^*$ vs. $r^*-i^*$ color-color diagram showing the
projection of the $ugri$ stellar locus rejection region in this plane.
See Figure~\ref{fig:fig2} for an explanation of the symbols
and lines.\label{fig:fig3}}
\end{figure}

In $ugri$ color-space, objects that have 1) survived the flag checks,
2) are outliers from the stellar locus (see Appendix A), and 3) are
not in the exclusion boxes mentioned below (\S~\ref{sec:exclusion})
are selected as quasar candidates to a magnitude limit of $i^*<19.1$.
As described in Appendix B, the primary selection is accomplished by
convolving the errors of a new object with the $4\sigma$ $ugri$
stellar locus depicted by the red outlines in Figures~\ref{fig:fig2}
and \ref{fig:fig3}, thereby creating a 3D error surface surrounding
the stellar locus.  If the colors of the object are inside this error
region, then the object is deemed to be consistent with the stellar
locus and the object is not selected.  If the object is outside of
this surface, it is considered to be a good quasar candidate by virtue
of its status as an outlier from the stellar locus.  The parameters
for the $ugri$ stellar locus are given in Table~\ref{tab:tab3}; see
also Appendix~A.  Since the stellar locus that we use is set a priori
and is not determined from the SDSS data for each new run, the cuts in
color space are uniform over the survey.  Note also that the boundary
of the stellar locus is determined from a series of cylinders, and no
attempt is made to make this boundary smooth where the cylinders
overlap.

During the $ugri$ color selection process, both extended and point
source objects are targeted as quasar candidates; we do not explicitly
differentiate between quasars and their lower-luminosity cousins that
are typically extended.  Note that by using PSF magnitudes throughout,
we isolate the colors of any point-like components of galaxies with
active nuclei.  However, not all extended objects are allowed to be
selected --- just those that have colors that are far from the colors
of the main galaxy distribution and that are consistent with the
colors of AGN.  At one point we had considered fitting an empirical
``galaxy locus'' much like our stellar locus however, in practice, we
found it easier to make some simple color cuts to keep from targeting
too many normal galaxies.

\begin{figure}[p]
\epsscale{1.0}
\plotone{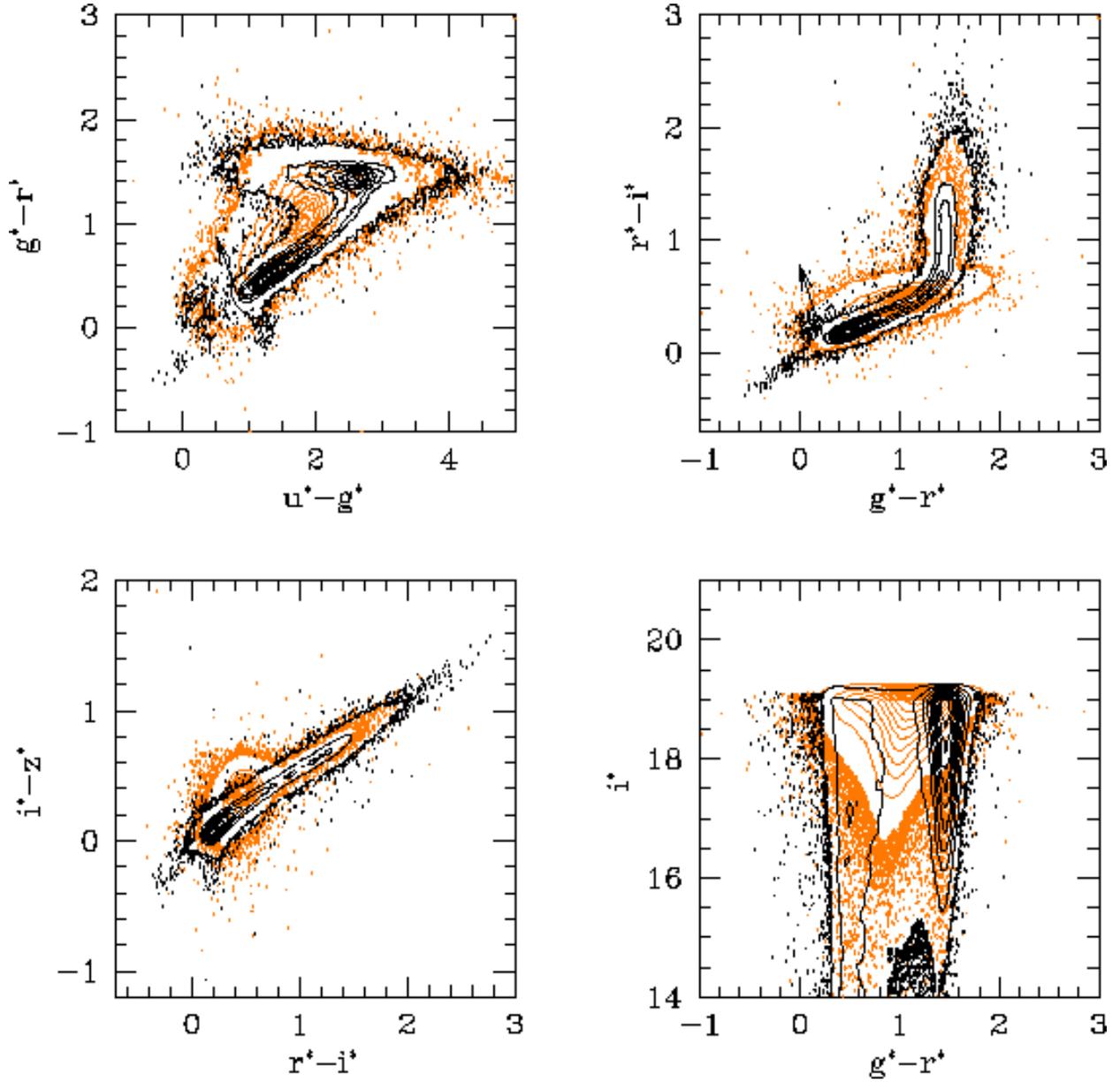}
\caption{Location of bright stars and galaxies in the SDSS photometric
system.  Black points and contours are stellar sources with
$i^*<19.1$.  Orange points and contours are extended sources with
$i^*<19.1$.  These data are all objects that did not have fatal or
non-fatal errors and were thus considered possible quasar candidates
from the testbed data (see \S~\ref{sec:commissioning}).  Note that
these plots look different than Figures~\ref{fig:fig2}
and~\ref{fig:fig3} because here we plot the colors of all of the
possible quasar candidates, regardless of the statistical errors in
their colors; Figures~\ref{fig:fig2} and~\ref{fig:fig3} only show
objects with small errors.  The black vector in the upper panels shows
the blue extent of the region where extended objects are
rejected.\label{fig:fig4}}
\end{figure}

In Figure~\ref{fig:fig4}, we show the color-color and
color-magnitude distribution of stars and galaxies that are brighter
than $i^*<19.1$.  These objects are all of the point and extended
sources in the testbed data (see \S~\ref{sec:commissioning}) that do
not have fatal or non-fatal errors according to the quasar target
selection algorithm and are otherwise good quasar candidates.  Black
points and contours show unresolved sources, orange points and
contours show the extended sources.  That the star/galaxy separation
algorithm is not perfect is obvious from the number of extended
sources along the branch of late-type stars ($r^*-i^* > 1$), a color
that low-redshift galaxies rarely have \citep{yfn+01,sik+01}.  Even
so, the separation is quite good.  Normal galaxy colors are clearly
distinct from the stellar locus; they are very concentrated in $riz$
color space, they extend redward of the stellar locus in $gri$ color
space, and they occupy roughly the space expected from a superposition
of stars in $ugr$ color space.

As Figure~\ref{fig:fig4} makes clear, galaxies can be outliers from
the stellar locus, so we use two color cuts to reject extended objects
unlikely to harbor an active nucleus.  First, extended objects that
are detected in both $u$ and $g$, that have errors less than $0.2$~mag
in each band, and that have $(u^*-g^*)>0.9$ are rejected.  Quasars
with colors redder than $(u^*-g^*)>0.9$ have redshifts that are
sufficiently large ($z\gtrsim2.6$) that they should not be resolved.

This cut misses the extension of the galaxy locus to somewhat bluer
$u^*-g^*$ colors, and thus we apply a second cut, rejecting extended
objects with $l>0$ and $k>0$ (in the notation of \citealt{ny97}; also
see Appendix A).  This second cut effectively removes all extended
objects that are ``above'' and to the ``left'' of the stellar locus in
the $(u^*-g^*),(g^*-r^*)$ color-color diagrams.  The blue edge of this
cut is illustrated by the vectors drawn perpendicular to the stellar
locus at the blue tip of the locus in the $ugr$ and $gri$ plots in
Figure~\ref{fig:fig4} and also Figures~\ref{fig:fig2}
and~\ref{fig:fig3}.  Note that the axes in these plots have
different scales, which causes the vector to appear as if it were not
perpendicular to the long direction of the stellar locus.

Objects that meet all of the above criteria are selected for follow-up
spectroscopy by setting the {\tt QSO\_CAP} bit in their target
selection flag (see Table~\ref{tab:tab1}).

\subsubsection{High-redshift ($griz$) Selection}
\label{sec:griz}

\begin{figure}[p]
\epsscale{0.8}
\plotone{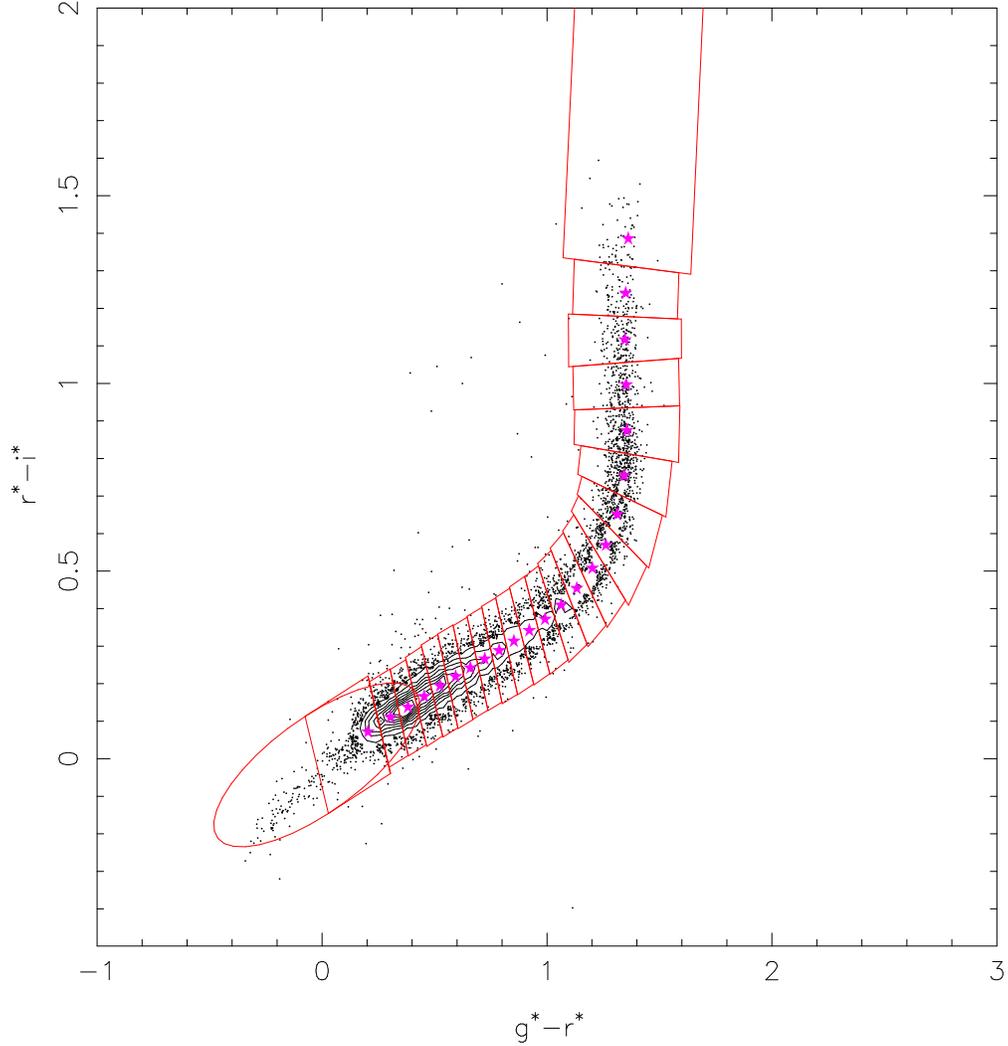}
\caption{$g^*-r^*$ vs. $r^*-i^*$ color-color diagram showing stars in
the $gri$-plane (black points and contours) and the projection of the
stellar locus region (red lines) in this plane for the $griz$
selection function (high-$z$).  This surface is not the same as the
surface in Figure~\ref{fig:fig3} since the locus is actually a
2-D projection of 3-D color space and the $griz$ selection uses a
different 3-D color space than the $ugri$ selection.  Objects with
zero error will be selected up to this surface, whereas non-zero
errors will cause this region to grow outward.  Magenta stars show the
trace of the stellar locus (see
Table~\ref{tab:tab4}). \label{fig:fig5}}
\end{figure}

\begin{figure}[p]
\epsscale{0.8}
\plotone{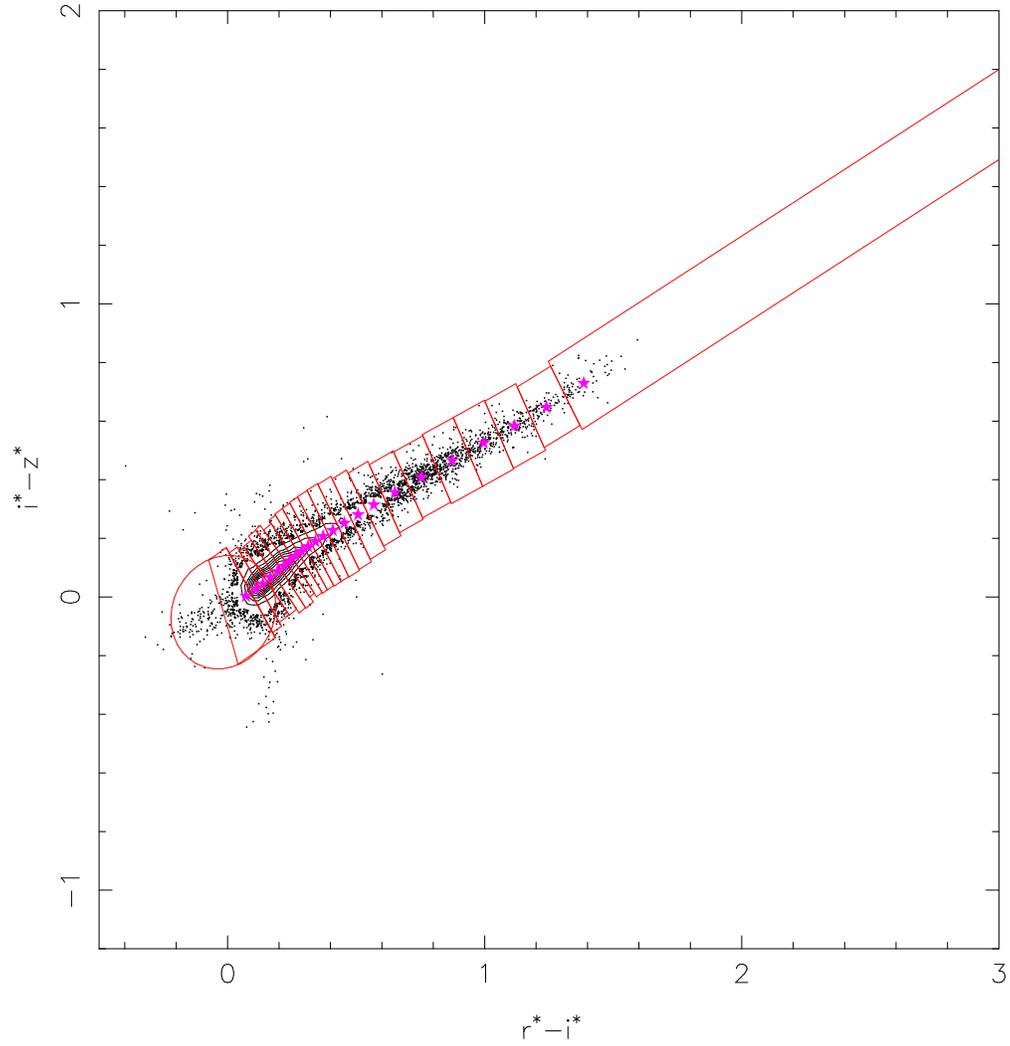}
\caption{$r^*-i^*$ vs. $i^*-z^*$ color-color diagram showing the
projection of the stellar locus rejection region in this plane for the
$griz$ selection function (high-$z$).  See
Figure~\ref{fig:fig5} for an explanation of the
symbols. \label{fig:fig6}}
\end{figure}

In a manner equivalent to that for the $ugri$ color-cube, outliers
from the stellar locus in the $griz$ color-cube with $i^*<20.2$ are
targeted for follow-up spectroscopy as quasar candidates.  These
objects will be flagged as {\tt QSO\_HIZ}.  The $4\sigma$ intrinsic
width of the $griz$ stellar locus is given by the red outlines in
Figures~\ref{fig:fig5} and \ref{fig:fig6}; the parameters for the
$griz$ stellar locus are given in Table~\ref{tab:tab4}, see also
Appendices A and B.  Targeted objects in the $griz$ color-cube must be
classified as stellar, as they will lie at redshifts above $z\sim3.5$,
and the majority of quasars with $z\ge0.6$ are classified as stellar
at the resolution of the SDSS.  Note that this means that we will be
biased against gravitational lenses with $z\gtrsim3$; however, we
cannot afford the level of contamination that would accompany the
inclusion of very red, extended sources.

The purpose of the $griz$ selection is specifically to select
high-redshift quasars.  However, some low-redshift quasars are
outliers from the $griz$ stellar locus as well; given that we target
the $griz$ color-cube 1.1 mag fainter than the $ugri$ color-cube, the
{\tt QSO\_HIZ} targets could be dominated by low-redshift quasars.  To
avoid this problem, objects are {\em not} selected as quasar
candidates in the $griz$ color-cube when the following conditions are
met:
\begin{eqnarray}
\begin{array}{rrclcrcl}
{\rm A})\;\, & g^*-r^* & < & 1.0 & & & & \\
{\rm B})\;\, & u^*-g^* & \ge & 0.8 & & & &\\
{\rm C})\;\, & i^* & \ge & 19.1 & {\rm OR} & u^*-g^* & < & 2.5.
\end{array}
\end{eqnarray}
These cuts are empirically defined to remove faint, low-redshift
quasars in addition to some faint objects that are clearly in the
stellar locus but that the automated code fails to exclude.

We have found that there are objects selected as high-redshift quasars
by simple color cuts (\citealt{fsr+01}) which are missed by this
color-selection algorithm.  The problem objects are most likely to be
objects that are just above the detection limit.  The primary
color-selection algorithm should work well for objects well-above the
detection limit where our assumption of symmetric magnitude errors is
roughly true.  It also should work well for objects below the nominal
detection limit (as long as they are detected in the $i$-band), since
we treat these objects specially.  However, our assumption of
symmetric magnitude errors (see Appendix~B) can break down for objects
just above the detection limit.

Consider an object to the lower right of the stellar locus in $gri$,
e.g., at $g^*-r^*=1.5$, $r^*-i^*=0$, which is barely detected in $g$.
It will thus have large photometric errors in $g$, large enough that
its $g^*-r^*$ color is consistent with the stellar locus at $4\sigma$.
However, there are two ways of asking whether such an outlier is
consistent with the stellar locus.  First is to ask whether the errors
of the object could make it consistent with the stellar locus.  Second
is to ask whether an object in the stellar locus could be scattered
into the region occupied by the outlier as a result of its errors.
Our algorithm is constructed from the first point of view.

For bright objects, these are essentially the same question; however,
for objects near the limiting magnitude, but that are still detected,
our assumption of symmetric magnitude errors causes these two
viewpoints to diverge.  If instead we took the second point of view,
the $g$ magnitude of the corresponding object in the stellar locus
would be substantially brighter, it would have substantially smaller
errors, and our outlier would be inconsistent with the stellar locus.
Thus, if instead of asking whether the errors of the outlier make it
consistent with the stellar locus, we asked whether the errors of a
stellar locus object could cause it to scatter to the position of the
outlier, we would get different answers (under the assumption of
symmetric magnitude errors).  

We attempt to correct for cases such as these where the color outlier
algorithm may miss good quasar candidates by implementing some simple
cuts in the $griz$ color cube, as described in the next section.  One
of these cuts is designed to recover $z \ge 3.6$ quasars, and the
second cut is designed to recover $z \ge 4.5$ quasars.

Finally, we wish to extend our {\tt QSO\_HIZ} sample to $i^*=20.2$ for
objects with $z > 3.0$.  These are outliers in the $ugr$ color diagram
and are selected with simple cuts in this diagram, as described in the
next section.

\subsection{Exclusion and Inclusion Regions\label{sec:hardcuts}}

Based on experience from previous multi-color quasar surveys and from
our own commissioning data, we know that there are regions of
color-space outside of the stellar locus which are dominated by
objects other than quasars.  There are also regions of color space
populated by quasars that are not selected by the color outlier
routines described above.  Therefore we have defined some regions of
color space where objects are either explicitly included or excluded.

\begin{figure}[p]
\epsscale{1.0}
\plotone{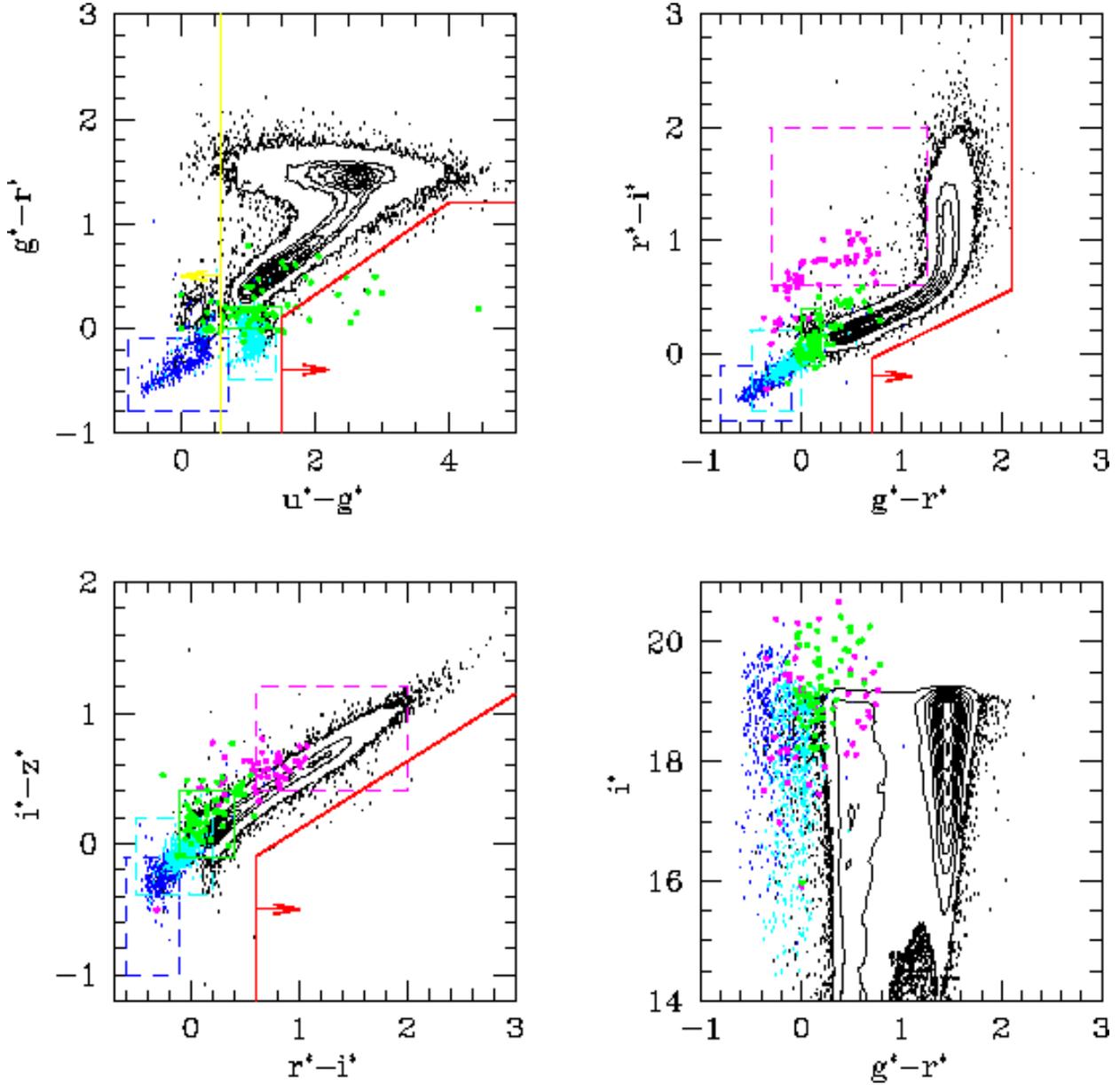}
\caption{Location of Exclusion and Inclusion Boxes.  Black points and
contours are stellar sources with $i^*<19.1$.  Dark blue (dashed)
lines indicate white dwarf exclusion regions; dark blue points
indicate spectroscopically confirmed white dwarfs.  Similarly for A
stars (light blue, dashed line), white dwarf + M star pairs (magenta,
dashed line), and mid-$z$ ($2.5<z<3.0$) quasars (green, solid line).
Note that the green box is an inclusion rather than an exclusion
region and that the green points are simply quasars with $2.5<z<3.0$,
not just those that lie in the mid-$z$ inclusion box.  Note also that
these boxes actually show projections onto 2D surfaces of what are
really 4D regions.  The yellow (solid) line shows the explicit UVX
color cut.  The red (solid) lines show the various high-redshift color
cuts; note that, unlike the ``boxes'', these high redshift cuts are
unique in each panel.\label{fig:fig7}}
\end{figure}

We reject objects which lie in the regions that are typically
dominated by white dwarfs (WD), A stars, and M star-white dwarf pairs
(WD+M).  See Figure~\ref{fig:fig7} and \S~\ref{sec:exclusion} for the
boundaries of these boxes.

In addition, as a result of the fact that, in the SDSS color space,
the ``quasar locus'' crosses the stellar locus for quasars with
redshifts between $z=2.5$ and $z=3.0$, we must explicitly include a
number of objects in this region in order to properly sample the
distribution of quasars in this redshift range.  This ``mid-$z$''
inclusion box is also shown in Figure~\ref{fig:fig7}.  The inclusion
of some of these objects as quasar candidates results in lower
efficiency, but without this approach we would not be able to
investigate this redshift range.

Finally, we have instituted some color cuts to aid in the selection of
high redshift quasars, as described in the previous subsection.  We
have also implemented a UVX color cut as an attempt at backwards
compatibility with previous UVX surveys for quasars.  Each of these
cuts in shown in Figure~\ref{fig:fig7}.

\subsubsection{Exclusion Regions\label{sec:exclusion}}

Objects in any of the three exclusion regions (WD, WD+M, A stars) are
flagged as {\tt QSO\_REJECT}, and are {\em not} targeted --- unless
they are also selected as FIRST targets.  After the {\tt QSO\_REJECT}
flag has been set, in the Main Survey, no further attempt is made to
target these objects based upon their colors.  However, we allow
ourselves the option of targeting {\tt QSO\_REJECT} objects in limited
areas of the sky (e.g., in the Southern Survey area) to explore the
nature of the objects we are missing in the Main Survey.

The white dwarf exclusion region, which is shown in dark blue in
Figure~\ref{fig:fig7}, includes objects which satisfy all the
following criteria
\begin{eqnarray}
\begin{array}{rrcccl}
{\rm A}) \;\, & -0.8 & < & u^*-g^* & < & 0.7 \\
{\rm B}) \;\, & -0.8 & < & g^*-r^* & < & -0.1 \\
{\rm C}) \;\, & -0.6 & < & r^*-i^* & < & -0.1 \\
{\rm D}) \;\, & -1.0 & < & i^*-z^* & < & -0.1. 
\end{array}
\end{eqnarray}
There are essentially no common astrophysical objects with colors
bluer than the blue limits given above, but by applying these limits,
we leave open the possibility of serendipitously discovering truly
unusual objects.

Next, we reject A stars, since their numbers are too small to have any
effect upon the definition of the stellar locus (see Appendix A), but
are found in sufficiently large numbers to significantly contaminate
the quasar sample.  The region that is rejected is defined by the
intersection of
\begin{eqnarray}
\begin{array}{rrcccl}
{\rm A}) \;\, & 0.7 & < & u^*-g^* & < & 1.4 \\
{\rm B}) \;\, & -0.5 & < & g^*-r^* & < & 0.0 \\
{\rm C}) \;\, & -0.5 & < & r^*-i^* & < & 0.2 \\
{\rm D}) \;\, & -0.4 & < & i^*-z^* & < & 0.2.
\end{array}
\label{eq:Astar}
\end{eqnarray}

The last region of color-space that we reject is that which is
occupied by unresolved red-blue star pairs, usually M star and white
dwarf pairs.  These objects appeared as significant contaminants in
the early target selection process when we attempted to target all
point sources that were outside of the stellar locus.  The color-space
that is rejected is the intersection of
\begin{eqnarray}
\begin{array}{rrcccl}
{\rm A}) \;\, & -0.3 & < & g^*-r^* & < & 1.25 \\
{\rm B}) \;\, & 0.6 & < & r^*-i^* & < & 2.0 \\
{\rm C}) \;\, & 0.4 & < & i^*-z^* & < & 1.2 \\
{\rm D}) \;\, & \sigma_{g^*} & < & 0.2, & &
\end{array}
\end{eqnarray}
where the additional restriction on the error in $g^*$ keeps us from
rejecting too many normal stars with large errors; such objects are
outliers according to our stellar locus definition and are properly
rejected by the stellar locus code.

We note that except for the requirement that $\sigma_{g^*}<0.2$ in the
WD+M box, there are no formal requirements that the error in the
colors be small for the objects in the exclusion regions.  That is, we
do not require that the objects be in the regions within their errors,
just within the regions based on the measured colors.

\begin{figure}[p]
\epsscale{1.0}
\plotone{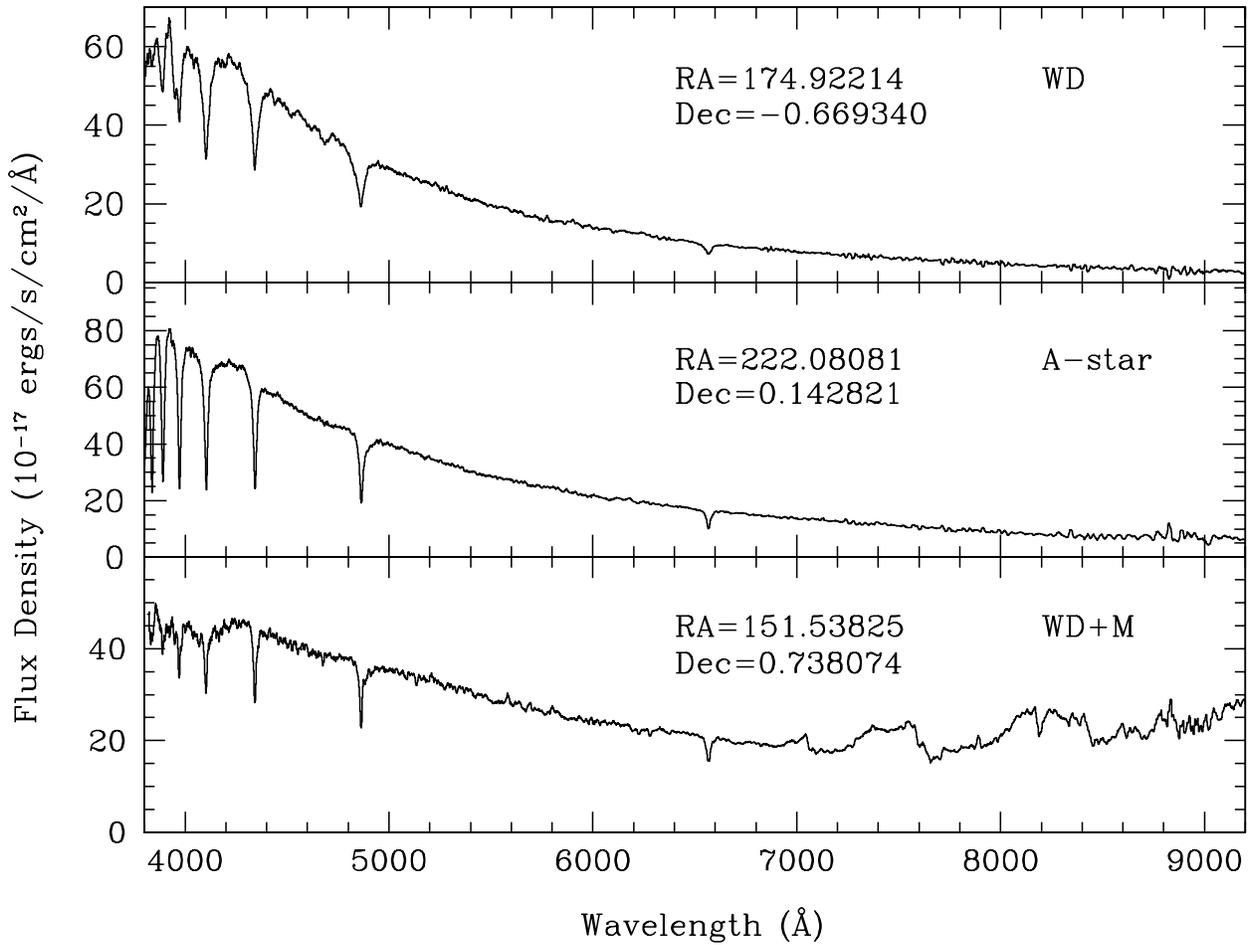}
\caption{Spectra of sample objects rejected by exclusion regions.
({\em Top}) A typical white dwarf (WD) spectrum.  ({\em Middle}) A
typical A star spectrum.  Note the narrower Balmer lines in this
spectrum as compared to the WD spectrum.  ({\em Bottom}) A typical
white dwarf + M star (WD+M) type object.  Note that although we refer
to these as WD+M pairs, this category is used to describe any blue+red
star pairing.\label{fig:fig8}}
\end{figure}

Figure~\ref{fig:fig8} displays the spectra of three objects that are
typical of the classes of objects that are excluded.  The top panel
shows a hot white dwarf, with its characteristic broad Balmer lines,
while the middle panel shows an A star lying in the box defined by
equation~(\ref{eq:Astar}).  The lower panel shows a blue/red pair;
note the rising spectrum in the blue with strong Balmer lines, and
rising spectrum in the red with characteristic TiO and VO bands.

\subsubsection{Inclusion Regions\label{sec:inclusion}}

In addition to objects that are explicitly rejected by these three
exclusion boxes, we also have a series of color cuts and boxes where
objects are explicitly included even if they do not meet the other
color-selection requirements.  We specifically target ``mid-$z$''
($2.5 < z < 3$) quasars in a small region of color-space where these
quasars cross the stellar locus in SDSS color space.  We also explore
three regions of color space for high redshift quasars that might
otherwise be missed by the color outlier selection routine
(\S~\ref{sec:colorselect}).

\begin{figure}[p]
\epsscale{1.0}
\plotone{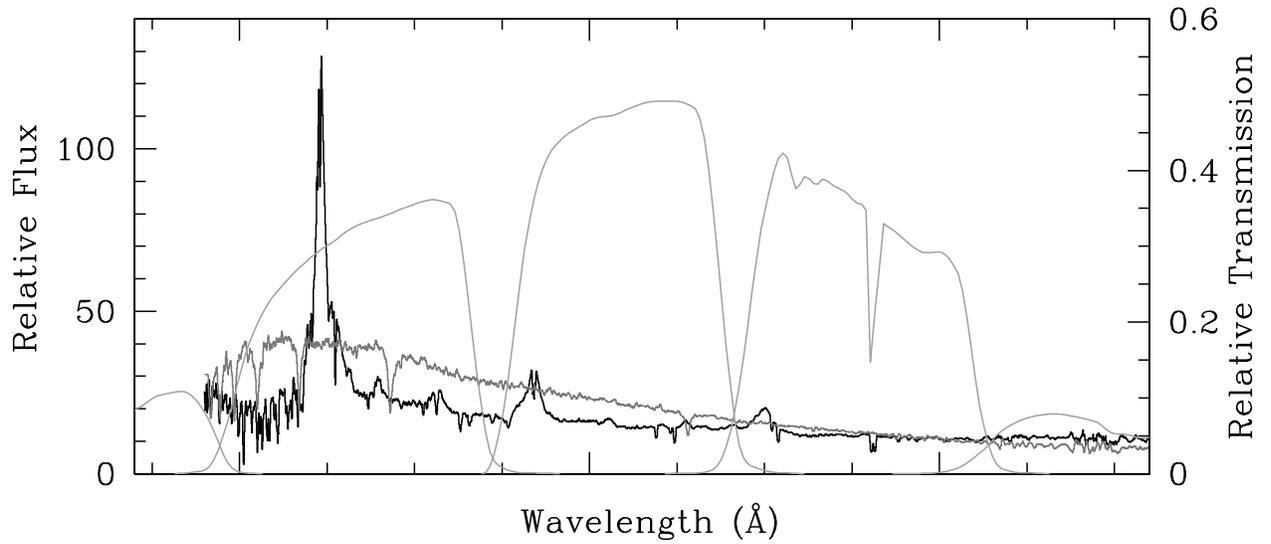}
\caption{Spectra of a sample mid-$z$ quasar ($z=2.67$) and a star with
similar colors superimposed upon the SDSS filter curves. Note that the
$g^*-r^*$ color is nearly the same for both objects.  See Figure 1 in
\citet{fan99} who used simulated spectra to demonstrate that this is
true for $u^*-g^*$ as well.\label{fig:fig9}}
\end{figure}

Figure~\ref{fig:fig9} illustrates the mid-$z$ problem, showing a
$z\sim2.7$ quasar spectrum, a star that inhabits the same region of
color space, and the SDSS filter curves.  Although the spectra do not
cover the $u$ filter, we can see that even though the continua of the
two spectra are different, the quasar emission lines cause the quasar
to have very similar broad-band colors to the star.  See Figure 1 of
\citet{fan99} for a similar figure using simulated spectra that shows
that the $u^*-g^*$ colors of the quasar and star are also similar.

The first inclusion region that we define is an attempt to deal with
this problem near $z\sim2.7$.  If we failed to target quasar
candidates in this region of color space, our completeness would be
very low for $2.5 < z < 3$, a range of tremendous importance for
quasar absorption line studies such as measurements of the primordial
deuterium abundance \citep[e.g.,][]{bkt99}.  In addition, this
redshift range covers the peak of the quasar co-moving number density
\citep{ssg95}.  However, if we were to explicitly target all of the
objects in this region of color space, our efficiency would become
unacceptably low.

Thus, we have decided on a hybrid approach to allow a statistical
study of the quasar population in this redshift range.  First, we
define a region in color space as follows:
\begin{eqnarray}
\begin{array}{rrcccl}
{\rm A}) \;\, & 0.6 & < & u^*-g^* & < & 1.5 \\
{\rm B}) \;\, & 0.0 & < & g^*-r^* & < & 0.2 \\
{\rm C}) \;\, & -0.1 & < & r^*-i^* & < & 0.4 \\
{\rm D}) \;\, & -0.1 & < & i^*-z^* & < & 0.4.
\end{array}
\end{eqnarray}

Second, the selected objects are further restricted by requiring that
they be point sources and that they lie outside of a $2\sigma$ region
surrounding the stellar locus (as opposed to the $4\sigma$ limit that
we normally use; see Figures~\ref{fig:fig2} and~\ref{fig:fig3} (green
lines), along with \S~\ref{sec:colorselect} and Appendix A).  Finally,
since this region of color space crosses the stellar locus, we choose
to explicitly target only 10\% of the objects in this mid-$z$ color
box in order to limit our reduction in efficiency.  This sparse
sampling is accomplished by targeting only those objects whose tenths
digit in decimal degrees right ascension is equal to seven, which is
as good of a way of defining a random sample as any.  The end result
is that we target enough objects in this region (flagged as {\tt
QSO\_CAP}) that we can correct for our incompleteness in a statistical
sense without wasting too many fibers taking spectra of normal stars.

Another inclusion region involves quasars with $z\le2.2$.  All objects
which are detected in both $u$ and $g$, that have errors less than
$0.1$ mag in both bands and $(u^*-g^*)<0.6$ are explicitly selected
(as {\tt QSO\_CAP}) so long as they are not in the white dwarf
exclusion box and satisfy the magnitude limits.  This cut is intended
to be the equivalent of previous UVX color-cuts
\citep[e.g.,][]{bfs+90,kk88} in the sense that this cut selects
objects with roughly the same upper limit in redshift (namely
$z\le2.2$).  In reality, this cut does not cause many objects to be
selected that are not otherwise selected as stellar locus outliers,
but most of the candidates selected only because of this color-cut are
indeed quasars or AGN.

The final series of color-cuts are designed to recover more high
redshift quasars than would be possible by simply targeting red
outliers from the stellar locus (see the discussion at the end of
\S~\ref{sec:griz}).  \citet{fsr+01} show that certain simple
color-cuts are very effective in the selection of high-redshift
quasars.  As a result, we also target objects (as {\tt QSO\_HIZ}) that
meet criteria similar to those used by \citet{fsr+01}.  In fact, the
cuts used herein are somewhat more inclusive than those used by
\citet{fsr+01}, since we can afford a slightly lower selection
efficiency as a trade-off for increasing our completeness.

Similar to \citet{fsr+01}, we create an inclusion region in $gri$
color-space, which is intended to recover quasars with $z\ge 3.6$
(which are problematic for the stellar locus outlier code).  The
criteria for the $gri$ inclusion region is the intersection of:
\begin{eqnarray}
\begin{array}{rrclcrcl}
{\rm A}) \;\, & \sigma_{i^*} &<& 0.2 & & & & \\
{\rm B}) \;\, & u^*-g^* &>& 1.5 & {\rm OR} & u^* &>& 20.6 \\
{\rm C}) \;\, & g^*-r^* &>& 0.7 & & & & \\
{\rm D}) \;\, & g^*-r^* &>& 2.1 & {\rm OR} & r^*-i^* &<& 0.44*(g^*-r^*) - 0.358 \\
{\rm E}) \;\, & i^*-z^* &<& 0.25 & & & & \\
{\rm F}) \;\, & i^*-z^* &>& -1.0. & & & &
\end{array}
\end{eqnarray}
Cuts C) and D), which involve $gri$ color-space, are depicted in the
upper right-hand panel of Figure~\ref{fig:fig7} (red line).  The
restriction on the $i^*$ errors, A), is to ensure the integrity of the
$r^*-i^*$ and $i^*-z^*$ colors.  The cuts in $u^*-g^*$ and $u^*$, B),
restrict the sample to $u$-band dropouts.  The vertical cuts in
$g^*-r^*$ and the diagonal cut, C) and D), keep the region from
getting too close to the stellar locus.  Finally, restriction E)
prevents most normal M stars from being selected and restriction F)
prevents objects that are too blue in $i^*-z^*$ to be high-$z$ quasars
from entering the selection region.

We similarly define a region in $riz$ color space that is designed to
select quasars with $z\ge4.5$.  These criteria are
\begin{eqnarray}
\begin{array}{rrcl}
{\rm A}) \;\, & \sigma_{i^*} &<& 0.2 \\
{\rm B}) \;\, & u^* &>& 21.5 \\
{\rm C}) \;\, & g^* &>& 21.0 \\
{\rm D}) \;\, & r^*-i^* &>& 0.6 \\
{\rm E}) \;\, & i^*-z^* &>& -1.0 \\
{\rm F}) \;\, & i^*-z^* &<& 0.52*(r^*-i^*) - 0.412.
\end{array}
\end{eqnarray}
In this case, the $u^*$ and $g^*$ magnitude limits, B) and C), are
designed to restrict the sample to objects that are both $u$- and
$g$-band drop-outs.  The $r^*-i^*$ cut, D), places a limit on the
amount of flux in the $r$-band in addition to keeping the selected
objects from getting too close to the stellar locus.  As above, the
blue limit on $i^*-z^*$, E), is designed to reject objects which are
too blue to be high-$z$ quasars.  Finally, the diagonal cut, F),
limits the number of normal stars that stray into the inclusion
region.  In the lower left-hand panel of Figure~\ref{fig:fig7}, we
show (as a red line) the cuts that are relevant to the $riz$ color
plane, D), E), and F).

Finally, we add an inclusion region in $ugr$ color space that was not
used by \citet{fsr+01}; this cut is designed to recover $z\ge3.0$
quasars --- especially those fainter than $i^* = 19.1$, which is the
magnitude limit of the low-redshift sample, and those $z\sim3.5$
quasars that may be undetected in $u$ but which are not outliers in
$griz$.  The selection criteria used in this region are
\begin{eqnarray}
\begin{array}{rrcl}
{\rm A}) \;\, & u^* &>& 20.6 \\
{\rm B}) \;\, & u^*-g^* &>& 1.5 \\
{\rm C}) \;\, & g^*-r^* &<& 1.2 \\
{\rm D}) \;\, & r^*-i^* &<& 0.3 \\
{\rm E}) \;\, & i^*-z^* &>& -1.0 \\
{\rm F}) \;\, & g^*-r^* &<& 0.44*(u^*-g^*) - 0.56.
\end{array}
\end{eqnarray}
The cuts that correspond to the $ugr$ region of color space are shown
(as a red line) in the upper left-hand panel of
Figure~\ref{fig:fig7}.  Both the magnitude limit on $u^*$, A), and
the color cut in $u^*-g^*$, B), are meant to ensure that the objects
are sufficiently red to be high-$z$ quasars.  The diagonal cut, F),
and the upper limit to $g^*-r^*$, C), keep the objects from getting
too close to the stellar locus (especially where the errors get large
for M stars).  The $r^*-i^*$ cut, D), is also designed to limit the
number of M stars making the cut.  Finally, as with the above high-$z$
regions, we exclude objects that are too blue in $i^*-z^*$, E), to be
real high-$z$ quasar candidates.

In practice, this $ugr$ high-$z$ cut is combined with another cut that
selects outliers in the $ugri$ color-cube that are sufficiently red.
These objects are required to be outside of the $ugri$ stellar locus
and must be detected in both $u^*$ and $g^*$, have $u^*$ and $g^*$
errors less than 0.2 mag and have $(u^*-g^*)>1.5$.  Objects that meet
these criteria or the $ugr$ high-$z$ criteria described above are
allowed to be selected as quasar candidates (flagged as {\tt
QSO\_HIZ}) to the fainter $i^*=20.2$ limit as long as they are not
extended sources.

\section{Diagnostics\label{sec:diagnostics}\label{sec:fansim}}

How well does the quasar selection algorithm perform?  As discussed
above, this is really two questions: what is the selection efficiency
and what is the completeness?  We need to know both of these
quantities as a function of redshift and magnitude (both apparent and
absolute).

Completeness and efficiency diagnostics were carried out using both
simulated quasar photometry and empirical results from early SDSS
imaging and spectroscopy.  As discussed in \S~\ref{sec:commissioning},
the empirical testbed contains all of the imaging data in runs 756,
1035, 1043, 1752, 1755 and some of run 752.  On the observational
side, completeness is defined as the fraction of previously known
quasars in the appropriate magnitude range that are selected by the
algorithm, as quantified in \S~\ref{sec:completeness}.  Efficiency is
defined as the fraction of quasar candidates that turn out to be
quasars when spectra are taken.  More spectroscopic data (using the
final version of the target selection code) is needed in order to
fully characterize the completeness and efficiency of the survey;
however, we have enough data to determine that the algorithm is
meeting the goals for completeness and efficiency.  In
\S~\ref{sec:efficiency}, we use data from over 100 square degrees with
spectra available for 90\% of the quasar candidates, and in
\S~\ref{sec:completeness}, we examine over 1500 known quasars in the
testbed area along with the colors of over 50,000 simulated quasars.

To help determine the survey completeness we also make use of
simulated quasar colors.  We calculate the simulated distribution of
quasar colors at a given redshift and magnitude, following the
procedures described in \citet{fan99} and \citet{fsr+01}.  The
intrinsic quasar spectrum model includes a power-law continuum and a
series of broad emission lines.  We use the same distributions of the
power-law index for the quasar continuum ($\alpha = 0.5 \pm 0.3$;
$f_{\nu} \propto \nu^{-\alpha}$) and the equivalent widths for the
emission lines as in \citet{fan99} (except for \ion{Fe}{2}, which now
has a larger equivalent width).  The synthetic quasar absorption
spectrum takes into account intervening \ion{H}{1} absorbers along the
line of sight, including Ly$\alpha$ forest systems, Lyman-limit
systems and damped Ly$\alpha$ systems, using distribution functions
similar to those used by \citet{fan99}.  Finally, we calculate the
SDSS magnitudes and the associated photometric error from the model
spectrum in each band, using the filter transmission curves and system
efficiency of \citet{sto+01}, assuming a median seeing of $1\farcs4$
FWHM.  The simulated colors are generated for a uniformly distributed
grid of redshift and $i$ magnitude.

\subsection{Completeness\label{sec:completeness}}

Completeness testing was accomplished in two ways: checking to see
that previously known quasars are recovered, and by evaluating the
results of target selection for simulated quasars.  We began by
seeking to maximize the completeness of previously known quasars.  To
facilitate this we created a catalog of previously known quasars; for
details see \citet{rfs+01}.  Objects in this catalog were matched to
the objects in the quasar target selection testbed data.

Table~\ref{tab:tab5} gives the results of this matching.  In short,
the SDSS quasar target selection algorithm targets 1456 of 1540
(94.5\%) of known quasars that should have been targeted, and
establishes an upper limit to the completeness of our algorithm with
respect to real AGN.  These can be broken down into a number of
categories.  Quasars from the NASA/IPAC Extragalactic Database (NED)
are recovered at the rate of 369 in 394 (93.7\%); 96.6\% (56 of 58) of
FIRST quasars are recovered (using both color and radio selection);
quasars discovered in the first 66 plates of SDSS spectroscopic
commissioning (using a more liberal, but less efficient selection
algorithm) are recovered by the final quasar target selection
algorithm at a rate of 1154 in 1210 (95.4\%); finally, 88.6\% of
high-$z$ quasars discovered during the early period of the SDSS
high-$z$ quasar follow-up campaign \citep{fsr+01} that satisfy the
magnitude requirements are recovered (39 of 44).  Note that the
completeness of the last category is somewhat lower than the others
because \citet{fsr+01} had less strict flag-checking requirements than
does the Main Survey for quasars; we cannot afford to be as lenient
during the Main Survey due to efficiency constraints.
Table~\ref{tab:tab5} also provides other information that may be
relevant to the completeness of the SDSS Quasar Survey.

We find that most of the known quasars not selected by our algoritm
are missing as a result of non-repeatable problems (cosmic rays, etc.)
and not because of shortcomings of our selection algorithm.  A known
quasar that is missed because of one or more cosmetic defects would
not necessarily be missed if new imaging data were obtained for that
object.

\begin{figure}[p]
\epsscale{1.0}
\plotone{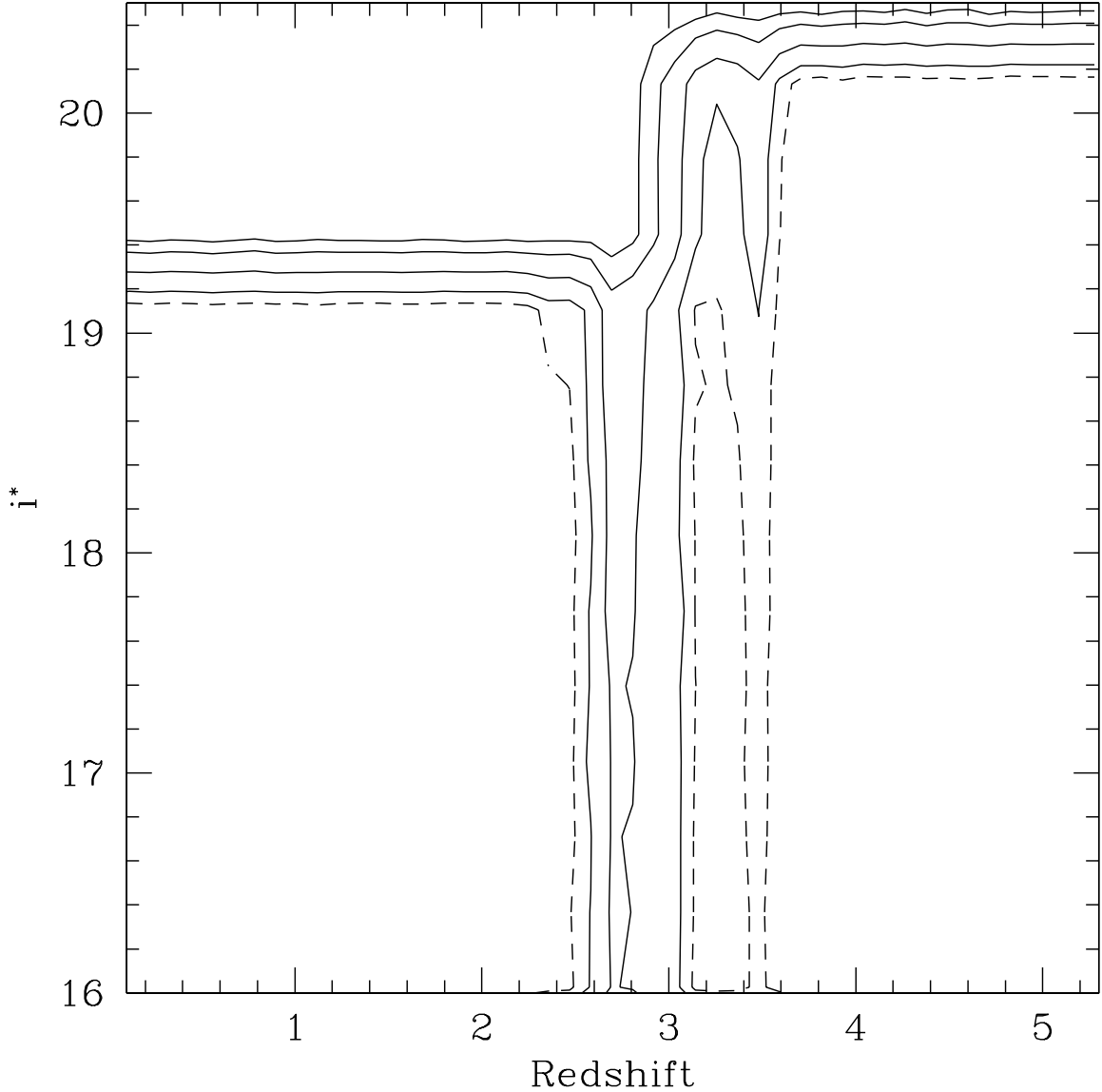}
\caption{Completeness as a function of redshift and $i^*$ as
determined from simulated quasars.  Contours are at the 10\%, 25\%,
50\%, 75\%, and 90\% completeness levels.  The 90\% contour level is
given by the dashed line; objects in the regions of parameter space
within the dashed lines (e.g., below the horizontal dashed lines near
$i^*=19.1$ and $i^*=20.2$) are targeted no less than 90\% of the time.
All of the parameter space shown is covered by the simulations.  Note
that the contours extend fainter than $i^*=19.1$ (for $z<2.2$) and
$i^*<20.2$ (for $z>3$) because we have dithered the simulations to
produce smoother contours.
\label{fig:fig10}}
\end{figure}

\begin{figure}[p]
\epsscale{1.0}
\plotone{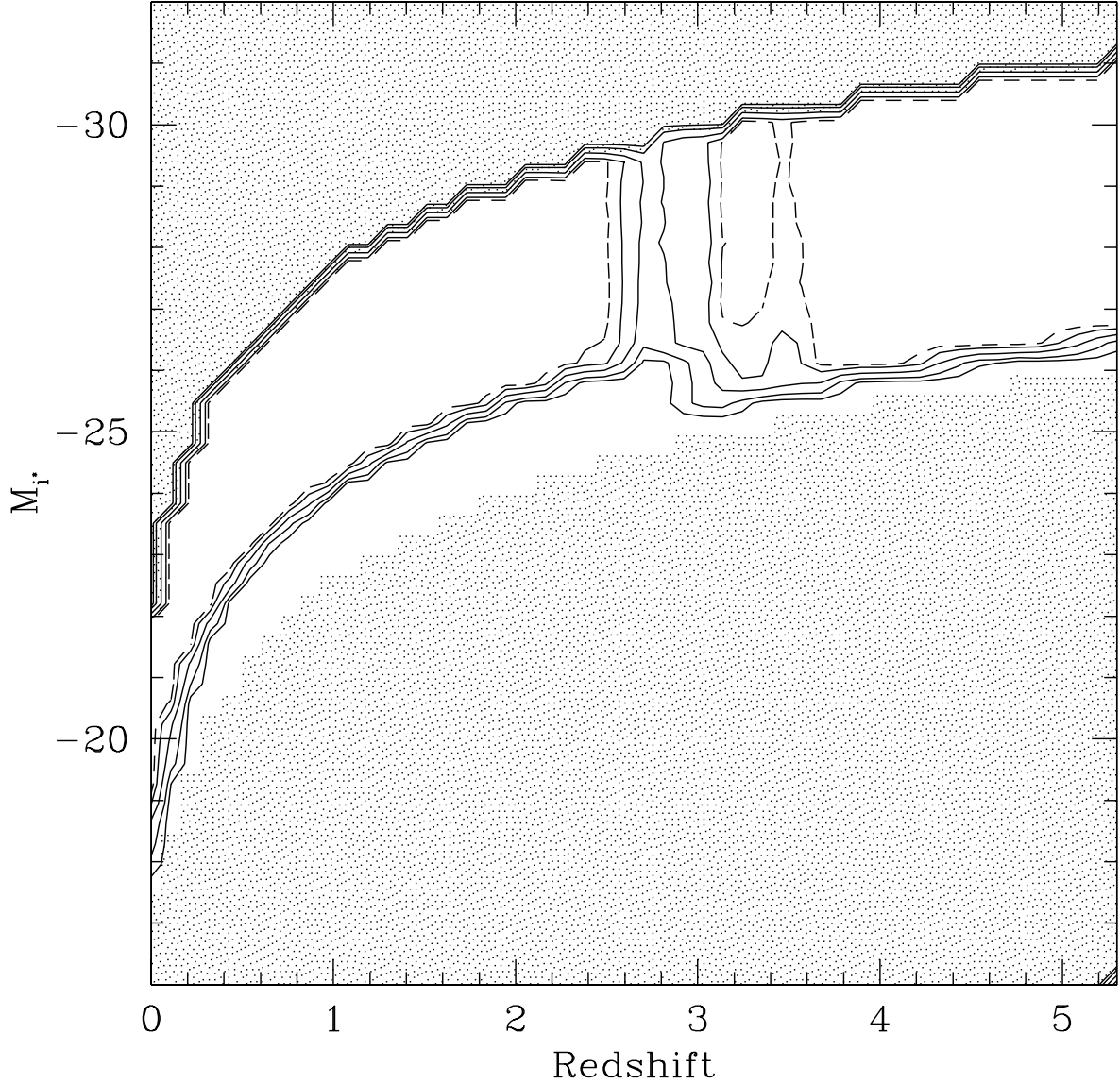}
\caption{Completeness as a function of redshift and $M_{i^*}$ as
determined from simulated quasars.  Contours are at the 10\%, 25\%,
50\%, 75\%, and 90\% completeness levels.  The 90\% contour level is
given by the dashed line; objects in the regions of parameter space
within the dashed lines are targeted no less than 90\% of the time.
The shaded regions are outside of the parameter space covered by the
simulations.  As with Figure~\ref{fig:fig10}, note that the contours
at the bright and faint magnitude limits are not vertical drop-offs
because we have dithered the simulations to produce smoother contours.
\label{fig:fig11}}
\end{figure}

We would also like to quantify the completeness as a function of
redshift and magnitude; we do so with the grid of model quasars
discussed above.  These simulated quasars have a much more homogeneous
distribution in redshift-magnitude space than do real quasars.
Figure~\ref{fig:fig10} shows the completeness of the selection
function in redshift-$i^*$ space, whereas Figure~\ref{fig:fig11} shows
the completeness as a function of redshift and $M_{i^*}$ for the same
simulated quasars.  The resulting selection function is tabulated in
Table~\ref{tab:tab6}, as a function of $i$ magnitude for $z\le5.3$ and
as a function of $z$ magnitude for $z>5.3$, where the quasars become
$i$-band drop-outs.  For the full grid of simulated quasars uniformly
distributed between $0 \le z \le 5.8$ and $16.0 \le i^* \le 19.0$, the
mean completeness is 94.6\%.

However, this fraction is somewhat misleading.  Our {\em true}
completeness must actually be smaller than this number.  This
completeness assumes a uniform distribution of quasars in redshift
space, which is unrealistic; the true, overall completeness of the
quasar survey will be different from the values quoted above.  In
particular, from Figures~\ref{fig:fig10} and~\ref{fig:fig11} and
Table~\ref{tab:tab6}, it is evident that the average completeness is a
function of quasar redshift and luminosity.  The survey completeness
is low in the redshift range $2.4 < z < 3.0$, because at these
redshifts the quasar locus crosses the stellar locus in the SDSS color
space, and we sparse sample in the mid-$z$ color box
(\S~\ref{sec:inclusion}) that selects these quasars. The survey
completeness is larger than 80\% for all other redshifts up to $z\sim
5.3$ for quasars $\gtrsim 0.5$ magnitude brighter than the survey
limit.  Note that the completeness is given in terms of ``true''
magnitude (without photometric error) in the simulation.  Therefore,
the completeness does not immediately reach zero for quasars fainter
(or brighter) than the survey limit ($i = 19.1$ for low-$z$ quasars,
and $i=20.2$ for $z>3$), as they are scattered into the selected range
due to the photometric error.  Note also that the SDSS is only
sensitive to relatively luminous quasars ($M_{i^*} \lesssim -25$) for
$z\gtrsim 2$.

In addition, the completeness based on the simulations is also an
over-estimate because none of the simulated quasars are affected by
cosmetic defects (unlike the known quasars).  We expect to miss a few
percent of quasars due to non-repeatable defects, independent of
redshift or apparent magnitude.  That is, if we observed the same
area of sky again, we would very likely recover these objects.  Of
course, such incompleteness is independent of redshift and largely
independent of color and brightness.

We note that our algorithm is quite sensitive to red quasars --- in
contrast to most previous optically selected surveys.  The reasons for
this sensitivity are threefold.  First, we are magnitude limited in
the $i$-band instead of the $B$-band and are thus less sensitive to
dust reddening effects which are a strong function of wavelength with
stronger absorption in the blue.  Second, the precision of our
photometry reduces the chance of a quasar being scattered into the
stellar locus, or for photometry errors to broaden the stellar locus,
which would shrink the region of color space in which we could select
quasars.  Finally, we are not biased against regions of color space
that are not known to harbor quasars; to the extent possible we
consider any source outside of the stellar locus to be a quasar
candidate.

One way to test the sensitivity of our color-selection algorithm is to
ask what fraction of the radio-selected quasars are also
color-selected.  As can be seen from Table~\ref{tab:tab5}, of the
known quasars that met our radio selection criteria, only 20 of the
162 (12.3\%) were not also selected according to their colors (see
also \S~\ref{sec:efficiency}).  Of the 3814 quasars in the Early Data
Release (EDR) quasar sample \citep{sch+02}, there were only 23 objects
which were radio-selected only; a composite of these 23 objects shows
that they are quite reddened and tend to be low-ionization BAL
quasars.

Understanding the completeness of the survey as a function of quasar
properties is crucial for the SDSS quasar Key Projects, the evolution
of quasar luminosity function and large scale distribution of quasars.
Here we only present the survey completeness in term of its average
value (both from matching with known quasars and from the simulation).
The completeness is also a function of the quasar SED, in particular
the continuum shape and emission line strength.  Detailed discussion
will be presented in a later work; for high-redshift quasars, see the
discussion in \citet{fsr+01}.  In the SDSS Southern Survey, for a
relatively small area, we plan to relax a number of color constraints
(including various exclusion boxes) in the quasar target selection
algorithm, and push the selection closer to the stellar locus in the
color space.  In a future paper, we will use these data to address the
completeness question in detail, especially for quasars at $z\sim3$.

Although the process of testing the completeness of the algorithm
using the Southern Survey is beyond the scope of this communication,
we have begun the process.  For example, we have further explored our
completeness by investigating those objects that are flagged as
QSO\_REJECT.  One of the goals of the Southern Survey is to take
spectra of all such objects that are otherwise good quasar
candidates\footnote{The QSO\_REJECT flag gets set before objects are
tested against the stellar locus, thus not all QSO\_REJECT objects
will be good quasar candidates.}.  As of this writing, we have taken
spectra of 1640 QSO\_REJECT objects from runs 94 and 1755 using a
modified version of the quasar target selection algorithm that 1)
allows QSO\_REJECT objects to be selected, 2) targets {\em all} point
sources in the mid-$z$ region (as opposed to 10\%), and 3) has fainter
magnitude limits.  Of these 1640, only 26 are AGN that are not also
selected by the Main Survey algorithm.  Of these, the majority are
weak LINERs \citep{hec80} with a strong stellar component, with
redshifts $0.1-0.3$.  There are only two quasars with higher redshifts
($z=2.25$ and $z=2.86$); one is a Type II quasar (e.g., see
\citealt{up95}).  The parent sample of objects selected from runs 94
and 1755 is about $11,500$ objects.  If 65\% of these are quasars, we
estimate that the WD, WD+M and A-star exclusion boxes
(\S~\ref{sec:exclusion}) that are used to set the QSO\_REJECT flag are
rejecting only $26/(0.65*11,500) = 0.3\%$ of quasars that are outside
of the stellar locus (assuming that the objects that we have spectra
of are representative of the full sample).  Thus, the exclusion boxes
have a negligible effect upon our completeness.

\subsection{Efficiency\label{sec:efficiency}}

To test the efficiency of the algorithm, we used a $\sim100$ square
degree region of sky which has been targeted by the final version of
the quasar target selection algorithm.  In particular, we examined the
1872 quasar candidates with $16.25^{\circ} < {\rm RA} < 57^{\circ}$,
$-1.25^\circ < {\rm Dec} < +1.25^\circ$ (runs 94 and 125 from the SDSS
Early Data Release; \citealt{sto+01}).  Of these objects, we have
spectra in hand for all but 185 objects, which we believe to be
largely spurious candidates due to the poor image quality of those
early SDSS runs.

Overall, we found that 1113 of the 1687 quasar candidates with spectra
are quasars or AGN.  This number corresponds to an efficiency of
66.0\%, which is consistent with the a priori requirements for the
algorithm.  Of the quasar candidates that are not quasars, 266 are
galaxies (15.8\%) and 294 are stars (17.4\%).
Table~\ref{tab:tab7} presents a summary of the efficiency
results for $\sim100$ square degrees of data; its columns are 1) the
number of quasar candidates, 2) the number of quasar candidates for
which we have obtained spectra, 3) the number in column 2 that turn
out to be quasars/AGN, 4) the ratio of column 3 to column 2, 5) the
number in column 2 that turn out to be galaxies, 6) the ratio of
column 5 to column 2, 7) the number in column 2 that turn out to be
stars, and 8) the ratio of column 7 to column 2.

In addition to the objects for which we obtained good spectra, there
were 11 candidates that were assigned fibers, but that either had no
spectra or were spectra of blank areas of sky.  We found that two of
these were the result of broken fibers in the spectrograph, one was a
supernova, and the others were mostly asteroids.

The contaminating galaxies come in three categories: star-forming,
blue, emission-line galaxies at low redshift; faint red objects at $z =
0.4\;{\rm to}\;0.5$; and compact E+A galaxies at $z = 0.4\;{\rm to}\;
0.8$ \citep[e.g.,][]{zzl+96}, whose strong Balmer break mimics the
onset of the Ly$\alpha$ forest in a high-redshift quasar in the SDSS
filter set.  The contaminating stars are largely cool M and L stars, a
few A, F, white dwarf, and white dwarf/M dwarf pairs, and a number of
interesting objects, such as carbon stars and low-metallicity subdwarf
M stars.

We have also broken down the efficiency according to category: low-$z$
($ugri$ color selected), high-$z$ ($griz$ color selected), and
FIRST-selected quasar candidates.  Low-$z$ targets include 1392 quasar
candidates, of which 1339 have spectra.  Of these 1339, 1005 are
quasars (75.0\%), 233 are galaxies (17.4\%), and 98 are stars (7.3\%).
Of the quasars, 116 are $z>2$, 16 are $z>3$.  Of the 1392 candidates,
1155 are not also targeted as high-$z$ or FIRST objects; 1110 of these
have spectra.  Of these 1110, 789 are quasars, of which 81 are $z>2$,
and 3 are $z>3$.

Similarly, for high-$z$ targets there were 663 quasar candidates, 529
of which have spectra.  Of these 529, 288 are quasars (54.4\%), 35 are
galaxies (6.6\%), and 194 are stars (36.7\%).  Of the 663 high-$z$
targets, 426 are exclusively high-$z$; 300 of these have spectra.  Of
those 300, only 72 are quasars; 61 have $z>2$, 55 have $z>3$, and 7
have $z>4$.

In addition to the low-$z$ and high-$z$ selected objects, there are 74
FIRST-selected quasar candidates, 69 of which have spectra.  Of these
69, 60 are quasars (87.0\%) and 9 are stars (13.0\%).  Of the 74 FIRST
targets, 22 were targeted only by matching with FIRST sources.  Ten of
these 22 are quasars, 9 are stars, and the remaining 3 don't have a
spectrum.  We examined the spectra of the ten AGN; two are unusual BAL
quasars \citep{hal02}, several are at $z\sim2.6$ where the quasar and
stellar loci cross, and the remaining few don't look particularly
unusual; they are perhaps a bit redder than a typical quasar.

\begin{figure}[p]
\epsscale{1.0}
\plotone{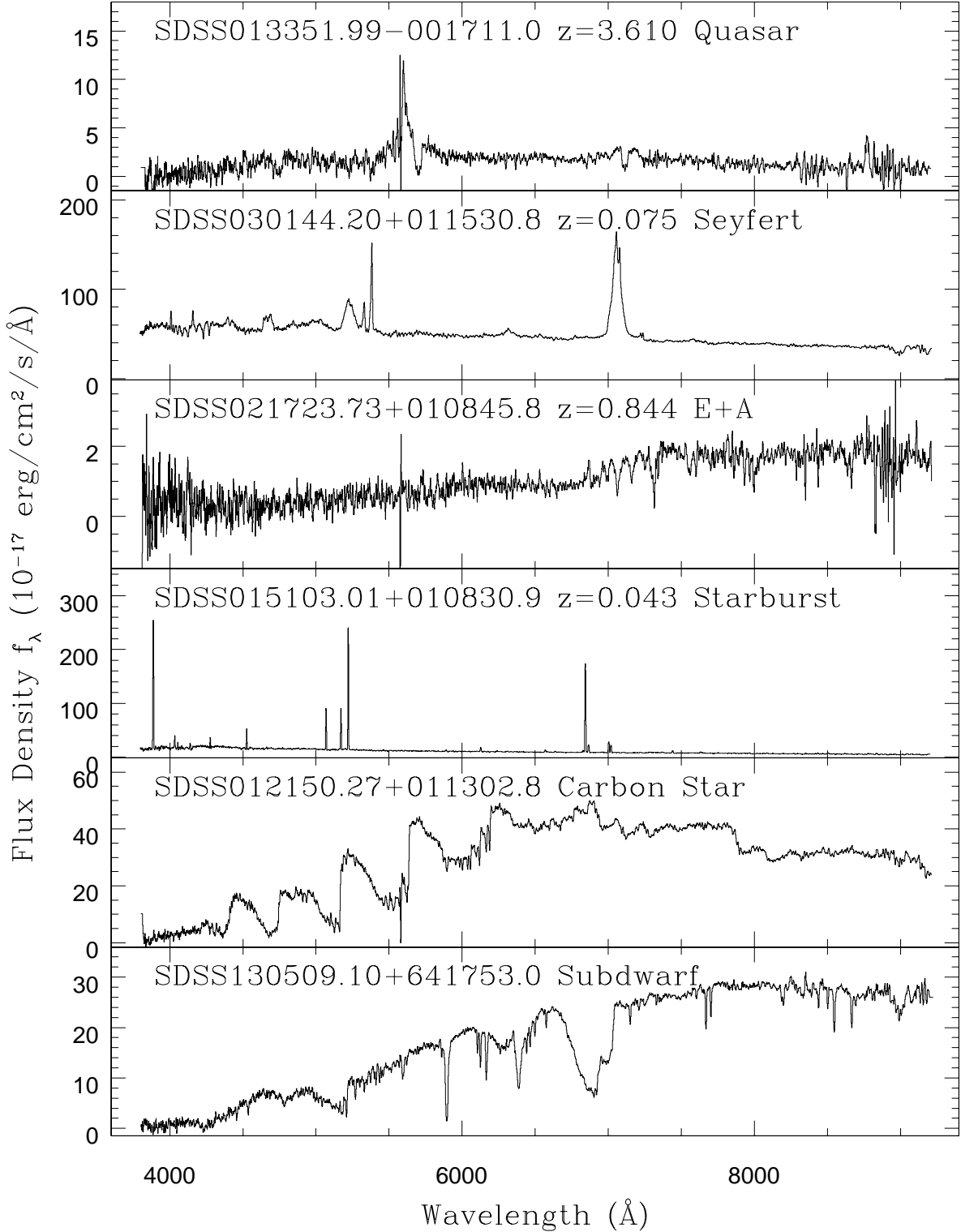}
\caption{Spectra of objects selected by the algorithm.  The top two
panels show representative quasars/AGN at high and low redshift,
respectively; note the stellar absorption lines in the latter.  The
quasars make up 65\% of all objects selected by the algorithm.  Also
shown are representative examples of other categories of contaminating
objects: a high-redshift E+A galaxy, a starburst galaxy, a carbon
star, and a low-metallicity subdwarf star.\label{fig:fig12}}
\end{figure}

Figure~\ref{fig:fig12} shows examples of two quasars from the sample:
one at z=3.6 which shows strong self-absorption in the \ion{C}{4} and
\ion{N}{5} lines, and one low-redshift Seyfert I galaxy (note the Ca H
and K stellar absorption lines).  Also shown are examples of some of
the non-quasars which are selected as quasar candidates: an E+A galaxy
\citep[e.g.,][]{zzl+96}, whose Balmer break gives colors reminiscent
of a quasar with the onset of the Lyman alpha forest at $z = 4.7$; an
emission-line dominated starburst galaxy; a carbon star (e.g., Margon
et al 2002, in preparation); and a low-metallicity subdwarf, whose
broad absorption lines give colors quite separate from the stellar
locus.

In summary, our current analysis shows that the algorithm is meeting
the a priori requirement of 65\% efficiency and we fully expect that
further observations will confirm this result.  This efficiency is a
function of the selection method with efficiencies of 75.0\%, 54.4\%
and 87.0\% expected for low-$z$ ($ugri$) color selection, high-$z$
($griz$) color selection, and FIRST radio selection, respectively.

\subsection{Density}

Based on the testbed data, the SDSS quasar target selection algorithm
selects an average of $18.7$ quasar candidates per square degree; the
$1\sigma$ error in the mean is $2.9$ quasars per square degree.  The
density as a function of category is given in Table~\ref{tab:tab8}
based on the $446.3952$ square degrees of sky in the testbed.  The
mean densities (quasars per square degree) as a function of category
are $13.0$, $7.7$, and $0.7$ for low-$z$, high-$z$, and FIRST targets,
respectively.  Note that the sum of the densities does not equal the
overall density due to overlaps between the categories.  Also, the
FIRST density is underestimated since not all of the runs were matched
to FIRST, yet we used the total area to compute the density, since it
is difficult to determine the exact area in common.  The true density
of FIRST sources is closer to one quasar per square degree; however,
only a small fraction of these are not also selected by the color
algorithm, so this will have little effect on our total density of
quasar candidates.

\subsection{Color and Redshift Distributions}

\begin{figure}[p]
\epsscale{1.0}
\plotone{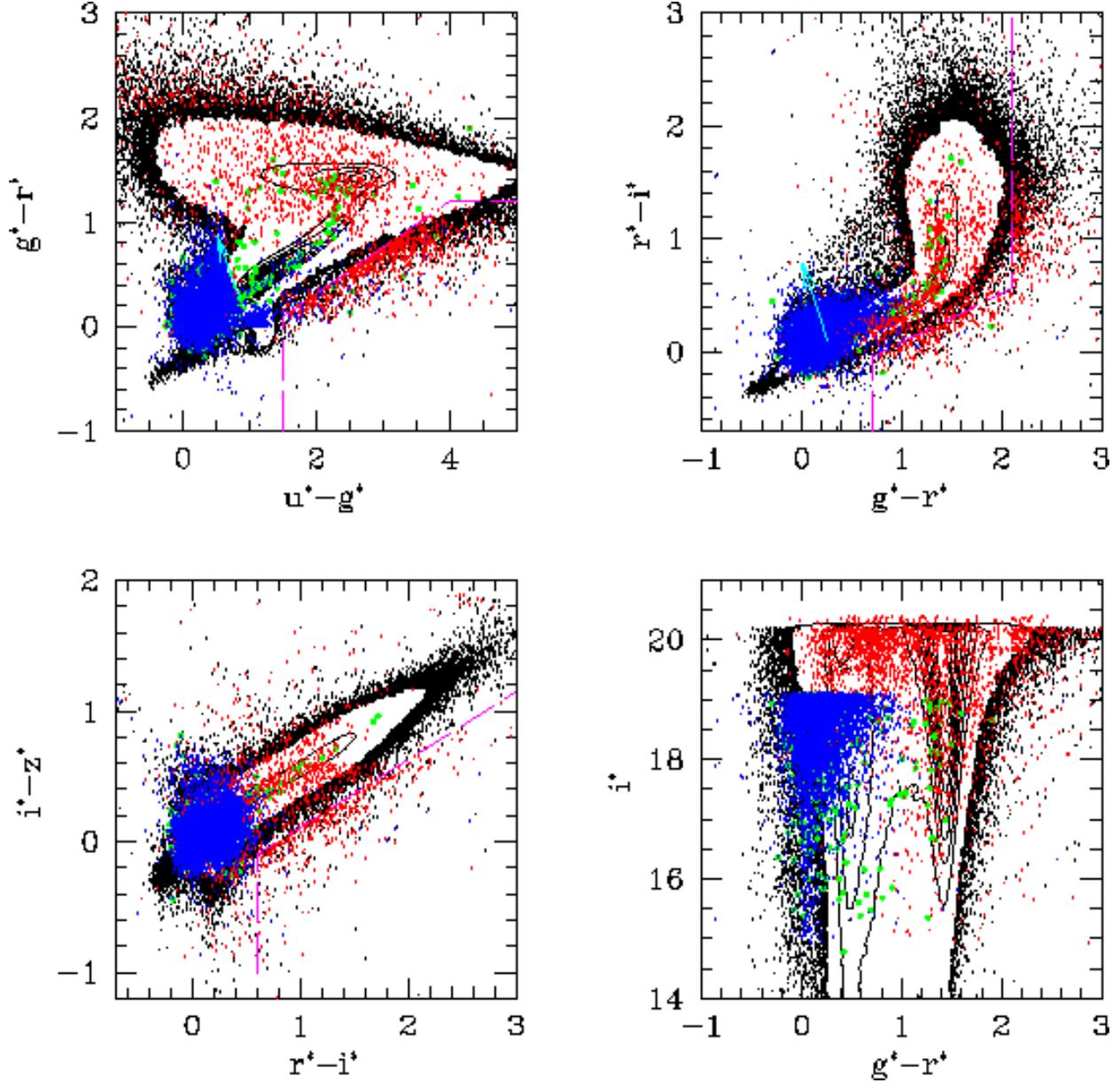}
\caption{Distribution of 8330 quasar candidates from the testbed.
Black points and contours define the stellar locus.  Blue points are
$ugri$-selected quasar candidates, red points are $griz$-selected
quasar candidates, whereas green points are FIRST-selected quasar
candidates.  The light blue lines show the blue end of the
$ugri$-selection extended object cut.  The magenta lines show part of
the three different high-$z$ quasar inclusion regions. The vectors are
as in Figure~\ref{fig:fig2}. \label{fig:fig13}}
\end{figure}

In Figure~\ref{fig:fig13} we show the color-color and color-magnitude
distributions of all 8330 quasar candidates from the testbed.  Blue
points show the distribution of low-$z$ ($ugri$-selected) quasar
candidates, whereas red points are high-$z$ ($griz$-selected), and
green points are FIRST-selected candidates.  The magenta lines show
the main parts of the three high-$z$ inclusion regions (see
\S~\ref{sec:inclusion}) and the light blue lines show the blue extent
of the extended object rejection for $ugri$-selected objects.

\begin{figure}[p]
\epsscale{1.0}
\plotone{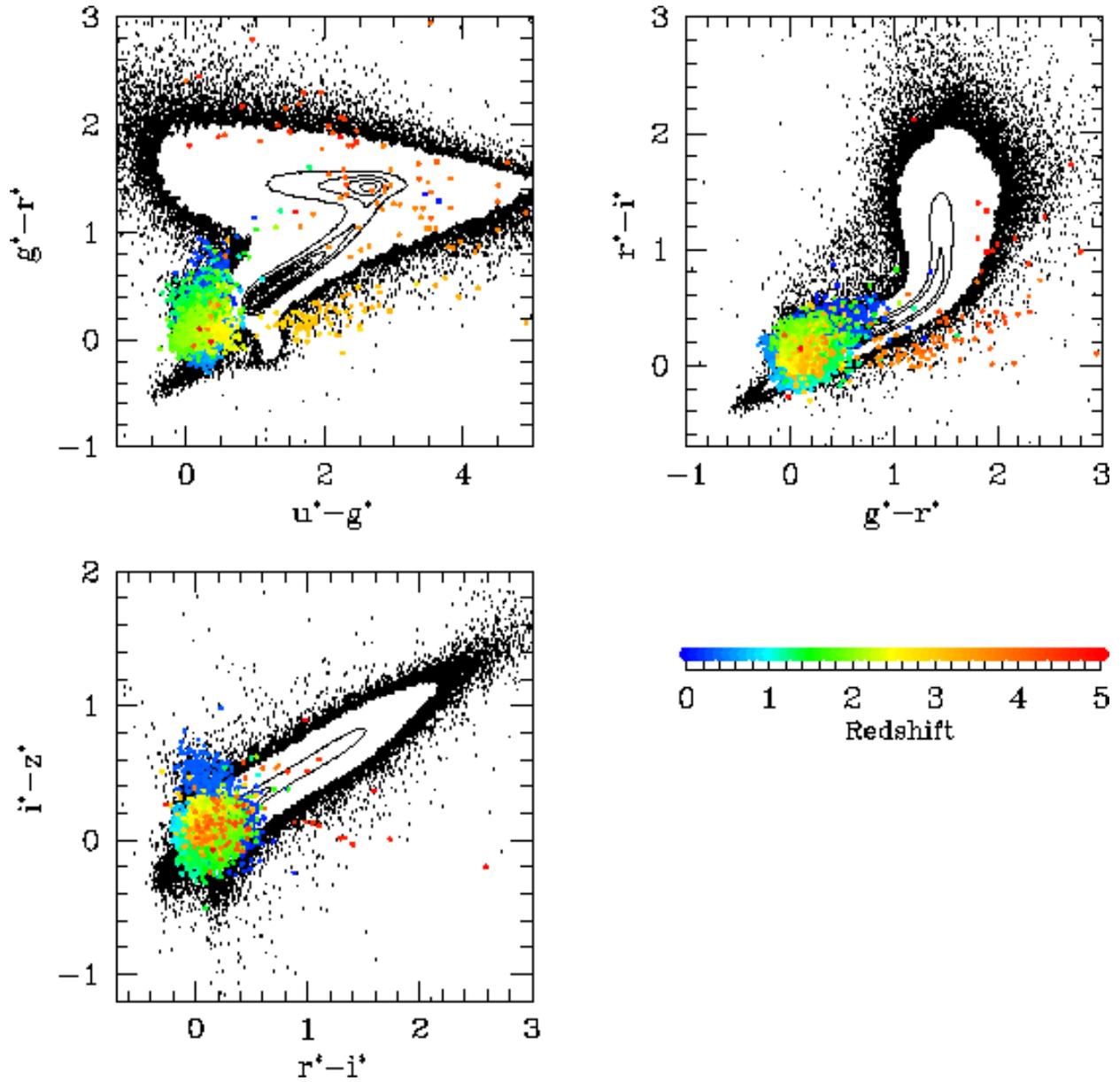}
\caption{Color-color plots of 3040 SDSS quasars from the sample of
8330 quasar candidates shown in Figure~\ref{fig:fig13}.  Quasars
are given by colored points where the colors are indicative of their
redshift. \label{fig:fig14}}
\end{figure}

\begin{figure}[p]
\epsscale{1.0} 
\plotone{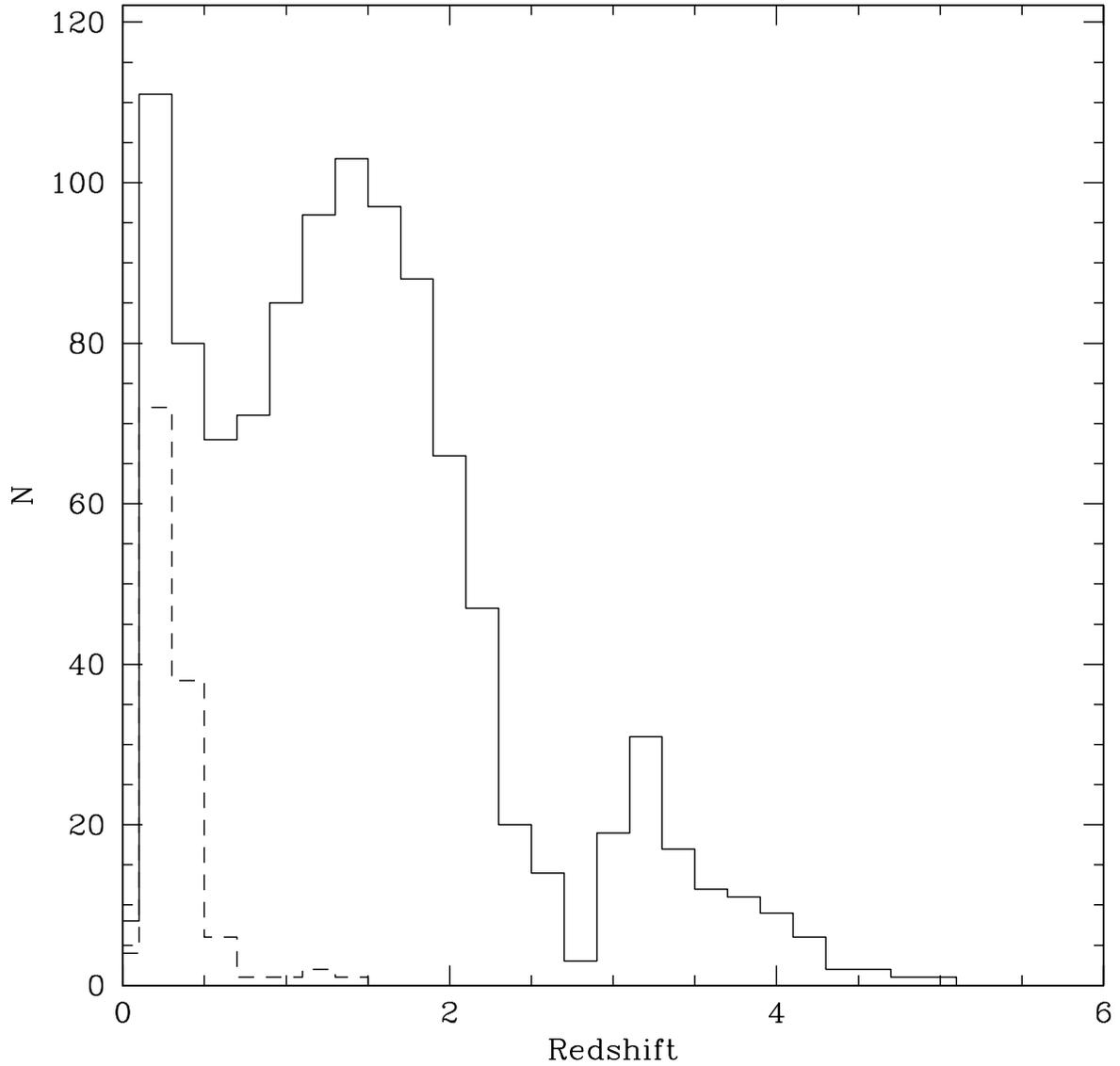}
\caption{Histogram of the first 1073 confirmed quasars that were
selected with the algorithm presented herein; the dashed histogram
indicates those objects that are classified as extended.  Note that
the peak at $z\sim0.25$ is caused by Seyfert galaxies (which are
likely to be classified as extended) and that the dip near $z\sim2.7$
is produced by the degeneracy of SDSS colors of stars and quasars for
quasars at or near this redshift. \label{fig:fig15}}
\end{figure}

For comparison, in Figure~\ref{fig:fig14}, we show the color-color
distribution of 3040 quasars out of 8330 quasar candidates in the
testbed\footnote{Note that many of the 8330 quasar candidates from the
testbed do not yet have spectra, thus the 3040 confirmed quasars do
not represent all of the true quasars in this sample.}.  In this
figure, the colors of the points indicate the redshift of the object
as given in the figure legend.  In Figure~\ref{fig:fig15} we show the
redshift distribution of the first 1073 confirmed quasars selected
using the algorithm described herein.  The color-redshift relation for
SDSS quasars was presented in \citet{rfs+01}.

\section{Discussion\label{sec:discussion}}

Multi-color quasar target selection is easy in principle as we have a
good idea of what regions of color space are inhabited by quasars.
However, it is difficult to estimate how well one's algorithm handles
more exotic objects and/or objects that occupy previously un- or
under-explored regions of parameter space.  This is one area where the
SDSS has an advantage over many previous quasar surveys.  The quality
and quantity of the data allow us to take full advantage of exploring
large regions of parameter space. In fact, the quasar algorithm has
allowed the discovery of a number of unusual objects, including
extreme BAL quasars; see \citet{hal02}.

The SDSS quasar selection algorithm was largely designed prior to the
start of the survey and was based on the idea that it should be
possible to roughly fit a multi-dimensional surface to the locus
defined by normal stars.  Using a small amount of commissioning data
and the work of \citet{ny97}, we defined the initial set of stellar
locus parameters.  The evolution of these parameters based on a much
larger set of data provides the backbone for the final version of the
algorithm that we have described herein.

With the knowledge that we have gained in constructing this algorithm
and with the plethora of SDSS data that is now available to us, we can
imagine how one might devise a superior algorithm, by making use of
probability density functions in multi-dimensional parameter space.
Such an algorithm might be designed as follows.  First, take a large
amount of SDSS data that has known spectral properties (or obvious
spectral properties that can be discerned purely from the colors of
the objects).  Place all of these objects on a 5D grid that includes
four colors and a magnitude.  At each of these grid points, we can
then compute the probability that the objects in the bin are quasars.
With this information, in principle, one could take a new object and
compare it to a look-up table of probabilities based on the control
sample.  The definition of quasar candidates would then involve
nothing more than setting a probability threshold.

The main difference between a probability density type of algorithm
discussed above and the algorithm that we have actually implemented is
that the algorithm described herein does not take into account the
fact that the stellar locus cross-section is not necessarily Gaussian,
nor that the density of stars along the locus is not the same.  Thus
this algorithm does not choose a threshold at constant stellar
density.  Nevertheless the algorithm does create a well-defined
stellar locus.

In practice, the size of the look-up table would make a probability
density type of algorithm impossible to implement as a result of the
amount of time it would take to run such an algorithm.  However, it is
possible to overcome this hurdle by implementing a Gaussian mixture
model based on the Expectation Maximization (EM) Algorithm
\citep{con+00}, which fits Gaussians to the distribution of objects in
multidimensional parameter space and has the effect of reducing an
otherwise unwieldy look-up table to a much smaller number of Gaussian
fits that describe the density of objects in multi-dimensional
parameter space.  We have begun work on just such an algorithm, which
we hope will eventually serve as a supplement to, and an independent
crosscheck on, the current algorithm.

\section{Conclusions\label{sec:conclusions}}

We have presented the SDSS Quasar Target Selection Algorithm.  Initial
testing shows that the algorithm is better than 90\% complete as
determined by both previously known quasars and simulated quasars ---
consistent with the 90\% completeness requirement that was established
prior to the development of the algorithm.  The expected efficiency of
the algorithm (in terms of what fraction of objects targeted as
quasars turn out to be quasars) is $\sim65$\% --- again, consistent
with the original requirement of 65\%.  This combination of
completeness and efficiency is unprecedented in quasar surveys.

The high-quality CCD photometry coupled with the stellar locus outlier
and FIRST selection techniques will aid in the exploration of new
regions of parameter space that have been unexplored in previous
surveys that have tended to rely on 2-3 color photometry from
photographic plates using multi-color selection techniques.
Furthermore, the high efficiency allows us to complete the goal of
obtaining spectra for 100,000 quasars during the SDSS Quasar Survey in
a remarkably short period of time, with relatively little
contamination from non-quasars --- all while maintaining a high degree
of completeness as a function of quasar parameter space.

\clearpage

\acknowledgements

This work was supported in part by National Science Foundation grants
AST99-00703 (GTR and DPS), PHY00-70928 (XF) and AST00-71091 (MAS).
MAS acknowledges additional support from the Princeton University
Research Board.  XF acknowledges support from a Frank and Peggy Taplin
Fellowship.  This research has made use of the NASA/IPAC Extragalactic
Database (NED) which is operated by the Jet Propulsion Laboratory,
California Institute of Technology, under contract with the National
Aeronautics and Space Administration.

The Sloan Digital Sky Survey (SDSS) is a joint project of The
University of Chicago, Fermilab, the Institute for Advanced Study, the
Japan Participation Group, The Johns Hopkins University, the
Max-Planck-Institute for Astronomy (MPIA), the Max-Planck-Institute
for Astrophysics (MPA), New Mexico State University, Princeton
University, the United States Naval Observatory, and the University of
Washington. Apache Point Observatory, site of the SDSS telescopes, is
operated by the Astrophysical Research Consortium (ARC).  Funding for
the project has been provided by the Alfred P. Sloan Foundation, the
SDSS member institutions, the National Aeronautics and Space
Administration, the National Science Foundation, the U.S. Department
of Energy, the Japanese Monbukagakusho, and the Max Planck
Society. The SDSS Web site is http://www.sdss.org/.

\appendix

\section{Construction of the Stellar Locus\label{sec:appendixA}}

\subsection{Overview}

Ideally, the region of color-space inhabited by stars would be defined
using a large, representative sample of stars with zero photometric
errors.  The algorithm to define the stellar locus would identify
regions of color space where the density was large enough to give us
unacceptable contamination in the quasar survey.  Then, the selection
algorithm would select objects which are inconsistent with being stars
at some threshold (say $4\sigma$).

The sample of stars we used to define the locus of stars in the SDSS
is large and representative, but is not free from photometric errors.
We use the algorithm of \citet{ny97} to define the stellar locus.
This algorithm is capable of generating a region of three-dimensional
color space which consists of a series of overlapping right elliptical
cylinders with half-ellipsoids on the ends.  Though it attempts to
place these cylinders on the densest parts of color space, it falls
short of describing a surface of constant stellar density.

We parameterize the locus in the regions of color space extending from
F stars through M stars.  A stars have low enough density relative to
the F-M stars that the \citet{ny97} algorithm does not fit them, so
these stars are eliminated with a separate color cut.  Very hot O and
B stars are rare enough that they can be ignored.  Stars cooler than M
(L and T dwarfs) are also sufficiently rare that we do not include
them in our stellar locus definition.

Since the SDSS data includes four colors, we defined two
three-dimensional loci; the $ugri$ locus uses ($u^*-g^*$),
($g^*-r^*$), and ($r^*-i^*$) colors, and the $griz$ locus uses
($g^*-r^*$), ($r^*-i^*$), and ($i^*-z^*$) colors.  The $ugri$ locus
does most of the work for the selection of $z \le 3$ quasars, whereas
the $griz$ locus is primarily responsible for higher redshift quasars.
Because our detection rate for $2.5 \le z \le 3.0$ quasars was
unacceptably low, we also define a ``mid-$z$'' locus which is the same
as the $ugri$ locus except that the width of every cylinder is
smaller, allowing us to dig further into the locus where we expect
these interesting quasars.  We refer the reader to
Figures~\ref{fig:fig2},~\ref{fig:fig3},~\ref{fig:fig5},
and~\ref{fig:fig6}, which show the locus parameterizations
for the three loci that are discussed herein.

\subsubsection{Locus Construction\label{sec:locusconstruct}}

The goal of locus construction is to define the series of right
elliptical cylinders so that it encloses the highest density regions
of stars in color space.  Each cylinder can be described by a center
in three-dimensional color space, a unit vector along the cylinder
axis ($\hat{k}$), the position angle of the major axis (defined as
$\cos(\theta) = \hat{l} \cdot [(\hat{k} \times \hat{z}) \times
\hat{k}] / |\hat{k} \times \hat{z}|$, where $\hat{z}$ is the reddest
color axis, $\hat{l}$ is the unit vector along the major axis of the
cross section, and $\hat{m}$ is the unit vector along the major axis
of the cross section), and the length of the major ($a_l$) and minor
($a_m$) axes.  The ends of each cylinder are naturally defined by the
planes perpendicular to, and bisecting, lines drawn between the center
of the cylinder and the cylinders which precede and follow it.  The
bare end of the first and last cylinders are terminated at the centers
of the half ellipsoids attached to the ends of the locus.  In
practice, we define the ellipsoid on the red end so that the reddest
cylinder effectively extends indefinitely.  Tables~\ref{tab:tab3}
and~\ref{tab:tab4} give the locus parameters.

The centers of the cylinders (``locus points'') are determined by a
stable, iterative process of adding locus points and moving them to
the centers of local high density regions \citep{ny97}.  The unit
vector, $\hat{k}$, points from the center of the preceding cylinder to
the center of the following cylinder.  The position angles of the
major and minor axes are determined from a principal components
analysis of the projected positions of stars within the cylinder onto
a plane perpendicular to $\hat{k}$.  The lengths of the major and
minor axes are four times the one-sigma deviations of the points from
the center of the cylinder.  Note that this procedure allows
discontinuities in the surface which separates the locus of stars from
the region in which quasars are selected, as is evident, e.g., in
Figure~\ref{fig:fig2}.  The blue end of the stellar locus is described
by a half ellipsoid which fits onto the end of the bluest cylinder.  A
single input parameter ($k_{\rm blue}$) determines the distance from
the center of the bluest cylinder, along its axis, to the center of
the end ellipsoid.  Two of the axis lengths are given by the major and
minor axis lengths of the bluest cylinder.  The length of the
ellipsoid axis along the stellar locus is given by a separate input
parameter, $a_{k {\rm blue}}$.  In the limit that $a_{k {\rm blue}} =
0$, the locus ends with a plane rather than an ellipsoid.

\subsubsection{Locus Specifics}

The input data used to define the three stellar loci consisted of
$400,000$ point sources from SDSS imaging runs 94, 125, 752, and 756
(see \citealt{sto+01}, Table~3 and Figure~1 for coordinates) using the
versions processed prior to 2000 November 3.  More recent processing
versions are available, but we found that constructing new loci with
the newer data did not change the locus parameters enough to justify
using the new parameters, given that many quasars had already been
selected using the older parameters.  These sources included only
those objects with errors smaller than $(0.1, 0.03, 0.03, 0.03, 0.06)$
in $(u^*, g^*, r^*, i^*, z^*)$.  In addition, objects with flags that
would cause them to be rejected by the quasar target selection
algorithm were removed from the sample (\S~\ref{sec:flags}).

In order to construct the loci, we must first define some input
parameters.  See \citet{ny97} for the definition of these quantities.
The algorithm from \citet{ny97} was extended to include ellipsoids at
the red and blue ends, following the definitions in
\S~\ref{sec:locusconstruct}.  For the construction of the $ugri$
stellar locus we first exclude point sources with $(u^*-g^*)<0.5$ or
$(g^*-r^*)<0$, which are likely to be quasars and A stars,
respectively (see Figures~\ref{fig:fig2} and~\ref{fig:fig3}).  Next we
chose endpoints of $\mathbf{r}_{\rm start} = (u^*-g^*, g^*-r^*,
r^*-i^*) = (0.75, 0.25, 0.1)$ and $\mathbf{r}_{\rm end} = (2.55, 1.3,
1.2)$.  The maximum distance from the locus point associated with a
star was chosen to be $d_x = d_y = d_z = 0.4$.  The locus width
spacing was defined by setting $N_{\sigma_{\rm spacing}} = 3$, such
that new locus points are not added if they are closer to each other
than $N_{\sigma_{\rm spacing}}$ times the major axis of the ellipse
fitted to the locus at a given point.  The algorithm was allowed to
iterate 10 times, and the maximum number of locus points was
constrained to be less than 20.  The ``width'' of the locus was set to
be $N_{\sigma_{\rm width}} = 4$ and the errors were convolved with
this width with $N_{\sigma_{\rm errors}} = 4$.  The end caps were
defined according to $k_{\rm blue} = -0.05$, $k_{\rm red} = 100$,
$a_{k\,{\rm blue}} = 0.2$, and $a_{k\,{\rm red}} = 0.0$.  The
parameterization of the $ugri$ stellar locus is given in
Table~\ref{tab:tab3}.

The mid-$z$ locus is exactly the same as the normal $ugri$ locus,
except that $N_{\sigma_{\rm width}} = 2$ for the mid-$z$ locus.  This
smaller locus width allows us to probe closer to the most dense
regions of the stellar locus when looking for $2.5 < z < 3.0$ quasars
that can have colors that are similar to normal stars.  The end caps
for the mid-$z$ locus were defined according to $k_{\rm blue} =
-0.05$, $k_{\rm red} = 100$, $a_{k {\rm blue}} = 0.1$, and $a_{k {\rm
red}} = 0.0$.

For the $griz$ stellar locus, $\mathbf{r}_{\rm start} = (g^*-r^*,
r^*-i^*, i^*-z^*) = (-0.1, -0.1, -0.1)$ and $\mathbf{r}_{\rm end} =
(1.4, 1.5, 0.8)$.  As with the $ugri$ locus, $N_{\sigma_{\rm spacing}}
= 3$, $d_x = d_y = d_z = 0.4$, $N_{\sigma_{\rm width}} = 4$,
$N_{\sigma_{\rm errors}} = 4$ and the number of iterations was set to
10; however, the maximum number of locus points was 25.  The end caps
were defined according to $k_{\rm blue} = -0.3$, $k_{\rm red} = 100$,
$a_{k {\rm blue}} = 0.5$, and $a_{k {\rm red}} = 0.0$.  The
parameterization of the $griz$ stellar locus is given in
Table~\ref{tab:tab4}, see also Figures~\ref{fig:fig5}
and~\ref{fig:fig6}.

\section{Details of the Outlier Rejection Algorithm\label{sec:appendixB}}

\subsection{Overview}

This appendix describes the algorithm which selects objects that,
within the measured photometric errors, do not fall within the stellar
locus parameterization (see Appendix~A).  It is intended that the
stellar locus parameterization describe a region of color space
inhabited by stars, with no photometric errors included.  If we had
exact three-color photometry for each SDSS object, then our color
selection routine would simply select all objects exterior to the
parameterized locus.  Simply stated, {\it we select all objects whose
photometric error ellipsoids (defined by $N_{\sigma_{err}}$ times the
one-sigma error ellipsoid) do not intersect the stellar locus.}  The
algorithm described in this appendix does nothing else.  It does not
use information about the locus of quasars, the object profiles, or
the properties of any other astronomical objects.  There are no
astrophysical decisions buried in the mathematics, which is far more
complicated than the statement of the problem which it addresses.
        
The mathematics solves this problem for three-dimensional color space.
These three colors are designated $(x, y, z)$.  In practice, they are
either $(u-g, g-r, r-i)$ or $(g-r,r-i,i-z)$.  To generate the error
ellipse for each object, we must additionally know the variances and
covariances in each color.  The target selection algorithm currently
assumes that the errors in each {\it filter} are uncorrelated and free
from systematics (and thus uncorrelated), so that $cov(a-b,b-c) =
-var(b)$, where $a, b,$ and $c$ are magnitudes in a single filter.
The variance for a given color is given by $var(a-b) = var(a) +
var(b)$.  This is a good estimate if the errors are not too large.

Objects with large errors are typically near the detection threshold,
and may be undetected in one or more filters.  This results in
incomplete knowledge of the position in color space.  For example, if
an object was detected in $g$, but was below the detection threshold
in $u$, then we know that $u-g > u_{lim} - g$.  If the $u$ measurement
was missed due to a defect in the image, then nothing would be known
about the $u-g$ color.  In general, we either know the color and
variances of an object, or we know a limiting color.  The limiting
color is calculated from the flux corresponding to $N_{\sigma_{err}}$
times the one-sigma error in flux.  In this case, the algorithm
determines if there is any allowed value of the unknown colors which
places the object in the stellar locus.  We require that at least one
color out of three be measured.

\subsection{Finding the Closest Locus Point}

Given the measured colors (but not the errors in the colors), we
determine which locus point is closest to the colors of the input
object.  We do this by looking at the Euclidean distance in color
space, not taking into account any variation in the width of the
locus, to simplify the computation.  

In the case that one or two of the colors were not measured, the
distance to each locus point is calculated after first moving the data
point as close as is allowed by the color limits.

If two locus points are equally distant from the datapoint, the redder
one is chosen.  If the closest locus point is not between the centers
of the ending ellipsoids, then the closest locus point is reassigned
to be the first one interior to the ellipsoid centers.

\subsection{Handling Errors in Individual Objects}

The closest locus point has an associated unit vector along the axis
of a right elliptical cylinder, and a defined elliptical cross section
and orientation, which define the local region of color space
inhabited by the locus of stars.  The catalog entry under
consideration is assumed to be a star if the extent of its error
ellipsoid overlaps any portion of this right elliptical cylinder which
is interior to the ellipsoids at the ends of the locus.  In practice,
we determine this by adjusting the size of the cylinder containing the
locus of stars according to the extent of the error ellipse, and then
asking whether the object's colors place it interior to this larger
right elliptical cylinder.  This is described in more detail below.

First, we estimate the error ellipse in the $(l, m)$ plane.  The
variance, covariance matrix for this plane is given by:
\[V_{ab} = \sum_{x} \sum_{y} \frac{\partial a}{\partial x} \frac{\partial b}{\partial y} S_{xy},\]
where ${\bf V}$ is the (two dimensional) covariance matrix in the
$l,m$ plane, ${\bf S}$ is the (three dimensional) covariance matrix in
color space, and the sums are over the three coordinates: $u-g, g-r,$
and $r-i$ (for the $ugri$ color-cube, and similarly for the $griz$
color-cube).  The error ellipse is given by:
\[{\bf \vec{l}^T V^{-1} \vec{l} } = N_{\sigma_{err}}^2,\]
where ${\bf \vec{l}} = l{\bf\hat{l}} + m{\bf\hat{m}}$.  Solving this
set of equations, the ellipse parameters are:
\[a_{err}^2 = N_{\sigma_{err}}^2 \frac{V_{ll} + V_{mm} \pm \sqrt{(V_{ll}-V_{mm})^2 + 4 V_{lm}^2}}{2},\]
\[ \tan \theta_{err} = \left\{ \begin{array}{ll}
                0       & V_{lm} = 0, V_{ll}>V_{mm} \\
                \infty  & V_{lm} = 0, V_{mm}>V_{ll} \\
                \frac{-(V_{ll}-V_{mm})+\sqrt{(V_{ll}-V_{mm})^{2}+4V_{lm}V_{lm}}}{2V_{lm}}       & V_{lm} \neq 0
                \end{array}
        \right. \]

Now that we have the parameters for the error ellipse, we will adjust
the stellar locus to reflect this information by ``convolving'' the
$N_{\sigma_{err}}$ error ellipse and the cross-sectional locus
ellipse. When convolving the two ellipses, we will make two important
assumptions: the error distribution around each data point is
Gaussian, and the distribution of stars within the stellar locus is
Gaussian. These assumptions allow us to represent each of the ellipses
as a Gaussian function, and, as such, we will convolve them.  If one
convolves two bivariate Gaussians, $G_1$ and $G_2$ (with elliptical
cross sections), one obtains a third bivariate Gaussian with
elliptical cross section. Here we will convolve the parameterized
locus ellipse for the stellar locus in the $l$-$m$ plane with the rotated
error ellipse, also in the $l$-$m$ plane. By convolving two Gaussians with
these ellipses as their cross sections, we will discover a third
ellipse which we will use to determine if the data point is within the
locus, or if it is a viable candidate for target selection.

\[G_{convolved} = \int_{-\infty}^{\infty}\int_{-\infty}^{\infty}{G_1(l-l', m-m')G_2(l', m')}dl'dm'\]
\[G_1 = \exp\left[-\left(\frac{l^2}{2a_{l}^2} + \frac{m^2}{2a_{m}^2}\right)\right]\]
\[G_2 = \exp\left(-\left[\frac{(l\cos \theta_{err} + m\sin \theta_{err})^2}{2a_{err maj}^2} + \frac{(-l\sin \theta_{err} + m\cos \theta_{err})^2}{2a_{err min}^2}\right]\right)\]

In this convolution, $G_1$ represents the parametrized locus and $G_2$
represented the error ellipse rotated to the $l$-$m$ plane.  After the
convolution we are left with a Gaussian which is given by:
\[K_{1}K_{2}\exp\left(\frac{-l^2}{2a_{l}^2} -  \frac{m^2}{2a_{m}^2} + \frac{l^2}{4Aa_{l}^4} + \frac{T^2}{4R}\right)\]

where:
\[A = \frac{ \cos^2 \theta_{err}^2}{2a_{err maj}^2} + \frac{\sin^2\theta_{err}^2}{2a_{err min}^2} + \frac{1}{2a_{l}^2}\]\
\[T = \frac{2l \sin \theta_{err} \cos \theta_{err} (a_{err maj}^{-2}-a_{err min}^{-2})}{4Aa_{l}^2}\]\
\[R=\frac{\sin^2 \theta_{err}}{2a_{err min}^2} + \frac{\cos^2 \theta_{err}}{2a_{err maj}^2} + \frac{1}{a_{l}^2} + \frac{\cos^2 \theta_{err} \sin^2 \theta_{err} (a_{err maj}^{-1} - a_{err min}^{-1})}{4A}.\]\

Here $K_{1}$ and $K_{2}$ are products of the integration, while the various terms within the exponent come from completing the square to facilitate the integration. Since we are only concerned with the elliptical cross section contained in the exponent, we will ignore the rest of the Gaussian and concentrate solely on the exponent. After considerable algebra the convolved ellipse looks like the following:
\[\alpha_{o}l^2-\beta_{o}lm+\gamma_{o}m^2=1\]
where:
\[\alpha_{o} = \frac{1}{4R}\left(\frac{4R}{4Aa_{l}^4} - \frac{4R}{2a_{l}^2} + \frac{\sin^2\theta_{err}\cos^2\theta_{err}(a_{err maj}^{-2} - a_{err min}^{-2})^2}{Aa_{l}^2}\right)\]\
\[\beta_{o} = \frac{1}{4R}\left(\frac{\sin\theta_{err}\cos\theta_{err}(a_{err maj}^{-2} - a_{err min}^{-2})}{4Aa_{l}^2}\right)\]\
\[\gamma_{o} = \frac{1}{4R}\left(\frac{1}{a_{m}^4} - \frac{4R}{2a_{m}^2}\right).\]\
 
After a great deal of simplification, the convolved ellipse is given
by:
\[\alpha l^2 - 2 \beta lm + \gamma m^2 = d,\]
where
\[\begin{array}{l}
d = a_{err maj}^2 a_{err min}^2 + (a_l^2 a_{err maj}^2 + a_m^2 a_{err min}^2) \sin^2 \theta_{err} + (a_{err maj}^2 a_m^2 + a_l^2 a_{err min}^2) \cos^2 \theta_{err} + a_l^2 a_m^2\\
\alpha = a_{err min}^2 \cos^2 \theta_{err} + a_{err maj}^2 \sin^2 \theta_{err} + a_m^2 \\
\beta = \sin \theta_{err} \cos \theta_{err} (a_{err maj}^2 - a_{err min}^2) \\
\gamma = a_{err min}^2 \sin^2 \theta_{err} + a_{err maj}^2 \cos^2 \theta_{err} + a_l^2 .
\end{array}\]

The major and minor axes of the convolved ellipse are given by the
eigenvalues of the matrix form of the ellipse and the angle of
inclination is the angle of the eigenvector pointed in the direction
of the major axis.  The variance/covariance matrix for the convolved
ellipse is:
\[V_{convolved} = \frac{1}{d}\left( \begin{array}{cc} \alpha & -\beta \\ -\beta & \gamma\end{array} \right)\\
\] 
The characteristic equation of this matrix is given by
\[\lambda^2 - \frac{\alpha + \gamma}{d}\lambda +\frac{\alpha\gamma - \beta^2}{d^2}\]
Solving this for the eigenvalues we get $\lambda_{1}=(a+b)^{-1}$ and
$\lambda_{2}=(a-b)^{-1}$. These eigenvalues are the values which
diagonalize the matrix, and thus they give the coefficients to any
point $(l,m)$ which happens to lie on either the major or minor
axis. Thus the major and minor axes are given by:
\[a_{tot maj}^2 = a + b,\]
\[a_{tot min}^2 = a - b,\]
The angle of rotation of the ellipse is found from the angle of
the eigenvector pointed towards the major axis:
\[\tan \theta_{tot} = \left\{ \begin{array}{ll}
		0			&\beta=0, \alpha=\gamma\\
		\infty			&\beta=0, \alpha\neq\gamma\\
		\frac{\alpha-\gamma+\sqrt{(\alpha-\gamma)^2+4\beta^2}}{2\beta} &\beta  \neq 0, \theta_{err} \ge 0\\
		-\frac{\alpha-\gamma+\sqrt{(\alpha-\gamma)^2+4\beta^2}}{2\beta} &\beta \neq 0, \theta_{err} < 0
		\end{array}
	\right. ,\]
where
\begin{eqnarray}
a &=& \frac{(\alpha+\gamma)}{2} = \frac{a_{err maj}^2 + a_{err min}^2 + a_l^2 + a_m^2}{2}, \nonumber\\
b &=& \frac{\sqrt{(\alpha + \gamma)^2 - 4 (\alpha \gamma - \beta^2)}}{2} =  \nonumber\\
  & & \frac{\sqrt{(a_{err maj}^2-a_{err min}^2)^2 + 2 (a_{err maj}^2-a_{err min}^2)(a_l^2-a_m^2)(\cos^2 \theta_{err} - sin^2 \theta_{err}) + (a_l^2-a_m^2)^2}}{2}. \nonumber
\end{eqnarray}
Here, $a$ and $b$ are positive by definition.  We need not worry about
taking the square root of a negative number when computing the angle,
since $a-b \leq \alpha \leq a+b$ for all cases.  If $b=0$, then the
resulting ellipse is circular, so we arbitrarily assign $\theta=0$

We now can redefine the ${\bf\hat{l}}, {\bf\hat{m}}$ axes associated
with the closest locus point to describe this new elliptical cross
section.  The value of $\theta_{tot}$ must be added to the original
angle $\theta$ describing the locus ellipse, since we have calculated
the angle with respect to the ${\bf\hat{l}}$ axis.  When adding, we
must be sure to keep the new value of $\theta$ in the allowed range:
$-\frac{\pi}{2} < \theta \leq \frac{\pi}{2}$.

In addition to increasing the width of the excluded locus, we increase
$a_k$ on the ends of the locus in a similar manner:
\[var(k) = \sum_x \sum_y  \frac{\partial k}{\partial x} \frac{\partial k}{\partial y} S_{xy},\]
\[a_{k tot} = \sqrt{a_k^2 + N^2_{\sigma_{err}} var(k) }, \]
where the k in this equation is not for the locus point which is closest to the datapoint, but instead 
for the locus point which is closest to the respective endpoints (subject to being between the two
endpoints).

In the remainder of this paper we will use $a_l, a_m, a_k,$ and
$\theta$ to refer to the derived $a_{tot maj}, a_{tot min}, a_{k tot}$
and $\theta + \theta_{tot}$.  Likewise, the ${\bf\hat{l}}, {\bf\hat{m}}$ unit
vectors are in the new coordinate system.

\subsection{Dealing with Non-detections in Some Filters}

If not all of the colors of the object are known, then we are free to
find the allowed point in color space that is most likely to be in the
locus.  This is {\it not} in general the same as finding the colors
that are closest to the locus point in the Euclidean sense, since the
parameterized region is an elliptical cylinder, not a sphere centered
at the locus point.  We instead wish to minimize the $r^*$ distance to
the locus, where $r^*$ is given by:
\[r^* = \sqrt{\left( \frac{l}{a_l} \right)^2 + \left( \frac{m}{a_m} \right)^2},\]
where $l \equiv \Delta {\bf\vec{r}} \cdot {\bf\hat{l}}, m \equiv \Delta {\bf\vec{r}} \cdot {\bf\hat{m}}$,
$\Delta {\bf\vec{r}} \equiv {\bf\vec{r}} - {\bf\vec{r}}_p$, and ${\bf\vec{r}}_p$ is the position in color space
of the closest point.

First, let's solve the case where only the $x$ color, is unknown.  To
make the equations easier to read, we will use the notation $\Delta
{\bf\vec{r}} \equiv (\Delta x, \Delta y, \Delta z), {\bf\hat{l}} = l_x
{\bf\hat{x}} +l_y {\bf\hat{y}} + l_z {\bf\hat{z}}, {\bf\hat{m}} = m_x
{\bf\hat{x}} + m_y {\bf\hat{y}} + m_z {\bf\hat{z}}$.  The solution is:
\[\Delta x = \left\{ \begin{array}{ll}
	-\frac{\{(a_m^2 l_x l_y + a_l^2 m_x m_y) \Delta y + (a_m^2 l_x l_z + a_l^2 m_x m_z) \Delta z \}}{a_m^2 l_x^2 + a_l^2 m_x^2}	& a_m^2 l_x^2 + a_l^2 m_x^2 \neq 0\\
	-\frac{l_y \Delta y + l_z \Delta z}{l_x}	& a_m^2 l_x^2 + a_l^2 m_x^2 = 0, l_x \neq 0\\
	-\frac{m_y \Delta y + m_z \Delta z}{m_x}	& a_m^2 l_x^2 + a_l^2 m_x^2 = 0, m_x \neq 0\\
	0	& a_m^2 l_x^2 + a_l^2 m_x^2 = 0, l_x = 0, m_x = 0
	\end{array}
	\right. .\]
The first case above is the result of the minimization process.  If
the denominator is zero, but $l_x \neq 0$, then it follows that $a_m =
0$.  Also, either $a_l = 0$ or $m_x = 0$.  In the first case, we will
only be able to find a color point in the locus if we can fortuitously
set $m=l=0$ by moving along the $x$ axis.  In the second case, we
cannot affect the magnitude of m.  Either case is optimized by setting
$l = 0$.  If the denominator is zero, but instead $l_x = 0$, then we
still need either $a_l = 0$ or $m_x = 0$.  In the first case we cannot
affect the distance along the major axis, so we set the distance along
the minor axis to zero.  The second case gives us ${\bf\hat{x}} =
{\bf\hat{k}}$, so we might as well set $\Delta x = 0.$

If the value of $x$ is completely unknown, then we use this calculated
$\Delta x$ to determine whether the object is consistent with being a
star.  If the value of the color, $x$, is a limit, then we must ask
whether the computed value is within the limits.  If it is not, then
we instead use the limiting $x$ value to determine whether the
datapoint is consistent with being a star.

Next, we tackle the case where both the $x$ color and $y$ color are
unknown.  When two of the colors are unknown, one can find values that
will put the point exactly on the line ${\bf\vec{r}}-{\bf\vec{r}}_p = \lambda
{\bf\hat{k}},$ where $\lambda$ is a free parameter.  We can solve for
$\lambda$, and then the two unknown colors using:
\begin{eqnarray}
\lambda &=& \left\{ \begin{array}{ll}
	\frac{\Delta z}{k_z},		& k_z \neq 0\\
	0,				& k_z = 0,\\
	\end{array}
	\right. \nonumber\\ 
\Delta y &=& \lambda k_y, \nonumber\\
\Delta x &=& \lambda k_x. \nonumber
\end{eqnarray}
If $k_z=0$, then the line down the center of the locus is in the $x,
y$ plane, so there are many values of $x$ and $y$ which will be on it.
We arbitrarily choose the one on the locus point, since we know that
this value is also within the $k$ limits of the stellar locus.

The existence of limiting values in $x$ and $y$ make this a little
trickier.  First, we calculate the value of $\Delta y$ which places
the point exactly on the center of the locus.  In the case that this
value is not consistent with the $y$ limits, we assign $\Delta y =
y_{lim} - y_p$.  Since we now have only one missing color, we can use
the procedure outlined above (the case where only the $x$ color is
unknown) to calculate the optimal value of $\Delta x$.  If a limit was
not reached, one can verify that the computed $\Delta x$ will be the
same as if we had used $\Delta x = \lambda k_x$.  If this computed
value of $\Delta x$ is within the allowed limits, then we have done
the best we can.  If it is not, then we set $\Delta x = x_{lim} - x_p$
and then compute the optimal value of $\Delta y$ given $\Delta x$ and
$\Delta z$.  The optimal value of $\Delta y$, given $\Delta x$ and
$\Delta z$, can be computed by reassigning the axes ($x \rightarrow y,
y \rightarrow z, z \rightarrow x$) in the equation that optimizes
$\Delta x$ given $\Delta y$ and $\Delta z$.

The above equations and their permutations ($x,y,z \rightarrow y,z,x$
and $x,y,z \rightarrow z,x,y$) are used to determine the values of
$\Delta x$, $\Delta y$, and $\Delta z$ which are most likely to be
included in the stellar locus, given the input limits on these
quantities.  Once we have this coordinate in three dimensions, we can
ask whether it is within the parameterized locus.  We must have $r^*
\leq 1$, where
\[r^* = \left\{ \begin{array}{ll}
	\sqrt{\left( \frac{l}{a_l} \right)^2 + \left( \frac{m}{a_m} \right)^2}	& a_l > 0, a_m > 0\\
	\frac{l}{a_l}					& a_l > 0, a_m=0, m = 0\\
	0									& l = m = 0\\
	\infty									& a_l = 0, l \neq 0\\
	\infty									& a_m = 0, m \neq 0
	\end{array}
	\right.  \]
The point must also be within the ellipsoids at the ends of the locus.
If $\Delta k <0$, then we already know we are within the red end
limit, since the locus point is guaranteed to be within the locus.
Similarly, if $\Delta k >0$, we are guaranteed to be within the blue
end limit.  So, we need only check one of the ellipsoids.  We ignore
the fact that the locus may curve around in $x, y, z$ coordinates, and
only look at the $l, m, k$ coordinates.  We are within the ends of the
locus if $k_{blue\_limit}(l,m) < k < k_{red\_limit}(l,m)$, where:
\[k_{blue\_limit} = k_{blue} - a_{k\ blue} \sqrt{1 - {r^*_{blue}}^2} \]
\[k_{red\_limit} = k_{red} + a_{k\ red} \sqrt{1 - {r^*_{red}}^2}. \]
Here, the values of $r^*$ are computed using $a_l$ and $a_m$ for the
end ellipses, but values of $l$ and $m$ are calculated from the closest
locus point.

\subsection{End Conditions}

If the previously calculated optimal values of $\Delta x, \Delta y,$
and $\Delta z$ put us within the locus, then we are done.  If we could
not find values for the coordinates that made $r^* \leq 1$, then we
are done.  However, if the values failed only the end conditions, then
there is a chance we could still put the datapoint in the locus by
minimizing the distance to the center of the end ellipsoid:
\[r^{**} = \sqrt{\left( \frac{l'}{a'_l} \right)^2 + \left( \frac{m'}{a'_m} \right)^2 + \left( \frac{k' - k'_{end}}{a_k} \right)^2}\]
rather than the center of the cylinder associated with the closest
locus point.  Here, $k_{end}$ is either $k_{blue}$ or $k_{red}$,
depending on whether $\Delta k<0$ or $\Delta k>0$.  The primes
indicate that the values are measured with respect to the position in
color space of the locus point which is closest to the center of the
end ellipsoid.  The $a_l$ and $a_m$ widths are for the end ellipsoid,
not the cylinder associated with the closest locus point.

We now calculate for the datapoint a new optimal color which minimizes
$r^{**}$.  If only $x$ is unknown, we compute $\Delta z'$ and $\Delta
y'$ using the definition $\Delta {\bf\vec{r}}' \equiv {\bf\vec{r}} - {\bf\vec{r}}_e
\equiv (\Delta x', \Delta y', \Delta z'),$ and ${\bf\vec{r}}_e$ is the
position in color space of the locus point closest to the center of
the end ellipsoid (subject to $k_{blue} < k_e < k_{red}$).  The value
of $\Delta x'$ which minimizes $r^{**}$ is
\[\Delta x' = \left\{ \begin{array}{ll}
	- \frac{{a'}_m^2 a_k^2 A l'_x + {a'}_l^2 a_k^2 B m'_x + {a'}_l^2 {a'}_m^2 C k'_x}{{a'}_m^2 a_k^2 {l'}_x^2 + {a'}_l^2 a_k^2 {m'}_x^2 + {a'}_l^2 {a'}_m^2 {k'}_x^2}
										& D_1 \neq 0 \\
	- \frac{a_k^2 A l'_x + {a'}_l^2 C k'_x}{a_k^2 {l'}_x^2 + {a'}_l^2 {k'}_x^2}
										& D_1 = 0, {a'}_m^2 = 0, m'_x = 0, {a'}_l^2 \neq 0 \\
	-\frac{B}{m'_x}				& D_1 = 0, {a'}_m^2 = 0, m'_x \neq 0, {a'}_l^2 \neq 0 \\
	- \frac{a_m^2 A l'_x + {a'}_l^2 B m'_x}{a_m^2 {l'}_x^2 + {a'}_l^2 {m'}_x^2}
										& D_1 = 0, {a'}_m^2 > 0 \\
	0									& D_1 = 0, {a'}_l^2 = 0, m'_x = 0, k'_x \neq 1 \\
	0									& D_1 = 0, l'_x = 1
	\end{array}
	\right. \]
where $A=(l'_y \Delta y' + l'_z \Delta z')$, $B=(m'_y \Delta y' + m'_z
\Delta z')$, $C=(k'_y \Delta y' + k'_z \Delta z' - k'_{end})$, and
$D_1 \equiv {a'}_m^2 a_k^2 {l'}_x^2 + {a'}_l^2 a_k^2 {m'}_x^2 +
{a'}_l^2 {a'}_m^2 {k'}_x^2$.  The first case is the formal result of
the minimization.  The second case is for $a_m^2 = 0$ and $m'_x =0$;
in this case we cannot affect the $m$ value, so we might as well find
the best position within the $l,k$ ellipse.  If instead we have $m'_x
\neq 0$, we instead put the object on the $m=0$ plane.  If $D_1=0$ but
${a'}_m^2>0$, then we must have $a_k^2=kx=0$, since
${a'}_l^2>={a'}_m^2$.  If ${a'}_l^2 = 0$, then the object is either on
the line of the locus or not, no change in x will give different
results than the previous calculated optimal $x$.  All cases where
$D_1=0$ and $l'_x=1$ reduce to $\Delta x' = 0$, and this sweeps up all
of the cases not covered by the other criteria.  Again, if this
computed value for $\Delta x'$ violates a limit in $x$, then we
reassign $x$ to the limit.

If both $x$ and $y$ are unknown, then we start by calculating the
values which minimize $r^{**}$.  We deal with limits in the identical
way as we did when we were minimizing $r^*$.  That is, we figure out
the optimal value of $x$ and $y$ together, but only assign the $y$
value.  Then we figure out the optimal value of $x$ given that $y$
value and the fixed value of $z$.  This way we can deal with the
limits in a sensible way.  The optimal value of $y$ is given by:
\[\Delta y' = \left\{ \begin{array}{ll}
	\frac{(k'_y k'_z a_k^2 + m'_y m'_z {a'}_m^2 + l'_y l'_z {a'}_l^2) \Delta z' - (l'_x m'_z {a'}_m^2 - m'_x l'_z {a'}_l^2) k'_{end}}{a_k^2 {k'}_z^2 + {a'}_m^2 {m'}_z^2 + {a'}_l^2 {l'}_z^2}
		& D_2 \equiv a_k^2 {k'}_z^2 + {a'}_m^2 {m'}_z^2 + {a'}_l^2 {l'}_z^2 \neq 0 \\
	k'_y k'_{end}					& D_2 = 0
	\end{array}
	\right. .\]
The first option is the formal minimization of $r^{**}$.  If $k'_z =
0$, then the center of the ellipsoid is in the $x,y$ plane.
Therefore, we try to place the object on that center (or as close as
we can get, given that $\Delta z'$ might not be zero).  If the
denominator is zero, then ${a'}_m^2=0$.  This is true because we
cannot have $k'_z=1, l'_z=0, m'_z=0$ due to the definition of our
system.  So, either ${a'}_l^2$ or ${a'}_m^2$ must be zero.  Either
way, ${a'}_m^2=0$.  If $k'_z \neq 0$, we must have ${a'}_k^2 = 0$ as
well.  Additionally, either ${a'}_l^2= 0$ (the endpoint is the only
hope) or $l'_z=0$ (${\bf\hat{l}}$ is in the $x,y$ plane, so the
endpoint is still the best choice, if we can get there).

As before, we then adjust $\Delta y'$ if it is outside the allowed
limits, then use our equations for only $\Delta x'$ missing to find
the optimal $x'$.  If the computed value of $x'$ is outside the
limits, then we set $x'$ to the limit, and recompute the optimal value
of $y'$.

We cannot think of a case in which the new position is in the interior
side of the endplane at $k_{end}$.  So, all we have to do is figure out
if the new position satisfies ${r'}^* \leq 1$ and is interior to the
$k$ limits on the outer surface of the ellipsoid.  If it is, then it
is consistent with being a star, otherwise it is not.

\clearpage


\clearpage

\clearpage

\clearpage

\clearpage

\clearpage

\clearpage

\clearpage

\clearpage

\begin{deluxetable}{lll}
\tabletypesize{\small}
\tablecaption{Quasar Target Selection Flags\label{tab:tab1}}
\tablewidth{0pt}
\tablehead{
\colhead{Flag Name} &  
\colhead{Hex Bit} &
\colhead{Description}
}
\startdata
{\tt TARGET\_QSO\_HIZ}             & 0x1          & high-redshift ($griz$-selected) QSO \\
{\tt TARGET\_QSO\_CAP}             & 0x2          & $ugri$-selected quasar at high Galactic latitude \\
{\tt TARGET\_QSO\_SKIRT\tablenotemark{a}}           & 0x4          & $ugri$-selected quasar at low Galactic latitude \\
{\tt TARGET\_QSO\_FIRST\_CAP}      & 0x8          & ``stellar'' FIRST source at high Galactic latitude \\
{\tt TARGET\_QSO\_FIRST\_SKIRT\tablenotemark{a}}    & 0x10         & ``stellar'' FIRST source at low Galactic latitude \\
{\tt TARGET\_QSO\_MAG\_OUTLIER\tablenotemark{b}}           & 0x2000000    & stellar outlier; too faint or too bright to target \\
{\tt TARGET\_QSO\_REJECT\tablenotemark{c}}          & 0x20000000   & object is in explicitly excluded region \\
\enddata
\tablenotetext{a}{At one point, we had considered separate selection
criteria in regions of high and low stellar density.  These regions
are referred to as the ``cap'' and ``skirt'' regions.  The ``cap'' and
``skirt'' divide the northern SDSS area according to the stellar
density.  The ``cap'' refers to the region of the Northern Galactic
Cap with less than 1500 stars per square degree according to the
Bahcall-Soneira model \citep{bs80}.  The ``skirt'' is then the region
outside of the ``cap'' that is still in the Survey area.  The original
intent for distinguishing between these two regions was to target
quasars to a higher density in the ``cap'' region where there is less
stellar contamination (see \S~\ref{sec:commissioning}), which would
result in a greater efficiency.  However, the selection efficiency was
found to be indistinguishable between cap and skirt, so targets are
selected exactly the same in both regions.  For historical reasons the
distinction was kept in the target selection flags.  Thus, the reader
should be aware that the selection of objects as QSO\_CAP and
QSO\_SKIRT (also QSO\_FIRST\_CAP and QSO\_FIRST\_SKIRT) is in fact
identical.  Furthermore, the final version of the code no longer uses
the QSO\_SKIRT and QSO\_FIRST\_SKIRT; we point out this distinction
only because some data that is already public uses this notation.}
\tablenotetext{b}{These objects are not targeted (unless another
``good'' quasar target flag is set).  These objects are flagged so
that it will be easier to explore slightly fainter (or brighter)
boundaries than are currently being used.  This information will be
useful for the Southern Survey and for any faint/bright quasar
follow-up projects.}
\tablenotetext{c}{Objects with these flags are not targeted (unless
they are also FIRST quasar targets).  They are in regions of
color-space that are explicitly rejected (WDs, WD+M, A-star), see
\S~\ref{sec:exclusion}.}
\end{deluxetable}

\begin{deluxetable}{lll}
\tabletypesize{\small}
\tablecaption{{\tt PHOTO} Flags Used By Quasar Target Selection\label{tab:tab2}}
\tablewidth{0pt}
\tablecolumns{3}
\tablehead{
\colhead{Flag Name} & 
\colhead{Hex Bit} &
\colhead{Description}
}
\startdata

\sidehead{\tt OBJC\_FLAGS}

{\tt BRIGHT}
& 0x2
& Object was detected in first, ``bright'' object-finding \\
& & step; generally brighter than $r^* = 17.5$ \\

{\tt EDGE}
& 0x4
& Object was too close to edge of frame \\

{\tt BLENDED}
& 0x8
& Object had multiple peaks detected within it; was \\
& & thus a candidate to be a deblending parent\\

{\tt CHILD}
& 0x10
& Object is the product of an attempt to deblend a {\tt BLENDED} object.\\

{\tt PEAKCENTER}
& 0x20
& Given center is position of peak pixel, rather than \\
& & based on the maximum-likelihood estimator \\

{\tt NODEBLEND}
& 0x40
& No deblending was attempted on this object, even though it is {\tt
BLENDED}.\\ 

{\tt SATUR}
& 0x40000
& The object contains one or more saturated pixels \\

{\tt NOTCHECKED}
& 0x80000
& There are pixels in the object which were not checked \\
& & to see if they included a local peak, such as cores of saturated stars\\

{\tt BINNED1}
& 0x10000000
& This object was detected in the $1\times 1$, unbinned image. \\

{\tt BINNED2}
& 0x20000000
& This object was detected in the $2\times 2$ binned image, \\
& & after unbinned detections are replaced by background.\\

{\tt BINNED4}
& 0x40000000
& This object was detected in the $4\times 4$ binned image. \\

\sidehead{\tt OBJC\_FLAGS2}

{\tt LOCAL\_EDGE}
& 0x80
& Center in at least one band is too close to an edge.\\

{\tt INTERP\_CENTER}
& 0x1000
& The object center is close to at least one interpolated pixel.\\

{\tt DEBLEND\_NOPEAK}
& 0x4000
& There was no detected peak within this child in at least one band. \\

{\tt NOTCHECKED\_CENTER}
& 0x4000000
& Center of the object is a {\tt NOTCHECKED} pixel \\

\enddata
\end{deluxetable}

\begin{deluxetable}{rrrrrrrrrrrr}
\tabletypesize{\small}
\tablewidth{0pt}
\tablecaption{$ugri$ Stellar Locus Points\label{tab:tab3}}
\tablehead{
\colhead{} &
\colhead{$k$} &
\colhead{$N$} &
\colhead{$u^*-g^*$} &
\colhead{$g^*-r^*$} &
\colhead{$r^*-i^*$} &
\colhead{$k_{u^*-g^*}$} &
\colhead{$k_{g^*-r^*}$} &
\colhead{$k_{r^*-i^*}$} &
\colhead{$a$} &
\colhead{$b$} &
\colhead{$\theta$}
}
\startdata
1 & 0.000 & 28955 & $0.855$ & $0.259$ & $0.094$ & $0.851$ & $0.492$ & $0.182$ & $0.282$ & $0.135$ & $-1.165$ \\
2 & 0.173 & 30041 & $1.002$ & $0.344$ & $0.126$ & $0.868$ & $0.467$ & $0.172$ & $0.247$ & $0.129$ & $-1.147$ \\
3 & 0.324 & 25980 & $1.136$ & $0.410$ & $0.150$ & $0.893$ & $0.422$ & $0.154$ & $0.221$ & $0.124$ & $-1.075$ \\
4 & 0.463 & 19313 & $1.262$ & $0.466$ & $0.170$ & $0.907$ & $0.396$ & $0.145$ & $0.219$ & $0.126$ & $-1.026$ \\
5 & 0.596 & 14991 & $1.382$ & $0.517$ & $0.189$ & $0.915$ & $0.379$ & $0.140$ & $0.216$ & $0.125$ & $-0.977$ \\
6 & 0.723 & 11847 & $1.499$ & $0.565$ & $0.207$ & $0.915$ & $0.379$ & $0.135$ & $0.217$ & $0.129$ & $-0.983$ \\
7 & 0.843 & 9305 & $1.609$ & $0.611$ & $0.223$ & $0.912$ & $0.387$ & $0.132$ & $0.224$ & $0.131$ & $-0.986$ \\
8 & 0.956 & 7609 & $1.712$ & $0.655$ & $0.238$ & $0.904$ & $0.402$ & $0.146$ & $0.227$ & $0.127$ & $-0.989$ \\
9 & 1.063 & 6313 & $1.808$ & $0.700$ & $0.255$ & $0.888$ & $0.430$ & $0.165$ & $0.233$ & $0.132$ & $-1.040$ \\
10 & 1.173 & 5790 & $1.904$ & $0.748$ & $0.273$ & $0.878$ & $0.449$ & $0.166$ & $0.248$ & $0.129$ & $-1.002$ \\
11 & 1.290 & 5540 & $2.007$ & $0.802$ & $0.293$ & $0.860$ & $0.478$ & $0.178$ & $0.266$ & $0.134$ & $-1.017$ \\
12 & 1.420 & 5544 & $2.117$ & $0.866$ & $0.317$ & $0.827$ & $0.521$ & $0.213$ & $0.278$ & $0.136$ & $-1.023$ \\
13 & 1.565 & 5738 & $2.234$ & $0.945$ & $0.351$ & $0.773$ & $0.573$ & $0.271$ & $0.309$ & $0.136$ & $-1.033$ \\
14 & 1.736 & 6466 & $2.361$ & $1.047$ & $0.403$ & $0.646$ & $0.650$ & $0.400$ & $0.382$ & $0.145$ & $-1.051$ \\
15 & 1.946 & 7661 & $2.478$ & $1.191$ & $0.502$ & $0.355$ & $0.634$ & $0.688$ & $0.463$ & $0.156$ & $-1.108$ \\
16 & 2.195 & 7110 & $2.518$ & $1.327$ & $0.707$ & $0.053$ & $0.278$ & $0.959$ & $0.484$ & $0.180$ & $-1.244$ \\
17 & 2.558 & 4092 & $2.510$ & $1.355$ & $1.068$ & $-0.022$ & $0.076$ & $0.997$ & $0.569$ & $0.212$ & $-1.669$ \\
\enddata
\tablecomments{For each locus point the columns are as follows: (1)
the number of the locus point; (2) the distance in magnitudes along
the stellar locus; (3) the number of sources associated with this
locus point, (4); (5); and (6); the $u^*-g^*$, $g^*-r^*$, and
$r^*-i^*$ position of the locus point; (7); (8); and (9); the
components of the $\hat{k}$ unit vector along the locus; (10) the
major axis; (11) the minor axis; and (12) the position angle (in
radians) of the ellipse fit to the cross section of the stellar locus.
See \citet{ny97} for more details.}
\end{deluxetable}

\begin{deluxetable}{rrrrrrrrrrrr}
\tabletypesize{\small}
\tablewidth{0pt}
\tablecaption{$griz$ Stellar Locus Points\label{tab:tab4}}
\tablehead{
\colhead{} &
\colhead{$k$} &
\colhead{$N$} &
\colhead{$g^*-r^*$} &
\colhead{$r^*-i^*$} &
\colhead{$i^*-z^*$} &
\colhead{$k_{g^*-r^*}$} &
\colhead{$k_{r^*-i^*}$} &
\colhead{$k_{i^*-z^*}$} &
\colhead{$a_l$} &
\colhead{$a_m$} &
\colhead{$\theta$}
}
\startdata
1 & 0.000 & 14876 & $0.204$ & $0.071$ & $0.003$ & $0.911$ & $0.351$ & $0.218$ & $0.207$ & $0.146$ & $0.067$ \\
2 & 0.110 & 29513 & $0.304$ & $0.110$ & $0.027$ & $0.916$ & $0.339$ & $0.213$ & $0.165$ & $0.126$ & $-2.907$ \\
3 & 0.194 & 31790 & $0.382$ & $0.137$ & $0.044$ & $0.910$ & $0.340$ & $0.237$ & $0.154$ & $0.128$ & $-2.990$ \\
4 & 0.274 & 26152 & $0.454$ & $0.166$ & $0.066$ & $0.895$ & $0.356$ & $0.268$ & $0.159$ & $0.134$ & $-0.029$ \\
5 & 0.354 & 21239 & $0.525$ & $0.194$ & $0.087$ & $0.905$ & $0.342$ & $0.253$ & $0.164$ & $0.133$ & $-0.194$ \\
6 & 0.429 & 15402 & $0.594$ & $0.219$ & $0.105$ & $0.913$ & $0.325$ & $0.246$ & $0.162$ & $0.133$ & $-0.315$ \\
7 & 0.501 & 10989 & $0.659$ & $0.242$ & $0.123$ & $0.911$ & $0.330$ & $0.246$ & $0.151$ & $0.133$ & $-0.610$ \\
8 & 0.571 & 8171 & $0.723$ & $0.265$ & $0.140$ & $0.915$ & $0.332$ & $0.231$ & $0.150$ & $0.127$ & $-0.858$ \\
9 & 0.641 & 6581 & $0.787$ & $0.288$ & $0.155$ & $0.916$ & $0.335$ & $0.220$ & $0.153$ & $0.128$ & $-0.935$ \\
10 & 0.713 & 5510 & $0.853$ & $0.313$ & $0.171$ & $0.906$ & $0.360$ & $0.222$ & $0.157$ & $0.124$ & $-0.917$ \\
11 & 0.789 & 4817 & $0.922$ & $0.341$ & $0.188$ & $0.897$ & $0.380$ & $0.227$ & $0.160$ & $0.125$ & $-0.921$ \\
12 & 0.867 & 4215 & $0.991$ & $0.371$ & $0.206$ & $0.876$ & $0.420$ & $0.237$ & $0.163$ & $0.123$ & $-0.898$ \\
13 & 0.951 & 3704 & $1.063$ & $0.409$ & $0.227$ & $0.832$ & $0.485$ & $0.267$ & $0.171$ & $0.123$ & $-0.949$ \\
14 & 1.036 & 3261 & $1.132$ & $0.454$ & $0.251$ & $0.778$ & $0.551$ & $0.301$ & $0.175$ & $0.125$ & $-1.033$ \\
15 & 1.129 & 3272 & $1.202$ & $0.507$ & $0.280$ & $0.704$ & $0.623$ & $0.342$ & $0.178$ & $0.127$ & $-1.127$ \\
16 & 1.222 & 3136 & $1.262$ & $0.569$ & $0.314$ & $0.566$ & $0.729$ & $0.386$ & $0.185$ & $0.135$ & $-1.323$ \\
17 & 1.327 & 3023 & $1.313$ & $0.651$ & $0.356$ & $0.362$ & $0.832$ & $0.420$ & $0.193$ & $0.129$ & $-1.423$ \\
18 & 1.446 & 2521 & $1.343$ & $0.754$ & $0.408$ & $0.168$ & $0.885$ & $0.434$ & $0.213$ & $0.131$ & $-1.554$ \\
19 & 1.579 & 1917 & $1.355$ & $0.874$ & $0.465$ & $0.035$ & $0.900$ & $0.435$ & $0.246$ & $0.137$ & $-1.628$ \\
20 & 1.715 & 1432 & $1.352$ & $0.996$ & $0.525$ & $-0.031$ & $0.899$ & $0.438$ & $0.250$ & $0.135$ & $-1.667$ \\
21 & 1.849 & 1058 & $1.347$ & $1.116$ & $0.583$ & $-0.008$ & $0.895$ & $0.446$ & $0.265$ & $0.133$ & $-1.647$ \\
22 & 1.988 & 755 & $1.350$ & $1.240$ & $0.646$ & $0.047$ & $0.879$ & $0.475$ & $0.246$ & $0.121$ & $-1.652$ \\
23 & 2.155 & 442 & $1.361$ & $1.385$ & $0.729$ & $0.067$ & $0.868$ & $0.493$ & $0.300$ & $0.139$ & $-1.530$ \\
\enddata
\tablecomments{Except for the colors, the column descriptions are the same as for Table~\ref{tab:tab3}}
\end{deluxetable}

\begin{deluxetable}{lrrrrr}
\tabletypesize{\small}
\tablewidth{0pt}
\tablecaption{Known Quasar Completeness\label{tab:tab5}}
\tablehead{
\colhead{} &
\colhead{All\tablenotemark{a}} &
\colhead{NED\tablenotemark{b}} &
\colhead{FIRST\tablenotemark{c}} &
\colhead{SDSS\tablenotemark{d}} &
\colhead{High-$z$\tablenotemark{e}}
}
\startdata
Known\tablenotemark{f} & 2096 & 682 & 66 & 1462 & 72 \\
Found\tablenotemark{g} & 1943 & 586 & 62	& 1404 & 59 \\
Bright\tablenotemark{h} & 1540 & 394 & 58 & 1210 & 44 \\
Target\tablenotemark{i} & 1456 & 369 & 56 & 1154 & 39 \\
\cutinhead{}
QSO\_GOOD\tablenotemark{j} & 1462 & 369	& 56 & 1159 & 40 \\
QSO\_HIZ\tablenotemark{k} & 414	& 114 & 16 & 295 & 40 \\
QSO\_LOWZ\tablenotemark{l} & 1375  & 360 & 48 & 1124 & 0 \\
QSO\_FIRST\tablenotemark{m} & 162 & 54 & 54 & 93 & 1 \\
QSO\_REJECT\tablenotemark{n} & 7 & 2 & 0 & 5 & 0 \\
QSO\_MAG\_OUTLIER\tablenotemark{o} & 332 & 158	& 3 & 159 & 14 \\
FIRST only\tablenotemark{p} & 20 & 6 & 8 & 10  & 0 \\
Color select FIRST\tablenotemark{q} & 159	& 49 & 48 & 95 & 4 \\
\enddata
\tablenotetext{a}{All four of the individual categories combined together.}
\tablenotetext{b}{NED quasars as of 2000 June 22.}
\tablenotetext{c}{FIRST quasars.}
\tablenotetext{d}{SDSS quasars from the first 66 plates of data.}
\tablenotetext{e}{SDSS quasars from high-$z$ follow-up searches prior to 2000 October 3 \citep{fsr+01}.}
\tablenotetext{f}{Known quasars in the area covered by the quasar target selection testbed.}
\tablenotetext{g}{Known quasars that had matches to SDSS objects within $3\arcsec$.}
\tablenotetext{h}{The set of ``found'' quasars that should have been bright enough to be targeted.}
\tablenotetext{i}{The number of ``bright'' quasars that were actually targeted.}
\tablenotetext{j}{A meta class including QSO\_HIZ, QSO\_LOWZ, and QSO\_FIRST, i.e. objects selected as ``good'' quasar candidates.  Note that the numbers in these columns need not equal the numbers in the ``Target'' columns because of the way that the two classes are defined.}   
\tablenotetext{k}{Objects flagged as QSO\_HIZ.}
\tablenotetext{l}{A meta class including QSO\_CAP and QSO\_SKIRT.}
\tablenotetext{m}{A meta class including QSO\_FIRST\_CAP and QSO\_FIRST\_SKIRT.}
\tablenotetext{n}{Objects explicitly rejected by the exclusion regions (\S~\ref{sec:exclusion}).}
\tablenotetext{o}{Objects that are too faint or too bright to be targeted.}
\tablenotetext{p}{Objects selected only because they are FIRST sources.}
\tablenotetext{q}{Color-selected objects that are also FIRST sources.}
\end{deluxetable}

\begin{deluxetable}{lllllllllll}
\tabletypesize{\small}
\tablewidth{0pt}
\tablecaption{Simulated Quasar Completeness\label{tab:tab6}}
\tablehead{
\colhead{} & \multicolumn{10}{c}{Apparent Magnitude\tablenotemark{a}} \\
\cline{2-11} \\
\colhead{Redshift} &
\colhead{16.0} &
\colhead{16.5} &
\colhead{17.0} &
\colhead{17.5} &
\colhead{18.0} &
\colhead{18.5} &
\colhead{19.0} &
\colhead{19.5} &
\colhead{20.0} &
\colhead{16.0--19.0\tablenotemark{b}}
}
\startdata
0.0--0.5 & 1.000 & 1.000 & 1.000 & 1.000 & 1.000 & 1.000 & 1.000 & 0.000 & 0.000 & 1.000 \\ 
0.5--1.0 & 1.000 & 1.000 & 1.000 & 1.000 & 1.000 & 1.000 & 1.000 & 0.000 & 0.000 & 1.000 \\ 
1.0--1.5 & 1.000 & 1.000 & 1.000 & 1.000 & 1.000 & 1.000 & 1.000 & 0.000 & 0.000 & 1.000 \\ 
1.5--2.0 & 1.000 & 1.000 & 1.000 & 1.000 & 1.000 & 1.000 & 1.000 & 0.000 & 0.000 & 1.000 \\ 
2.0--2.5 & 0.966 & 0.964 & 0.966 & 0.972 & 0.970 & 0.968 & 0.934 & 0.000 & 0.000 & 0.963 \\ 
2.5--3.0 & 0.598 & 0.596 & 0.594 & 0.586 & 0.570 & 0.546 & 0.514 & 0.114 & 0.100 & 0.572 \\ 
3.0--3.5 & 0.914 & 0.916 & 0.910 & 0.906 & 0.902 & 0.896 & 0.852 & 0.742 & 0.642 & 0.899 \\ 
3.5--4.0 & 0.998 & 0.998 & 0.998 & 0.998 & 0.998 & 0.998 & 0.996 & 0.984 & 0.970 & 0.998 \\ 
4.0--4.5 & 1.000 & 1.000 & 1.000 & 1.000 & 1.000 & 1.000 & 1.000 & 1.000 & 0.992 & 1.000 \\ 
4.5--5.0 & 1.000 & 1.000 & 1.000 & 1.000 & 1.000 & 1.000 & 1.000 & 1.000 & 0.992 & 1.000 \\ 
5.0--5.3 & 1.000 & 1.000 & 1.000 & 1.000 & 1.000 & 1.000 & 1.000 & 1.000 & 0.997 & 1.000 \\ 
0.0--5.3 & 0.951 & 0.950 & 0.950 & 0.949 & 0.947 & 0.944 & 0.934 & 0.419 & 0.405 & 0.946 \\ 
5.4 & 1.000 & 1.000 & 1.000 & 1.000 & 1.000 & 1.000 & 1.000 & 0.860 & 0.316 & 1.000 \\ 
5.5 & 1.000 & 1.000 & 1.000 & 1.000 & 1.000 & 1.000 & 0.980 & 0.918 & 0.143 & 0.997 \\ 
5.6 & 1.000 & 1.000 & 1.000 & 0.980 & 0.980 & 0.900 & 0.780 & 0.608 & 0.021 & 0.949 \\ 
5.7 & 1.000 & 1.000 & 1.000 & 0.980 & 0.760 & 0.620 & 0.480 & 0.000 & 0.000 & 0.834 \\ 
5.8 & 1.000 & 0.980 & 0.900 & 0.720 & 0.500 & 0.020 & 0.000 & 0.000 & 0.000 & 0.589 \\ 
5.9 & 1.000 & 1.000 & 0.940 & 0.660 & 0.060 & 0.000 & 0.000 & 0.000 & 0.000 & 0.523 \\ 
6.0 & 0.920 & 0.780 & 0.660 & 0.180 & 0.000 & 0.000 & 0.000 & 0.000 & 0.000 & 0.363 \\ 
6.1 & 0.900 & 0.620 & 0.400 & 0.060 & 0.000 & 0.000 & 0.000 & 0.000 & 0.000 & 0.283 \\ 
6.2 & 0.640 & 0.240 & 0.100 & 0.000 & 0.000 & 0.000 & 0.000 & 0.000 & 0.000 & 0.140 \\ 
6.3 & 0.100 & 0.040 & 0.000 & 0.000 & 0.000 & 0.000 & 0.000 & 0.000 & 0.000 & 0.020 \\ 
\enddata
\tablenotetext{a}{For $z\le5.3$ columns 2 through 11 refer to the
$i$-band magnitude, whereas for $z\ge5.4$ columns 2 through 11 refer
to the $z$-band magnitude.}
\tablenotetext{b}{The total values for completeness given in the last row
and column of the two tables average over a uniform grid in apparent
magnitude and redshift.  True quasars are not uniformly distributed in
these quantities; these value do not represent the true completeness
for a realistic distribution of quasars.}
\end{deluxetable}

\begin{deluxetable}{lrrrrrrrr}
\tabletypesize{\small}
\tablewidth{0pt}
\tablecaption{Quasar Target Selection Efficiency\label{tab:tab7}}
\tablehead{
\colhead{} &
\colhead{N} &
\colhead{N w/ Spec.} &
\colhead{N} &
\colhead{\%} &
\colhead{N} &
\colhead{\%} &
\colhead{N} &
\colhead{\%} \\
\colhead{} &
\colhead{QSO Cand.} &
\colhead{QSO Cand.} &
\colhead{QSO/AGN} &
\colhead{QSO/AGN} &
\colhead{Gal.} &
\colhead{Gal.} &
\colhead{Star} &
\colhead{Star} 
}
\startdata
All & 1872 & 1687 & 1113 & 66.0 & 266 & 15.8 & 294 & 17.4 \\
Low-$z$ & 1392 & 1339 & 1005 & 75.0 & 233 & 17.4 & 98 & 7.3 \\
Low-$z$ Only & 1155 & 1110 & 789 & 71.1 & 230 & 20.7 & 89 & 8.0 \\
High-$z$ & 663 & 529 & 288 & 54.4 & 35 & 6.6 & 194 & 36.7 \\
High-$z$ Only & 426 & 300 & 72 & 24.0 & 32 & 10.7 & 185 & 61.7\\
FIRST & 74 & 69 & 60 & 87.0 & 0 & 0.0 & 9 & 13.0 \\
FIRST Only & 22 & 19 & 10 & 52.6 & 0 & 0.0 & 9 & 47.3 \\
\enddata
\end{deluxetable}

\begin{deluxetable}{lccccc}
\tabletypesize{\small}
\tablewidth{0pt}
\tablecaption{Quasar Target Selection Density \label{tab:tab8}}
\tablehead{
\colhead{} &
\colhead{Low-$z$} &
\colhead{High-$z$} &
\colhead{FIRST} &
\colhead{All} &
\colhead{All ($i^*<19.1$)}
}
\startdata
N & 5813 & 3454 & 302 & 8330 & 6540 \\
Density\tablenotemark{a} & 13.02 & 7.74 & 0.68\tablenotemark{b} & 18.66 & 14.65\\
Error\tablenotemark{c} & 1.81 & 2.57 & 0.42 & 2.91 & 1.65\\
$i^*_{max}$\tablenotemark{d} & 19.1 & 20.2 & 19.1 & \nodata & 19.1\\
\enddata
\tablenotetext{a}{Quasars per square degree; the area covered was 446.3952 square degrees of sky.}
\tablenotetext{b}{The FIRST density (and therefore the overall density) should be taken as a lower limit since not all of the runs in the testbed were matched to FIRST sources, but the density is computing under the assumption that all areas have been covered.  The true density is closer to one quasar per square degree.}
\tablenotetext{c}{$1\sigma$ error in the mean density, based on the deviations between the camera columns in each of the runs in the testbed.}
\tablenotetext{d}{$i^*$ magnitude limit to which objects are selected
in each of the categories.}
\end{deluxetable}

\end{document}